\documentclass[11pt,a4paper]{article}
\pdfoutput=1
\usepackage{jcappub}
\usepackage{lineno}
\usepackage{epstopdf}
\usepackage{rotating}
\usepackage{float}
\usepackage{units}
\newcommand\ion[2]{#1$\;${\small \uppercase\expandafter{\romannumeral #2}}}%  
\newcommand\ionalt[2]{#1$\;${\scriptsize \uppercase\expandafter{\romannumeral #2}}}%  

\newcommand{\mpch}{h^{-1}{\rm Mpc}}

\newcommand{\vt}{\mathbf{t}}

\newcommand{\aiso}{\alpha_{\rm iso}}
\newcommand{\apar}{\alpha_\parallel}
\newcommand{\aperp}{\alpha_\perp}

\newcommand{\lyaf}{Lyman-$\alpha$ forest\ }
\newcommand{\lya}{Lyman-$\alpha$\  }
\newcommand{\vk}{\mathbf{k}}
\newcommand{\vx}{\mathbf{x}}
\newcommand{\lr}{\lambda_{{\rm rest}}}
\newcommand{\Tr}{\mathrm{Tr}}

\newcommand{\caiso}{$100\times(\aiso-1)=-1.6^{+2.0\ +4.3\ +7.4}_{-2.0\ -4.1\ -6.8}$\,(stat.) $\pm1.0$\,(syst.)}
\newcommand{\capar}{$100\times(\apar-1)=-1.3^{+3.5\ +7.6\  +12.3}_{-3.3\ -6.7\ -10.2}$\,(stat.) $\pm2.0$\,(syst.)}
\newcommand{\caper}{$100\times(\aperp-1)=-2.2^{+7.4\ +17}_{-7.1\ -15}$\,(stat.) $\pm3.0$\,(syst.)}

\title{Measurement of Baryon Acoustic Oscillations in the
  Lyman-$\alpha$ Forest Fluctuations in BOSS Data Release 9}

\author[a]{An\v{z}e Slosar,}
\author[b]{Vid Ir\v{s}i\v{c},}
\author[c]{David Kirkby,}
\author[d]{Stephen~Bailey,}
\author[e]{Nicol\'as G. Busca,}
\author[f]{Timoth\'ee Delubac,}
\author[f]{James Rich,}
\author[e]{\'{E}ric Aubourg,}
\author[e]{Julian E. Bautista,}
\author[g,d]{Vaishali Bhardwaj,}
\author[c]{Michael~Blomqvist,}
\author[h]{Adam S. Bolton,}
\author[i]{Jo Bovy,}
\author[h]{Joel Brownstein,}
\author[d]{Bill Carithers,}
\author[j]{Rupert A.C. Croft,}
\author[h]{Kyle S. Dawson,}
\author[k,d]{Andreu Font-Ribera,}
\author[f]{J.-M. Le Goff,}
\author[j]{Shirley Ho,}
\author[l]{Klaus Honscheid,}
\author[m]{Khee-Gan Lee,}
\author[c]{Daniel~Margala,}
\author[d]{Patrick McDonald,}
\author[n]{Bumbarija Medolin,}
\author[o,p]{Jordi Miralda-Escud\'{e},}
\author[q]{Adam D. Myers,}
\author[r]{Robert C. Nichol,}
\author[s]{Pasquier Noterdaeme,}
\author[f]{Nathalie Palanque-Delabrouille,}
\author[s,t]{Isabelle P\^aris,}
\author[s]{Patrick Petitjean,}
\author[r]{Matthew M. Pieri,}
\author[n]{Yodovina Pi\v{s}kur,}
\author[d]{Natalie A. Roe,}
\author[d]{Nicholas P. Ross,}
\author[f]{Graziano Rossi,}
\author[d]{David J. Schlegel,}
\author[u,v]{Donald P. Schneider,}
\author[d]{Nao Suzuki,}
\author[a]{Erin S. Sheldon,}
\author[d]{Uro\v{s} Seljak,}
\author[w,x]{Matteo Viel,}
\author[y]{David H. Weinberg,}
\author[f]{Christophe Y\`eche}

\affiliation[a]{Brookhaven National Laboratory, Blgd 510, Upton NY 11375, USA}
\affiliation[b]{Faculty of Mathematics and Physics, University of Ljubljana, Jadranska 19, 1000 Ljubljana, Slovenia}
\affiliation[c]{Department of Physics and Astronomy, University of California, Irvine, CA 92697, USA}
\affiliation[d]{Lawrence Berkeley National Laboratory, 1 Cyclotron  Road, Berkeley, CA 94720, USA}
\affiliation[e]{APC, Universit\'{e} Paris Diderot-Paris 7, CNRS/IN2P3, CEA,  Observatoire de Paris, 10, rueA. Domon \& L. Duquet,  Paris, France}
\affiliation[f]{CEA, Centre de Saclay, IRFU,  F-91191 Gif-sur-Yvette, France}
\affiliation[g]{Department of Astronomy, University of Washington,  Box 351580, Seattle, WA 09195, USA}
\affiliation[h]{Department of Physics and Astronomy, University of Utah, 115 S 1400 E, Salt Lake City, UT 84112, USA}
\affiliation[i]{Institute for Advanced Study, Einstein Drive, Princeton, NJ 08540, USA}
\affiliation[j]{Bruce and Astrid McWilliams Center for Cosmology, Carnegie Mellon University, Pittsburgh, PA 15213, USA}
\affiliation[k]{Institute of Theoretical Physics, University of Zurich, 8057 Zurich, Switzerland }
\affiliation[l]{Department of Physics and Center for Cosmology and Astro-Particle Physics, Ohio State University, Columbus, OH 43210, USA}
\affiliation[m]{Max-Planck-Institut f\"ur Astronomie, K\"onigstuhl 17, D69117 Heidelberg, Germany}
\affiliation[n]{7020 108th St, Forest Hills, NY 11375, USA}
\affiliation[o]{Instituci\'{o} Catalana de Recerca i Estudis  Avan\c{c}ats, Barcelona, Catalonia}
\affiliation[p]{Institut de Ci\`{e}ncies del Cosmos, Universitat de Barcelona/IEEC, Barcelona 08028, Catalonia}
\affiliation[q]{Department of Physics and Astronomy, University of Wyoming, Laramie, WY 82071, USA}
\affiliation[r]{Institute of Cosmology and Gravitation, Dennis Sciama Building, University of Portsmouth, Portsmouth, PO1 3FX, UK}
\affiliation[s]{Universit\'e Paris 6 et CNRS, Institut d'Astrophysique de Paris, 98bis blvd. Arago, 75014 Paris, France}
\affiliation[t]{Departamento de Astronom\'ia, Universidad de Chile, Casilla 36-D, Santiago, Chile}
\affiliation[u]{Department of Astronomy and Astrophysics, The Pennsylvania State University, University Park, PA 16802, USA}
\affiliation[v]{Institute for Gravitation and the Cosmos, The Pennsylvania State University, University Park, PA 16802, USA}
\affiliation[w]{INAF, Osservatorio Astronomico di Trieste, Via G. B. Tiepolo 11, 34131 Trieste, Italy}
\affiliation[x]{INFN/National Institute for Nuclear Physics, Via Valerio 2, I-34127 Trieste, Italy.}
\affiliation[y]{Department of Astronomy, Ohio State University, 140 West 18th Avenue, Columbus, OH 43210, USA}

\emailAdd{anze@bnl.gov}

\abstract{ We use the Baryon Oscillation Spectroscopic Survey (BOSS) Data
  Release 9 (DR9) to detect and measure the position of the Baryonic
  Acoustic Oscillation (BAO) feature in the three-dimensional correlation
  function in the \lyaf flux fluctuations at a redshift $z_{\rm
    eff}=2.4$. The feature is clearly detected at significance between
  3 and 5 sigma (depending on the broadband model and method of error
  covariance matrix estimation) and is consistent with predictions of
  the standard $\Lambda$CDM model.  We assess the biases in our
  method, stability of the error covariance matrix and possible
  systematic effects. We fit the resulting correlation function with
  several models that decouple the broadband and acoustic scale
  information. For an isotropic dilation factor, we measure \caiso\
  (multiple statistical errors denote 1,2 and 3 sigma confidence
  limits) with respect to the acoustic scale in the fiducial
  cosmological model (flat $\Lambda$CDM with $\Omega_m=0.27$,
  $h=0.7$). When fitting separately for the radial and transversal
  dilation factors we find marginalised constraints \capar\ and
  \caper.  The dilation factor measurements are significantly
  correlated with cross-correlation coefficient of $\sim -0.55$.
  Errors become significantly non-Gaussian for deviations over $3$
  standard deviations from best fit value.  Because of the data cuts
  and analysis method, these measurements give tighter constraints
  than a previous BAO analysis of the BOSS DR9 \lyaf sample, providing an
  important consistency test of the standard cosmological model in a
  new redshift regime. }


\begin{document}
% \linenumbers
\maketitle

\section{Introduction}
Discovery of the accelerated expansion of the Universe at the end of
20$^{\rm th}$ century was one of the largest surprises in cosmology
\cite{RIFIET98,PEADET99}. Since then, much effort has gone into
measuring the properties of the dark energy that is believed to drive
the acceleration of the Universe using a variety of techniques (see
extensive review \cite{2012arXiv1201.2434W} and references therein).
Measurements that directly probe the expansion history of the Universe
reveal how the energy density of its constitutive components changes
with expansion and thus allow inferences to be made about their
nature. The two important techniques in this field are the type Ia
supernovae (SN Ia) and baryonic acoustic oscillations (BAO)
\cite{SEEI03,2005ApJ...633..560E,SEEI05,2012arXiv1202.0090P,2012arXiv1203.6594A,2012MNRAS.425..405B}. The
former rely on the fact that SNIa are standarizable candles. Measuring
supernovae fluxes and redshifts thus provide information about the
luminosity distance as a function of redshift. The BAO technique,
which is the focus of this paper, relies on the fact that the baryonic
features in the correlation properties of tracers of the large scale
structure of the Universe can act as a comoving standard ruler.

BAO are acoustic oscillations in the primordial plasma before
decoupling of the baryonic matter and radiation. These oscillations
imprinted a characteristic scale into the correlation properties of dark
matter, which exhibit themselves as a distinct oscillatory pattern in
the power spectrum of fluctuations or equivalently a characteristic
peak in the correlation function. The scales involved are large ($\sim
150$ Mpc) so the feature remains in the weakly non-linear regime
even today.  Any tracer of the large-scale structure can measure this
feature and use it as a standard ruler to infer the distance to the
tracer in question (when the ruler is used transversely) or the expansion
rate (when the ruler is used radially).

The main attraction of BAO technique is that one is measuring the
position of a peak and that many systematic effects are blind about
the preferred scale and thus influence the two-point function only in
broadband sense. Such systematic effects can be efficiently
marginalized out with little or no signal loss. The caveat in the case
of the \lyaf analysis is that systematic effects that appear at
certain pairs of wavelengths can produce a peak-like feature in the
correlation function in the radial direction.

Traditionally, the tracer used to measure BAO consisted of galaxies,
and measuring BAO with galaxies is a mature method
\cite{2012arXiv1203.6594A,2012MNRAS.425..405B}.  However, the current
measurements of the BAO with galaxies are restricted to redshifts
$z<1$, due to difficulty in measuring redshifts of sufficient numbers
of objects efficiently. Even the proposed future BAO projects such as
BigBOSS \cite{2011arXiv1106.1706S} will measure the BAO only to
redshifts of less than $z\sim 1.5-2$ using galaxies as a tracer of
cosmic structure. However, there is strong scientific motivation for
measuring the expansion history of the Universe at higher
redshifts. Dark energy suffers from many tuning problems that can be
alleviated by introducing an early dark energy component
\cite{2012ApJ...749L...9R, 2010ApJ...714.1460A, 2009JCAP...04..002X,
  2008JCAP...06..004L, 2006JCAP...06..026D,
  2001PhRvD..64l3520Dt}. Such a component can only be measured by a
technique sensitive to the expansion history at high redshift. In this
paper we present such a measurement using the \lyaf as a tracer of
dark-matter fluctuations at $z\sim 2.4$.

This ``\lyaf'' denotes the absorption features in the spectra of distant
quasars blue-ward of the \lya emission line
\cite{1971ApJ...164L..73L}. These features arise because the light
from a quasar is resonantly scattered by the presence of neutral
hydrogen in the intergalactic medium. Since the quasar light is
constantly red-shifting, hydrogen at different redshifts absorb at
different observed wavelength in the quasar spectrum. The amount of
scattering reflects the local density of neutral hydrogen. Neutral
hydrogen is believed to be in photo-ionization equilibrium and
therefore the transmitted flux fraction at position $\vx$ is given by
\cite{1997MNRAS.292...27H}

\begin{equation}
  F (\vx) = \exp [-\tau (\vx)] \sim C \exp\left[-A(1+\delta_b(\vx))^p\right],
\end{equation}
where $\tau$ is optical depth for \lya absorption at position
$\vx$. This optical depth scales roughly as a power law in the baryon
over-density $\delta_b$. The coefficient $p$ would be two for an
idealized two-body process, but is in practice closer to $\sim 1.8$
due to the temperature dependence of the recombination rate.  The \lyaf thus
measures some highly non-linear but at the same time very local
transformation of the density field. Any such tracer will, on
sufficiently large scales, trace the underlying dark matter field in
the observed redshift-space as
\begin{equation}
  \delta_F(\vk) = (b + b_v f \mu^2) \delta_m(\vk),
  \label{eq:3}
\end{equation}
where $\delta_F(\vk)$ are the flux and matter over-densities in
Fourier space respectively and $\delta_m$ is the matter over-density
in the Fourier space.  The quantity $\mu=k_\parallel/|\vk|$ is the
cosine of the angle of a $\vk$ vector with respect to the line of
sight and the parameter $f=d \ln g / d \ln a$ is the logarithmic
growth factor.  In the peak-background split picture, the two bias
parameters can be derived as a response of the smoothed transmitted
flux fraction field $F_s$ to a large scale change in either
overdensity $\delta_l$ or peculiar velocity gradient
$\eta_l=-H^{-1}dv_\parallel/dr_\parallel$ \cite{2003ApJ...585...34M}:
\begin{eqnarray}
  b  &=&\frac{\partial \log F_s}{\partial \delta_l},\\
  b_v&=&\frac{\partial \log F_s}{\partial \eta_l}.
\end{eqnarray}

Both bias parameters ($b$ and $b_v$) can in principle, be determined
from numerical simulations of the intergalactic medium, but they can
also be determined from the data as free parameters to be fit.  The
dimensionless redshift-space distortion parameter $\beta= f b_v/b$
determines the strength of distortions \cite{1987MNRAS.227....1K}.
Conservation of the number of galaxies requires that $b_v=1$ in that
case and hence $\beta$ is completely specified by the value of $f$
and density bias $b$.  Unfortunately, there is no such conservation
law for transmitted flux fraction and hence the two parameters are
truly independent for the \lyaf 2-point function.  In the rest of this
paper we will work with density bias parameter $b$ and redshift-space
distortion parameter $\beta$ and treat them as free parameters to be
constrained by the data.

Numerical simulations have shown that for plausible scenarios, the
\lyaf is well in the linear biasing regime of equation (\ref{eq:3}) at
scales relevant for BAO. Fluctuations in the radiation field can, if
sufficiently large, invalidate the equation (\ref{eq:3}), but it has
been shown that these variations are unlikely to produce a sharp
feature that could be mistaken for the acoustic feature
\cite{2009JCAP...10..019S,2010ApJ...713..383W}.  Therefore, measuring
three-dimensional (3D) correlations in the flux fluctuations of the \lyaf provides an
accurate method for detecting BAO correlations
\cite{MCDON03,2003dmci.confE..18W,2007PhRvD..76f3009M}.

Using the \lyaf to measure the 3D structure of the
Universe became possible with the advent of the Baryonic Oscillation
Spectroscopic Survey (BOSS, \cite{2012arXiv1208.0022D,Bolton2012}), a
part of the Sloan Digital Sky Survey III (SDSS-III) experiment
\cite{2006AJ....131.2332G,1998AJ....116.3040G,2000AJ....120.1579Y,1996AJ....111.1748F,
  2012arXiv1208.2233S,2011AJ....142...72E}. This survey was the first
to have a sufficiently high density of quasars to measure correlations
on truly cosmological scales \cite{2011JCAP...09..001S}. This paper
was also first to point out difficulties in measuring the large-scale
fluctuations in the \lyaf data; and many design decisions in this
paper were chosen to alleviate them.  The BAO in the \lyaf was first
detected in \cite{FPG} using the same dataset as this work. They found
a dilation scale $\aiso=1.01\pm0.03$. Both works were done within the
same working group in the BOSS collaboration, but use different and an
independently developed pipelines.  We assess these differences more
thoroughly in Section \ref{sec:comparison-busca-et}.  We refer the
reader to these publications for a more thorough review of the \lyaf
physics and the BOSS experiment.

This paper is focused on the methodology and results of measuring the 3D
correlation function from the flux transmission fluctuations. The
companion paper \cite{peanutboy} focuses on the methodology of
constraining the BAO parameters. The final results on the BAO
parameters are presented in this paper.

This paper is dedicated purely to detecting the BAO feature in the
correlation function of \lyaf fluctuations as measured by BOSS Data
Release 9 \cite{2012arXiv1207.7137S}.  We do not attempt to provide
robust measurements of bias and $\beta$ parameters; we therefore adopt
aggressive inclusion of data that is tuned to maximize the likelihood
of a significant BAO peak detection. We stress that in the early
stages of analysis, the peak between 100-130$\mpch$ was blinded and
these cuts were decided on and fixed before ``opening the box''. No
cuts were changed after that, but the data analysis code has been
improved.

The paper is structured as follows. In section \ref{sec:data} we
describe our data samples and basic selection criteria. In section
\ref{sec:data-analysis} we explain the complicated process of
measuring the 3D correlation function using an optimal estimator, its
errors and biases, while presenting intermediate results.  Section
\ref{sec:results-bary-acoust} discusses application of the method to
the real and synthetic data and discusses potential errors. Finally,
we convert our many noisy measurements into a plot of the BAO bump in
Section~\ref{sec:visualizing-peak}.  We conclude in Section
\ref{sec:conclusions}. The mathematical details of optimal estimators
are elaborated in Appendices.

\section{Instrument, Data \& Synthetic data}
\label{sec:data}

\subsection{DR9Q Sample}
In this work we use BOSS quasars from the Data Release 9 (DR9,
\cite{2012arXiv1207.7137S}) sample in order to facilitate
reproducibility by outside investigators. The quasar target selection
for the DR9 sample of BOSS observations is described in detail in
\cite{2012ApJS..199....3R} and \cite{2011ApJ...729..141B}. A similar
dataset is released in \cite{2012arXiv1211.5146L} including continua
for normalization and default recommendations for data masking.

The main approach in this work has been to aggressively use as much
data as possible, to maximize the likelihood of a significant
detection of the BAO peak and the accuracy of the determination of its
position. The argument that we are making, and which underpins the
robustness of many BAO measurements, is that systematic effects that
affect the measured two-point statistics will in general produce a
broadband contribution and thus not affect the position of the peak
itself, since broadband contributions are marginalized out in the
process of fitting the BAO position. At the same time, however, adding
poorly understood data might lead to an increase in the noise that
might outweigh the benefit of additional information. We settled on a
set of cuts described below.

\begin{figure}[h!]
  \centering
  \includegraphics[width=\linewidth]{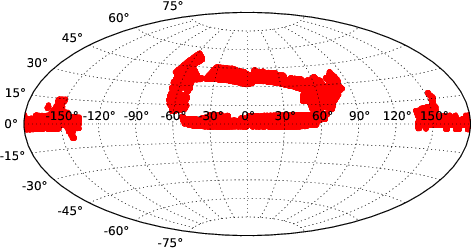}
  \caption{\label{fig:proj} Distribution of 58,227 quasars used in this work
    on the sky in J2000 equatorial coordinates, shifted by $180\deg$
    to more clearly show the main survey area. }
\end{figure}

We select quasars from the DR9Q version of the Value Added Catalog
(VAC) of the French Participation Group (FPG) visual inspections of
the BOSS spectra \cite{2012arXiv1210.5166P}. The position of the
quasars on the sky is plotted in figure \ref{fig:proj}. The total
number of quasars in our analysis after the cuts performed below is
58,227. The covered area is optimized through survey strategy
considerations and will become more uniform as the survey approaches
completion in 2014.

\begin{figure}[h!]
  \centering
  \includegraphics[width=\linewidth]{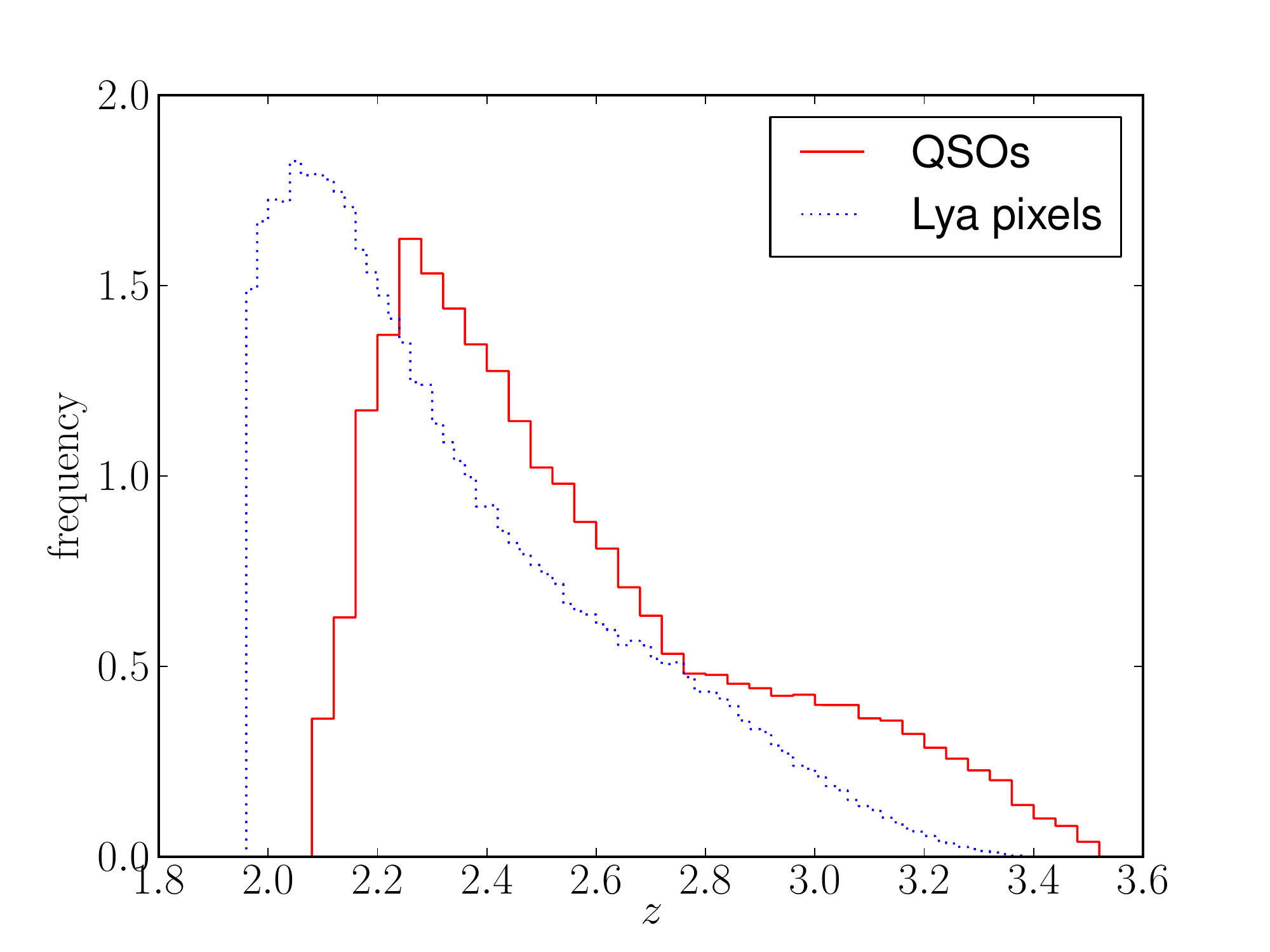}
  \caption{\label{fig:qsos} Distribution of quasar redshifts (red,
    solid) and \lyaf pixels (blue, dotted). }
\end{figure}

We limit our analysis to quasars with redshifts
$2.1<z<3.5$. Throughout this work we use the visual redshifts
(\texttt{Z\_VI}) from the DR9Q.  The low number density of quasars at
redshifts larger than 3.5 renders them not useful for measuring the BAO
signal, which is the main goal of this paper. We show the distribution
of quasar redshifts and the \lyaf pixels in Figure
\ref{fig:qsos}.

For repeated observations of a given object the DR9Q catalog contains
all the listed observations, as well as the observation deemed to be
optimal. We performed the analysis either by restricting the sample to
the subset of best observations as well as using all observations. In
the latter case, we simply co-add the observations on the nearest
pixel point using the pipeline-reported inverse variances. We do not
use the noise correction (discussed later) when performing this
co-add, but since the correction is a function of wavelength it would
rescale all contributions equally and thus have no effect when
stacking in observed wavelength.

We apply several cuts to the data. First, we treat quasars that are
considered Broad Absorption Line (BAL) quasars differently. The DR9Q
catalog contains a measured value of \emph{balnicity index}
\cite{1991ApJ...373...23W}.  We remove quasars with
$\texttt{BI\_CIV}>2000$\ km/s. We additionally drop quasars that DR9Q
visual inspection flagged as highly unusual ($\texttt{BI\_CIV}=-1$).

Second, we mask regions around bright sky lines. The mask is obtained
by the following procedure: We subtract the sky model from the sky fibers. The
model's prediction at the position of the sky fiber does not match the
spectrum of the sky fiber, since the model fits a smooth polynomial
model over all sky fibers and thus sky subtracted sky fibers do not
have zero flux. We stack all sky-subtracted sky fibers and measure the
root mean square defined over $\pm25$ standard BOSS pixels ($\Delta
\log_{10} \lambda=10^{-4}$) boxcar of this
stack. Next we mask all pixels that exceed this r.m.s. by a factor of
1.25. The masked pixels are excluded from the r.m.s.  measurement and
the process is iterated until the mask converges. This mask excludes very
few pixels in the blue end of spectrograph, simply because there are
not many sky-line in that wavelength region.

Third, we exclude portions of the data affected by the Damped
Lyman-$\alpha$ (DLA) systems using the so-called concordance DLA
sample \cite{billinprep}. Briefly, this dataset was established by
three groups within the BOSS \lyaf working group, two of which
developed algorithms for detecting DLAs in data automatically and the
third inspected all quasars visually. All three groups have provided
catalogs of systems. One of these three samples is published in
\cite{2012A&A...547L...1N}. The concordance catalog DLAs were
identified by at least two of the three groups. We do not use data
within 1.5 equivalent widths from the center of a DLA.

When measuring the correlation in the \lya forest, we use a wide
region of the forest, spanning $1036$-$1210$~\AA~for most quasars and
$1036$-$1085$~\AA~for the remaining BAL quasars ($0<BI\le2000$). We
also exclude pixels with observed wavelength less than $3600$~\AA,
beyond which the signal-to-noise of the spectrograph and
spectro-photometric calibration become very poor.

We assume that the measurement noise does not correlate between
pixels, but we apply a correction to the BOSS spectral extraction
pipeline estimates of the noise.  To derive this correction, we fit a
cubic polynomial to the smooth region of quasar spectra between
rest-frame 1420 and 1510~\AA~and compared the fit residuals with the
pipeline error estimates.  We find that the co-added spectra have an
observer-frame wavelength dependent mis-estimation of the noise, which
follows the square root of the ratio of pixel sizes between the
co-added and individual exposure spectra.  This value is a $0-10\%$
under-estimate of the noise variance below observer-frame
$\sim$4750~\AA~and a $0-15\%$ over-estimate above $\sim$4750~\AA; this
correction is not valid beyond $\sim$6000~\AA~where both arms of the
BOSS spectrograph contribute to the spectrum.  We correct for this
mis-calibration by adjusting the pipeline noise by the square root of
the co-add-to-individual exposure pixel size ratio.

\subsection{Synthetic data}
\label{sec:synthethic-data}
Data analysis procedures in this work were tested on the synthetic
data that was generated in nearly the same manner as for the previous
work \cite{2011JCAP...09..001S} and we briefly review it here for
completeness.
\newcommand{\vr}{\mathbf{v}}

We want to generate a transmitted flux-field with desired two-point
function and approximately correct probability distribution
function. For a general transformation $F=T(\delta_G)$, where
$\delta_G$ is a Gaussian auxiliary field, the 2-point correlation
functions of the flux field, $\xi_F(|\vr|) = \left<\delta_F(\vx)
  \delta_F(\vx+\vr) \right>$, and that of the auxiliary Gaussian
field, $\xi(|\vr|) = \left<\delta_g(\vx) \delta_g(\vx+\vr) \right>$,
can be related by
\begin{align}
   \xi_F(r_{12}) = & \left< F_1 F_2 \right> \nonumber \\
               = & \int_0^1 dF_1  \int_0^1 dF_2 p(F_1,F_2) F_1 F_2
               \nonumber \\
               = & \int_{-\infty}^\infty d\delta_1  \int_{-\infty}^\infty d\delta_2 p(\delta_1,\delta_2) T(\delta_1) T(\delta_2) \nonumber\\
               = & \int_{-\infty}^\infty d\delta_1  \int_{-\infty}^\infty d\delta_2 \frac{ e^{-\frac{\delta_1^2 
               + \delta_2^2-2\delta_1\delta_2\xi^2(r_{12})}{2(1-\xi^2(r_{12}))}} }{2 \pi \sqrt{1-\xi^2(r_{12})}} T(\delta_1) T(\delta_2) ~ .
\end{align}

We assume that $\tau=\log F$ is log-normally distributed so that
specifying the mean transmitted flux fraction $\bar{F}$, in addition
to the desired target power spectrum completely specifies the required
properties of the auxiliary Gaussian field. This inversion is done
numerically. The Gaussian field is initially generated assuming
parallel \lyaf lines of sight and stationary (non-evolving) field: the
effects of non-parallel lines of sight and field evolution are then
obtained by interpolating between redshifts. The assumption of
parallel lines of sight and stationary fields significantly simplifies
the problem. Decomposing the line of sight flux fluctuations of a
quasar $q$ into its Fourier components $\delta_k(q,k_{\rm 1D})$, the
cross-correlation vanishes for different Fourier modes:
\begin{equation}
  \left< \delta_k(q,k_{\rm 1D}) \delta_k(q',k'_{\rm 1D}) \right> = \delta(k-k') P_x (k_{\rm 1D}, r_{\perp}(q,q')).
\end{equation}
This allows one to generate the field one parallel Fourier mode at a
time. For each such-mode, one thus needs to generate a correlated
Gaussian field vector of the size $N_q$, where $N_q$ is the number of
quasars. We perform this using standard methods for generating
correlated Gaussian fields with a desired covariance matrix by
Cholesky decomposing the matrix and multiplying with a vector of
uncorrelated unit-variance Gaussian field.  Technical aspects of
synthethic data generation and tests that demonstrate the validity of
the generated fields are described in great detail in
\cite{2012JCAP...01..001F}.

Our present implementation of synthethic data generation is still
numerically extremely demanding. To be able to generate a full
synethetic dataset, we split it into four geometrically
self-contained ``chunks'', ignoring correlations across chunk
borders.  We have tested explicitly that this approximation introduces
no bias.  We have thus generated fifteeen synthethic realizations of
our dataset. These synethetic datasets were also used in \cite{FPG}.

Compared to \cite{2011JCAP...09..001S}, mock data were created using
slightly different bias parameters $b=-0.14$ and $\beta=1.4$.  There
were made more realistic in this work by adding several effects that
are present in the spectrograph and were not accounted for in the
previous work.  We add noise to the mocks based upon an
object-by-object fit to the observed photon noise in the real data.
We then generate a noisy and mis-calibrated estimate of the true noise
to model the pipeline reported noise and use that noise to generate
a noise realization for the analysis.  The mock spectra fluxes were
multiplied by a polynomial in $\lambda$ to match the real data to
simulate QSO flux mis-calibration and a small amount of sky spectrum
is added to model a sky subtraction bias in the BOSS data.  These
additions will be described in greater detail in a future publication
\cite{baileymocks}.  In one respect, the synthetic data were less
realistic that the similar synthetic datasets used in
\cite{2011JCAP...09..001S}, namely, we did not contaminate the data
with associated forest metals in this work.

\section{Data Analysis}
\label{sec:data-analysis}

In this section we describe the steps taken in the data analysis. The
same procedure is adopted both to the real and the synthetic data
described in section \ref{sec:synthethic-data}.

The overview of data analysis is as follows:

\begin{itemize}
\item \emph{Continuum fitting.}  (Section \ref{sec:continuum-fitting})
  First we perform the basic continuum fitting, where the pixels are
  treated as independent. Continuum fitting can be described as a
  process of fitting a model for what the spectrum of a given quasar
  would look like in a completely homogeneous universe, i.e. there
  would be no forest, except the mean hydrogen absorption. We fit for
  the mean continuum, the redshift dependence of the mean transmission
  in the forest and the intrinsic variance including its redshift
  dependence. We also fit for the per-quasar amplitude and slope
  across the forest.

\item \emph{One-dimensional Pixel Power Spectrum measurement.}  (Section
  \ref{sec:1d-pixel-power}) In the second step we generalize the
  variance in the pixels to the full one-dimensional (1D) power spectrum of flux
  fluctuations as a function of redshift. We refit the individual
  quasar forest amplitudes and effective spectral slopes, the mean
  transmission as a function of redshift and the effective 1D power
  spectrum. This power spectrum is not appropriate for cosmological
  analysis, as it does not deconvolve the instrument PSF and the
  instrument noise is treated coarsely. This co-variance is only
  required for the weighting in the measurement of the 3D correlation
  function.

\item \emph{Estimation of the three-dimensional correlation function and its
    covariance matrix.} (Section \ref{sec:estim-3d-corr}) The next
  step is estimation of the 3D correlation function. This is done by
  weighting quasar data with a per-quasar inverse covariance matrix
  and performing optimal quadratic estimation of the correlation
  function. This analysis is done in a coordinate system that is
  composed exclusively of observable quantities, i.e.\ difference in
  logarithm of the observed wavelength, angular separations on sky and the pixel
  pair redshift (which is just a proxy for the mean redshift of a
  given pixel pair). We use several methods to estimate the covariance
  matrix coming from the optimal estimator.

\item \emph{Estimation of BAO position and significance.} (Section
  \ref{sec:fitt-bary-acoust}) Finally, we measure the position of the
  BAO peak from the measured 3D correlation function. The
  uncertainties on the BAO parameters of interest using different
  covariance matrices and broadband models.

\end{itemize}

\subsection{Continuum fitting}
\label{sec:continuum-fitting}

\newcommand{\lre}{\lambda_r}
\newcommand{\lo}{\lambda_o}
\newcommand{\dll}{\delta \log \lambda}
\newcommand{\PSF}{{\rm PSF}(q,\lo)}

A reasonably general model of the observed flux can be written as follows
\begin{equation}
  f (q,\lo) =  \PSF \star \left[ A(q,\lre) \bar{F}(\lo) C(\lre)
    T(\lo) (1+\delta_F (q,\lo)) + S(\lo) \right] + \epsilon(q,\lo).
  \label{eq:1}
\end{equation}

Here we use the index $q$ to denote a particular quasar (probing a
direction on the sky), $\lo$ is the observed wavelength and $\lre=\lo
/ (1+z_q)$ is the rest-frame wavelength of the quasar. Since hydrogen
absorbing through \lya at redshift $z$ produces a decrement at
observed $\lo = (1+z)\lambda_{\alpha}$, we can use $\lo$
interchangeably with absorption redshift.  The $\PSF\star$ term
denotes convolution, i.e.,  smoothing by the point-spread function of
the spectrograph and the final term in equation (\ref{eq:1}) is
the spectrograph noise contribution. The pipeline provides an estimate
of the pixel noise under the assumption that it is independent from
pixel to pixel (but we correct for this as described earlier).

The mean unabsorbed quasar spectrum, also referred to as the continuum,
is given by $C(\lre)$ and we absorb the diversity in both the quasar
amplitude and quasar shape into $A(q,\lre)$. In this work, we model
$A(q,\lre)$ as linear function of $\log \lambda$ 
\begin{equation}
  A(q,\lre) = A(q,\lambda_{ol}) + (A(q,\lambda_{oh})-A(q,\lambda_{ol})) \frac{\log \lo-\log
    \lambda_{ol}}{\log \lambda_{oh}-\log \lambda_{ol}},
  \label{eq:2}
\end{equation}
where $\lambda_{oh}$ and $\lambda_{ol}$ are the highest and lowest
wavelengths in the forest.

The $\bar{F}(\lo)$ term in Equation \ref{eq:1} is the mean quasar
transmission in the forest and can be distinguished from the residual
variations in the spectrograph transmissivity, $T(\lo)$, by the virtue
that the former is unity outside the forest. Finally, the
$S(\lre,\lo)$ term describes an additive systematic component, likely
to be dominated by the sky. The $\delta_F$ term inside brackets are
the fluctuations in the transmission, whose correlation properties we
are attempting to infer from the data.

With expediency in mind, we make several simplifying
assumptions. First, we ignore the point-spread function of the
spectrograph, since it affects exclusively small scales and can be
completely neglected for purely 3D analysis. Second, we
set the sky contribution $S(\lre,\lo)=0$ and the spectrograph
transmissivity $T(\lo)=1$. This choice means that $C(\lre)$ and
$\bar{F}(\lo)$ are really just parametrization of continuum model
under the assumption that it is factorisable as $C(\lre)\bar{F}(\lo)$
after the individual quasar mean luminosity and spectral slope have
been fitted (Equation (\ref{eq:2})). The fitted $\bar{F}(\lo)$ will
also absorb the contribution from the variation of spectrograph
properties as a function of wavelength which have not properly been
calibrated by the pipeline.

Finally, in the initial parts of the fitting process, we model
fluctuations as Gaussian and uncorrelated. Therefore, our likelihood
model in the first step of continuum fitting can be written as

\begin{equation}
  \log L = \sum_i \left[ -\frac{\left(f (q,\lo) - A(q,\lo) C(\lre)
        \bar{F}(\lo)\right)^2}{2 (N(q,\lo) + \sigma^2(\lo))} -
    \frac{\log [N (q,\lo) + \sigma(\lo)^2]}{2} \right],
\end{equation}
where the sum is over all forest pixels in all quasars, $\sigma^2$
denotes the intrinsic variance of the flux transmissions (i.e., variance
in the cosmological $\delta_F$ field) and $N=\left<\epsilon^2\right>$
the pipeline noise.

We measure two amplitude parameters for each quasar
($A(q,\lambda_{ol})$ and $A(q,\lambda_{oh})$, defining amplitude and
slope), $C(\lre)$ in 95 bins between $\lre=1036$\AA\ and
$\lre=1210$\AA\ and further 167 bins extending to $\lre=1600$
\AA. The intrinsic variance $\sigma^2$ and $\bar{F}$ are measured in 16
bins between redshifts $1.95$ and $3.45$ (linearly interpolating
between bins for the case of $\bar{F}$).

This approach contains a considerable number of parameters. We
maximize the likelihood by brute-force maximization, employing a
Newton-Raphson algorithm with numerical derivatives. We do not attempt
to derive errors on these parameters as these errors will not dominate
the errors on the final quantities that we are attempting to derive.

To gracefully approach a good solution, we start by fitting just $C$,
$\bar{F}$, intrinsic variances and constant amplitudes across rest wavelength
$1036-1600$\AA. Next we focus on the forest region and refit individual
amplitudes and slopes together with refitting variances. As a third
step we refit variances and slopes together with measuring the 1D power
spectrum as described in the next section.

\begin{figure}[h!]
  \centering
  \includegraphics[width=1.0\linewidth]{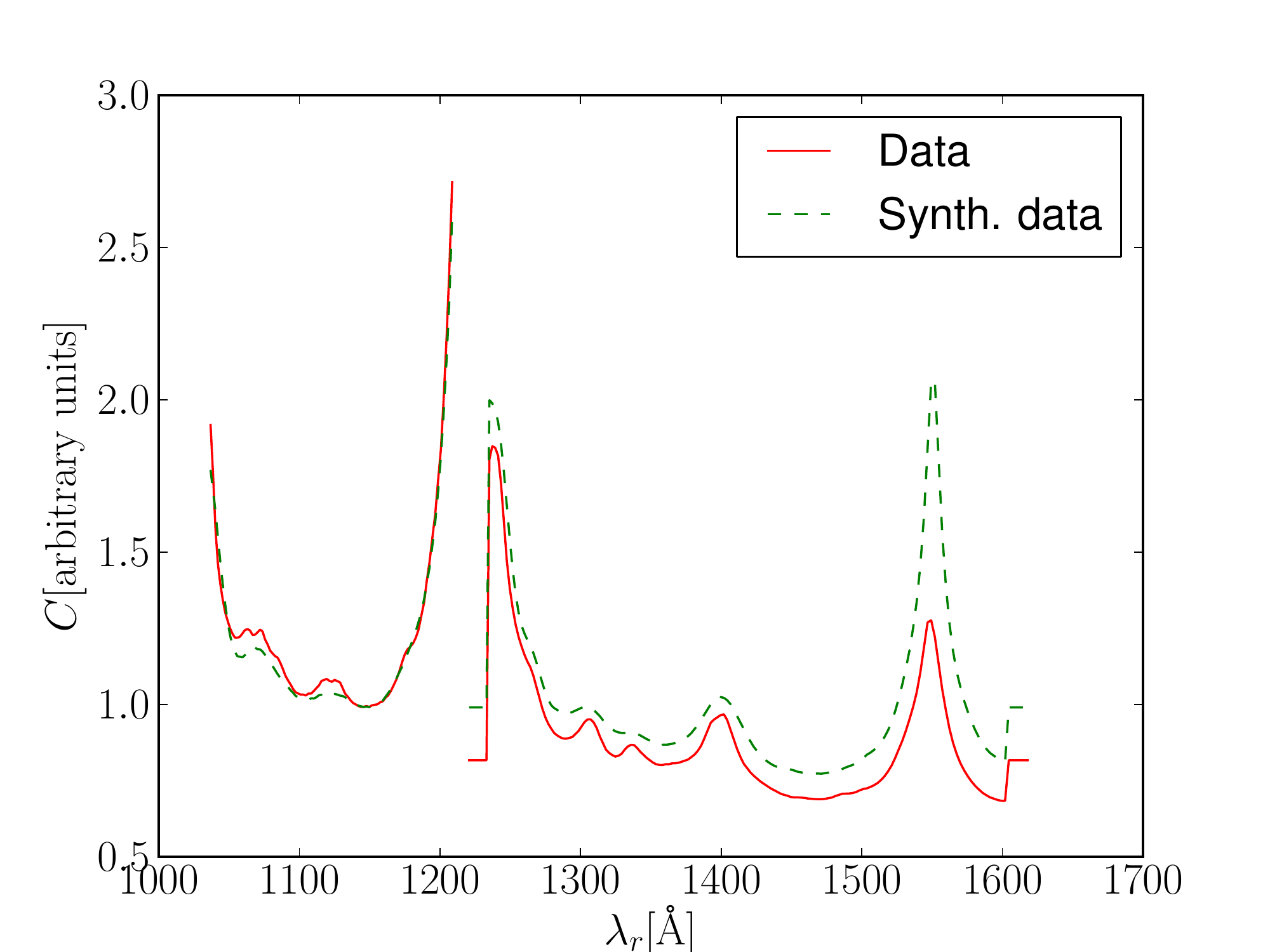}

  \caption{Fitted mean continuum, $C(\lr)$ in arbitrary units per unit
    wavelength.  Solid red curve corresponds to data, while the dashed
    green line is for synthetic data. }
  \label{fig:CF}
\end{figure}

\begin{figure}[h!]
  \centering
 
  \includegraphics[width=1.0\linewidth]{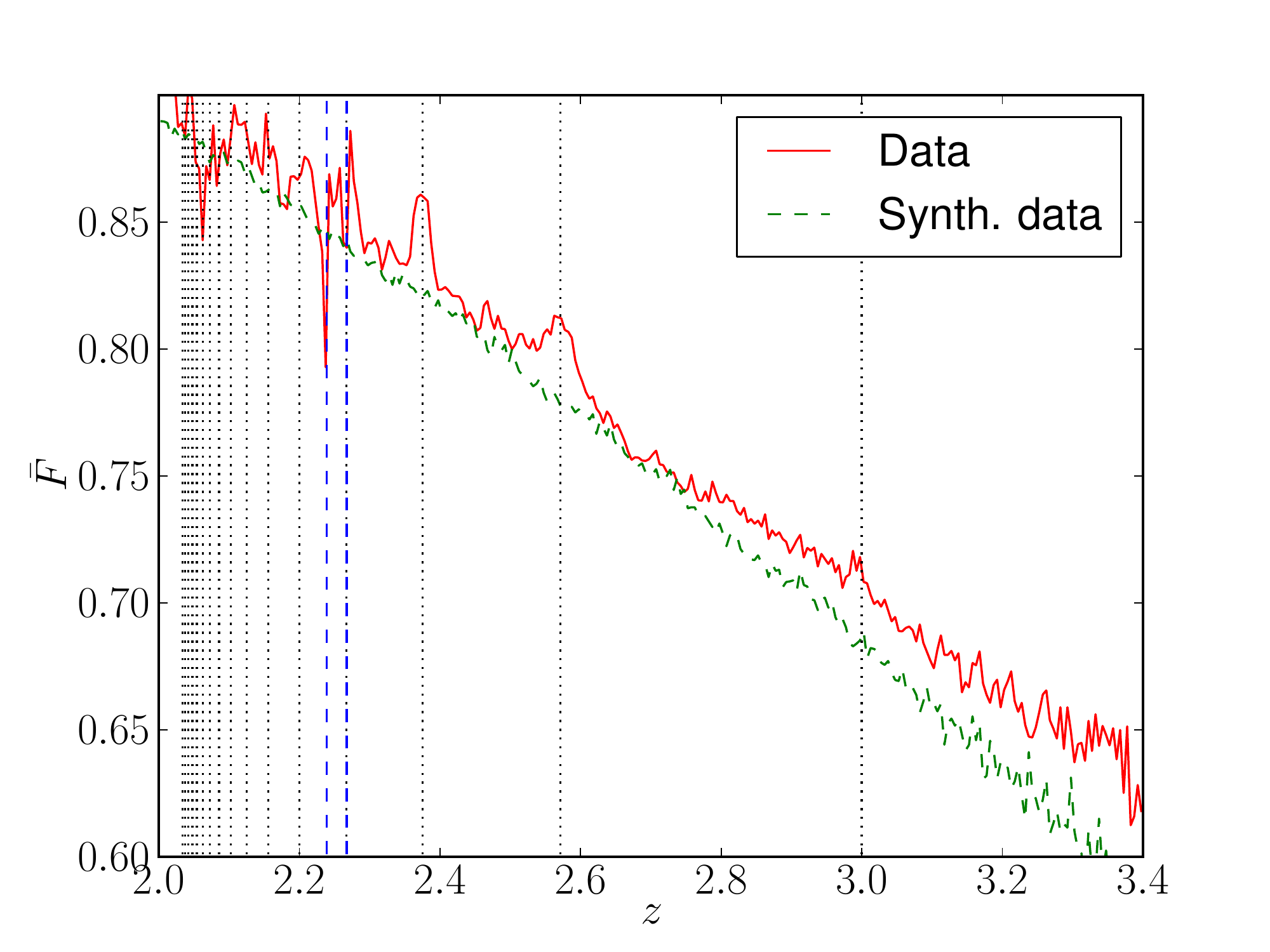}
  \caption{Fitted mean transmission in the forest ($\bar{F}$) with
    arbitrary normalization. Solid red curve corresponds to data,
    while the dashed green line is for synthetic data.  Dotted lines
    show the Balmer series. The blue dashed lines are positions of
    Galactic calcium extinction.}
  \label{fig:CF2}
\end{figure}

Figures \ref{fig:CF} and \ref{fig:CF2} show the mean continuum
and $\bar{F}$ for data and synthetic data. The measurement of
$\bar{F}$ shows the expected structure: there is more absorption
at higher redshifts. However, we also see structure at the position of
rest-frame Balmer lines. This structure is an artifact of the
pipeline and arises due to imperfect interpolation of masked Balmer regions
in the calibration stars. These features are essentially completely
absorbed into $\bar{F}$ and we do not see any residual structure
associated with Balmer lines in other tests. We also see features at
the positions of CaH and CaK absorptions.

\begin{figure}[h!]
  \centering
  \includegraphics[width=0.49\linewidth]{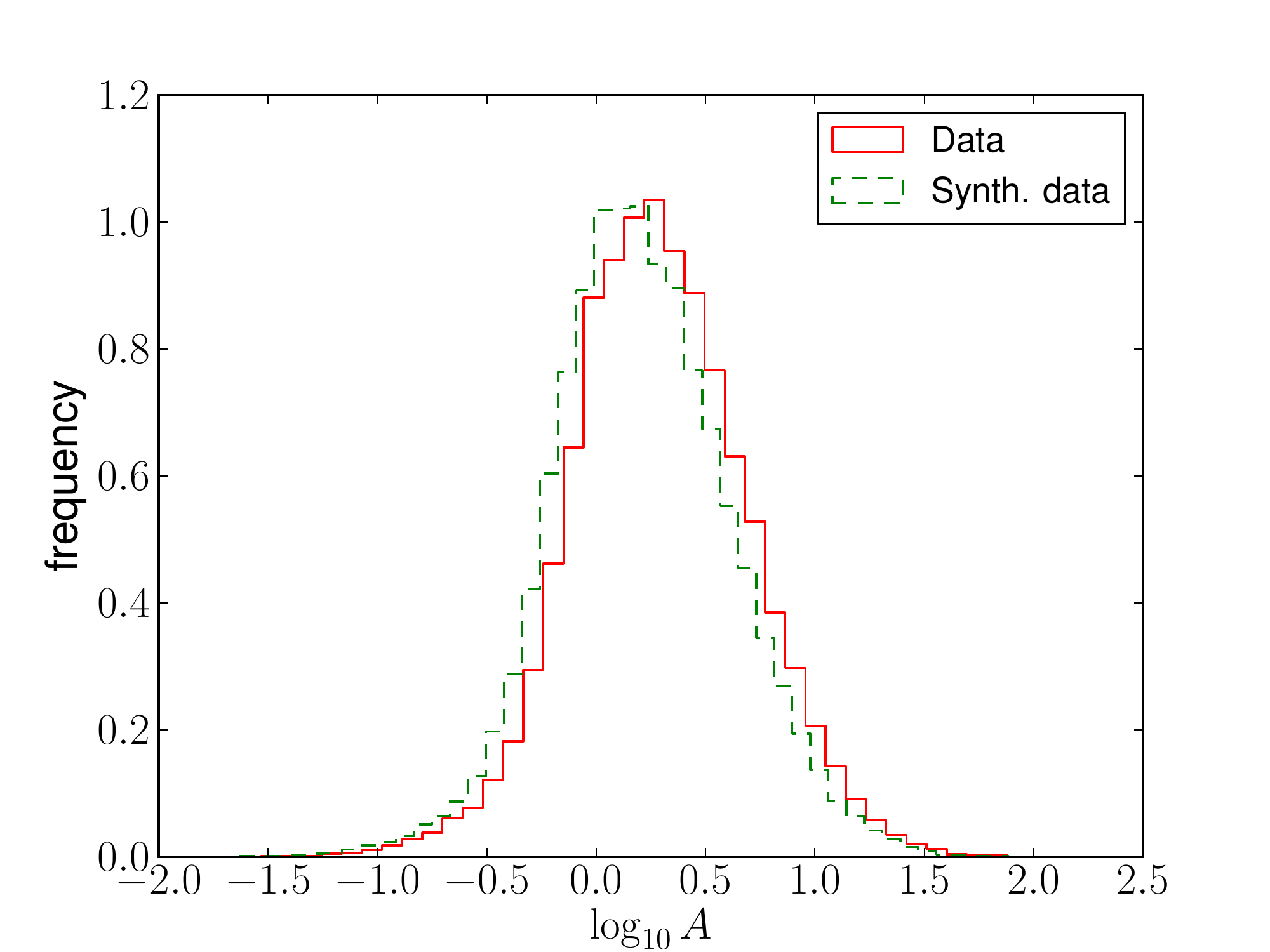}
  \includegraphics[width=0.49\linewidth]{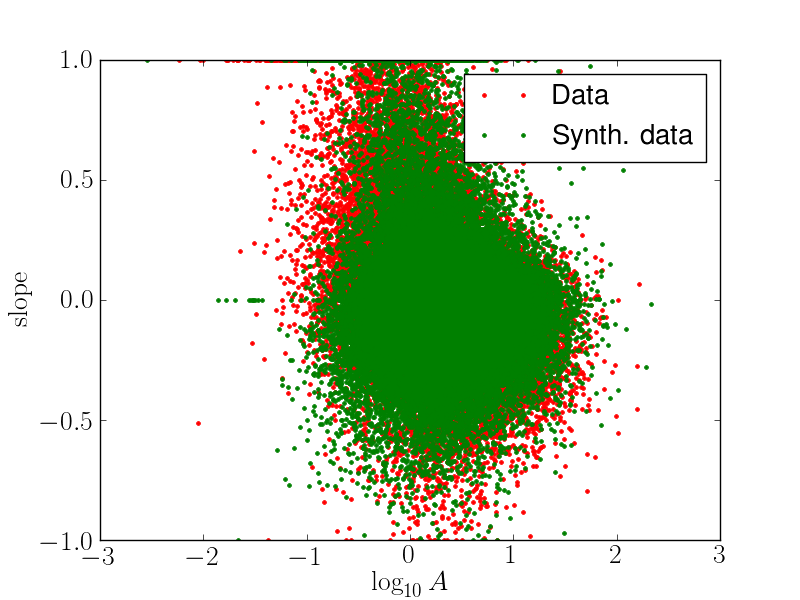}
  \caption{The amplitude and slope distributions
    when fitting data and synthetic data. The left-hand slide plot
    shows the histogram of fitted amplitudes for data (red,solid) and
    synthetic data (green, dashed). The right-hand side plot shows the
    distribution of amplitudes and slopes for data (red points) and
    synthetic data (green points). See text for further
    discussion. }
  \label{fig:As}
\end{figure}

In Figure \ref{fig:As} we show the distribution of amplitudes and
effective spectral slopes as fitted to the data and to the synthetic
data.  The histogram of amplitudes for the data matches that of the
synthetic data very well. This result arises essentially by
construction, since our synthetic quasars are matched in flux to real
quasars. The right-hand panel considers the spectral slope. The slope
is defined so that it is $\pm 1$ if any one of the two individual
fitted amplitudes at two wavelengths is zero. We fit for the slope at
two fixed observed wavelengths at the edge of the forest even though
there might be no data at those edges. Therefore the amplitude
measurement can be quite uncertain and produce a value outside a
physically acceptable range. In general, the diversity of spectra in
synthetic data is a good representation of that in real data,
although, as expected the real data are slightly more diverse.

\begin{figure}[h!]
  \centering
  \includegraphics[width=\linewidth]{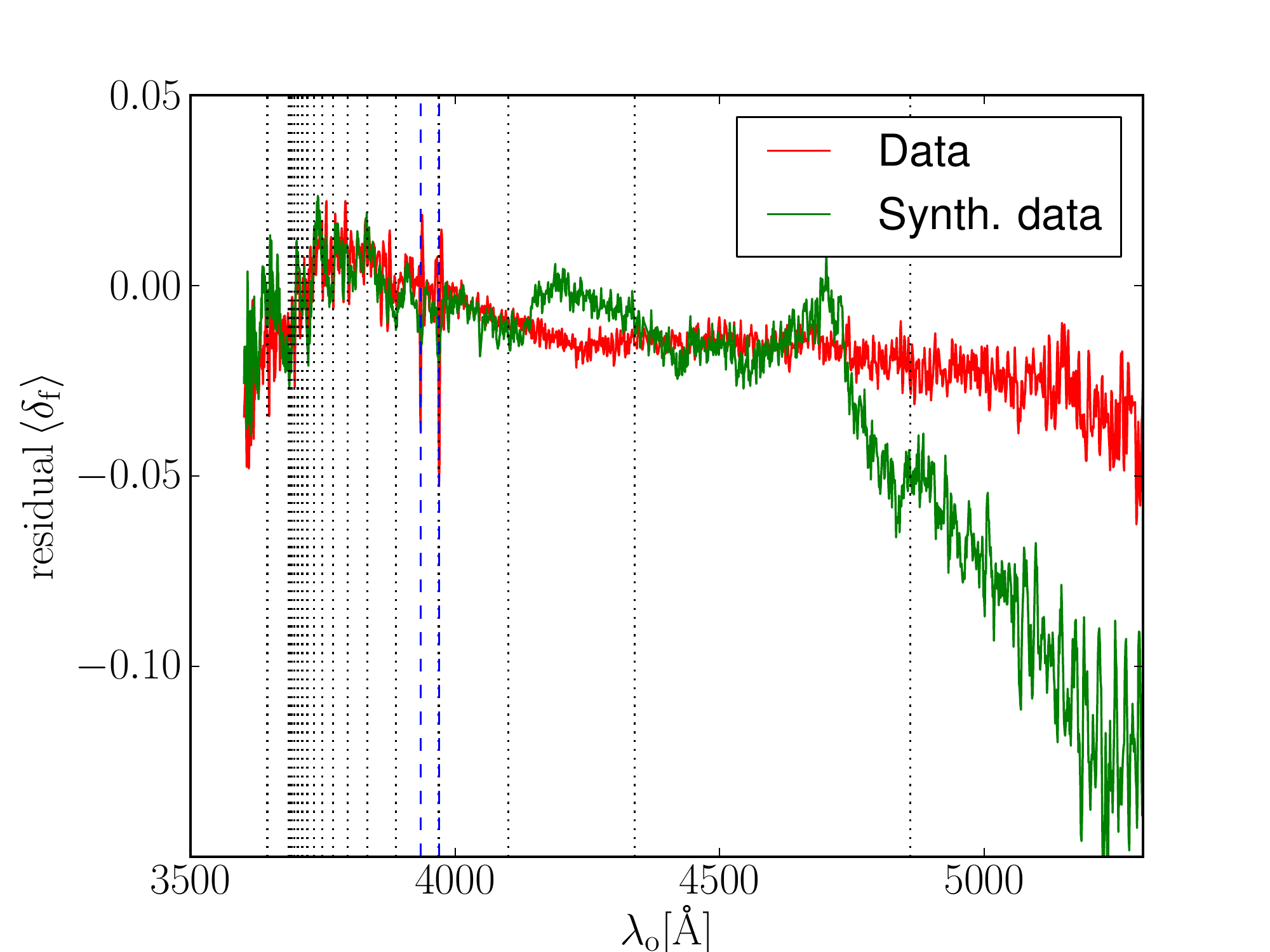}
  \caption{Residual estimates of $\delta_F$ averaged over bin in the
    observed frame wavelength. Green is for one realization of the
    synthetic data, while red is for the real data.  Dotted lines show
    the Balmer series that is clearly seen in the $\bar{F}$ fitting in
    the Figure \ref{fig:CF2}. The blue dotted lines are positions of
    Galactic calcium extinction. }
  \label{fig:sky}
\end{figure}

Our model for the continuum in Equation \ref{eq:1} is not perfect and
there will be systematic residuals between true $\delta_F$ and our
estimates in addition to noise. A useful check for these residuals is
to stack our estimates of $\delta_F$ in bins of observed wavelength
(or redshift): for unbiased estimates these should vanish, but in
practice this is not exactly the case.  In Figure \ref{fig:sky} we
show the residuals from our continuum fitting which are obtained by an
inverse covariance weighted average of all data in bins of observed
wavelength. These residuals are removed from the data before measuring
the 3D correlation function, but they essentially imply a calibration
uncertainty on the final correlation function of a couple of percent
(although this is irrelevant for BAO fits).  Since we are fitting for
the amplitude of quasars in flux and then dividing by a noisy estimate
to derive $\delta_F$, we expect that the mean of $\delta_F$ will not
be zero (essentially because $1/\left<A\right> \neq \left<1/A
\right>$).  We also notice that these residuals are considerably worse
for the synthetic data. This result arises partially because the
synthetic data is somewhat noisier, especially at higher redshift (see
the next section) and partially because addition of some systematic
effects in the synthetic data might be introducing more contamination
than exists in the real data.  Most importantly, however, we see no
effects of structure at the position of Balmer lines, indicating that
these features are nearly perfectly removed with the $\bar{F}$
fit. The caveat is, of course, that one could have structure
associated with Balmer features that would be, for example, correlated
from plate to plate and produce contaminating signal while still
averaging to zero. This concern is discussed further in Section
\ref{sec:disc-syst-effects}.

\subsection{One-dimensional Pixel Power Spectrum measurement}
\label{sec:1d-pixel-power}

As the final step in the continuum fitting process, we measure the 1D
power spectrum of pixels. This power spectrum enables a nearly optimal
weighting to be used when measuring the 3D two-point
function. This step essentially generalizes the model of the
covariance of the flux transmission $\delta(q,\lo)$ from uncorrelated
pixels to pixels correlated across a single quasar.

\newcommand{\Pod}{P_{\rm 1D}}
\newcommand {\zeff}{z_{\rm eff}}
\begin{equation}
  \left <\delta (q,\lo) \delta (q,\lo')\right> = \xi
  (\zeff, \Delta \log \lo) = \frac{1}{2\pi} \int
  \Pod(\zeff,k) e^{ik\Delta \log \lo} dk
\end{equation}

In practice, we measure the power spectrum in flat power bands in $k$
(i.e., the 1D power spectrum is approximated as a set of
top-hat bins) and interpolated in redshift. For any pair of pixels, the
effective redshift $\zeff$ is taken to be given by the geometrical
mean of $(1+z)$ of two pixels (or equivalently geometrical mean of
their $\lr$ or mean of their $\log \lr$). The distance is measured in
$\Delta \log \lo= |\log \lo' - \log \lo|$ converted to the standard
units of ${\rm km}/{\rm s}$ (by multiplying by speed of light).  The
integral of flat power band can be trivially calculated analytically.

%  giving
% \begin{equation}
%   \left <\delta (q,\lo) \delta (q',\lo')\right> =
%   \delta_{qq'}\frac{1}{\pi} 
%   \sum_i \frac{P_i(\zeff)}{\Delta \log \lo} \left[
%     \sin (k_{\rm up,i}\Delta  \log \lo)-\sin (k_{\rm down,i}\Delta  \log \lo)\right],
% \end{equation}
% where $k_{\rm up,i}$ and $k_{\rm down,i}$ denote limits of the flat
% band-power corresponding to bin $i$.

To estimate the 1D power spectrum we use the standard optimal
quadratic estimator
\cite{1998PhRvD..57.2117B,2000ApJ...533...19B,1998ApJ...503..492S} and
has been applied to \lyaf data in before in \cite{MCSEET06}.  The
formalism behind optimal quadratic estimators generalizes the
signal-to-noise weighting of the data from scalar weights proportional
to the inverse variance applied to each data-point individually, to
multiplying the entire data-vector by the inverse of the covariance
matrix. Optimal quadratic estimator methodology can improve the
signal-to-noise of the measurement and provides good (but not perfect)
error-estimates. We review the methodology in the Appendix
\ref{sec:append-optim-quadr}.  

We measure the 1D power spectrum in 20 bins $k$ logarithmically spaced
between $10^{-0.35}$ ${\rm s/km}$ (center of the first bin) and
$10^{-1.45}$ ${\rm s/km}$ (center of the last bin) in steps of $0.1$
dex and 9 bins in redshift, uniformly spaced between $1.8$ and
$3.4$. Although we have very few pixels with $z<2.0$, they contribute
to the power spectrum measurements extrapolated to the $z=1.8$ bin (to
keep the binning in redshift uniform). This results in 180 1D power
spectrum overall bins.  The lower $k$ range for the lowest $k$-bin is
extended to $k=0$.

We do not attempt to deconvolve the point spread function or carefully
understand the effect of noise -- we are only attempting to estimate
the correct pixel-pixel correlations to use in weighting in the 3D
correlation estimation. For example, if noise is underestimated, then
one measures a power spectrum that has a constant offset with respect
to the true power spectrum, but the total signal-plus-noise power will
remain unchanged. The same is true for the effect of the point-spread
function. If we deconvolved the effect of the PSF, we would be putting
exactly the same PSF back into the correlation matrix when making a
prediction for the covariance matrix of a single quasar forest. In
summary, in this paper, the 1D power spectrum is treated as an
intermediate product used for weighting the 3D two-point function
estimate and is not intended to be useful as a measurement of the
physical 1D \lyaf power spectrum.  Any error in 1D power
spectrum results in a less-optimal 3D weighting, but will not bias the
3D measurement.

When measuring the 1D power spectrum, we iterate between measuring the
1D power spectrum and refitting all per-quasar amplitude and slope
parameters. The process converges in approximately five iterations. When
measuring the 1D power spectrum we consistently marginalize out modes
that were fitted by the continuum fitting as described in Appendix
\ref{sec:append-marg-out} and discussed later.

\newcommand{\sit}{Si\ III}

\begin{figure}[h!]
  \centering
  \includegraphics[width=1.0\linewidth]{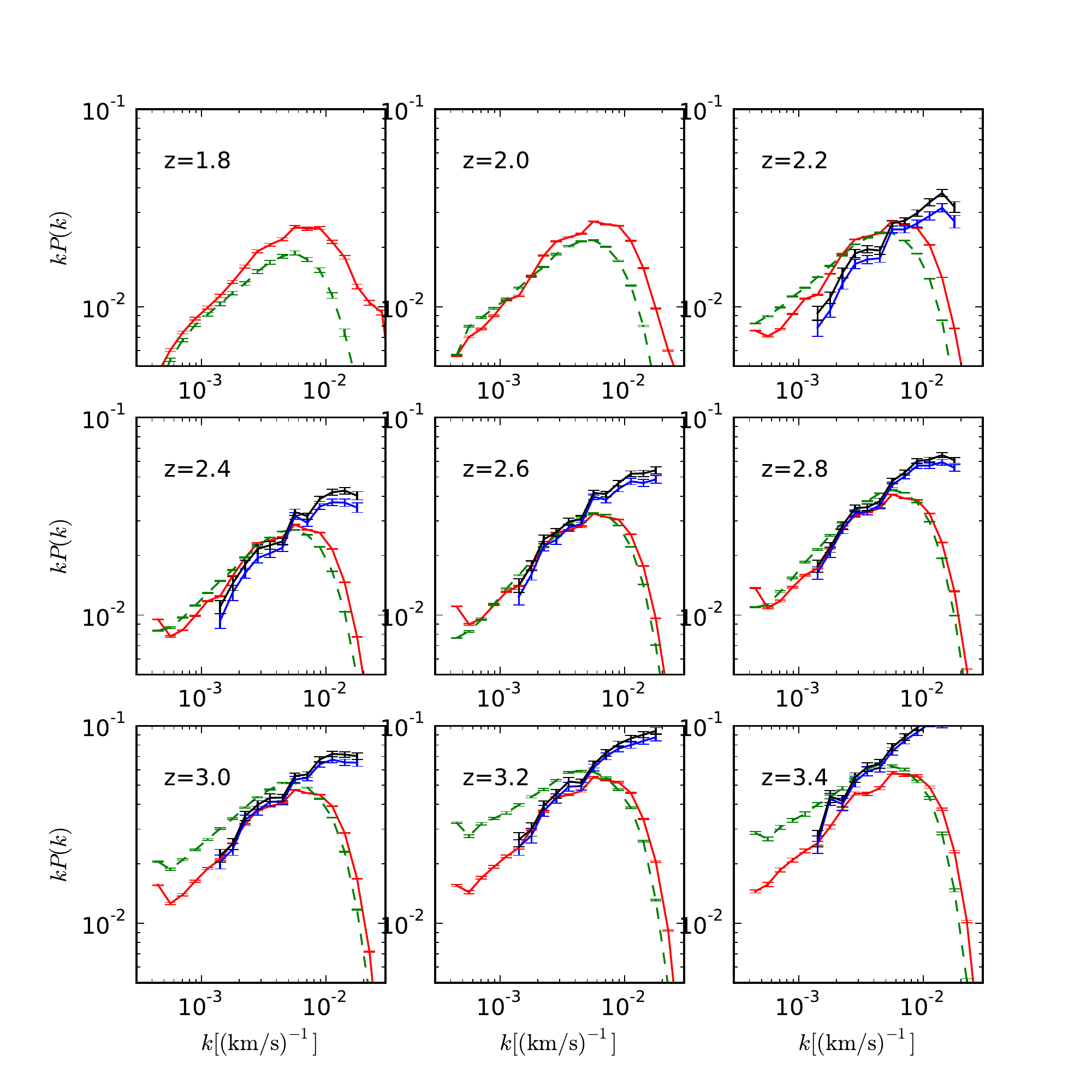}
  \caption{One-dimensional power spectra for data (solid red curve) and synthetic
    data (dashed green curve). Panels correspond to redshifts bins
    indicated. Noise has been subtracted, but the PSF has not been
    corrected for. Blue and black curves are measurements from
    \cite{MCSEET06}, before and after corrections for background
    power. See text for further discussion.}
  \label{fig:pk1d}
\end{figure}

In Figure \ref{fig:pk1d} we plot the 1D power spectra for real and
synthetic data. The data and the synthetic data show a good overall
agreement, although they differ considerably in the high redshift
bins. The edge redshifts are essentially extrapolations from the
measured data and should not be taken too seriously. In addition, the
data in the highest redshift bins come from the region close to the
quasar (we use data all the way to $1210$\AA\ and do not use quasars
with $z>3.5$). Finally, in this work we use only a  rough
correction for the pipeline mis-estimation of the noise, which can be
thought of as a constant additive uncertainty for the data.  The
errors reported by the synthetic datasets are biased (on purpose, to
model the pipeline errors) which we did not attempt to correct, so
it is natural that the synthetic data will not measure just the
intrinsic power spectrum. However, when doing inverse covariance
weighting for the 3D measurement, the relevant quantity is the sum of
noise and intrinsic power, so noise mis-estimation will result in
extra (or less) intrinsic power, but will affect the sum only at
second order.  We also plot the measured values from \cite{MCSEET06},
which have been corrected for the effect of resolution and thus do not
show the beam suppression at high $k$ values.  The wiggles in the
measured power spectra due to presence of \sit\ are clear in both
our dataset and \cite{MCSEET06} power spectra, while absent from the
synthetic data.

\subsection{Quasi-optimal estimation of the three-dimensional correlation function}
\label{sec:estim-3d-corr}

Once the 1D correlation properties of quasars are measured, we proceed
to measure the 3D correlation function.

As in the case 1D power spectrum measurement, we use the optimal
quadratic estimator.  Applying the estimator blindly would involve
inverting matrix the size of the number of pixels. Since this is
numerically prohibitive, we instead approximate it as a block-diagonal
matrix, where each block corresponds to a single quasars. In other
words, \emph{for the purpose of data-weighting}, we assume quasars to
be uncorrelated. This approach is optimal in the limit of small
cross-correlations between neighbouring quasars. Technical details are
outlined in the Appendix \ref{sec:append-optim-quadr}, in particular,
we numerically implement equations \ref{eq:est2}, \ref{eq:est21} and
\ref{eq:est22}.  This approach is always unbiased, but the resulting
error-covariance matrix will ignore contributions to the error
covariance matrix stemming from correlations between pixels in
neighboring quasar pairs.

We never measure the 3D correlations using pixel-pairs that reside in
the same quasar since continuum fitting errors are expected to be
correlated within the same quasars, but significantly suppressed
across quasars (true quasar continua are thought to be completely
uncorrelated across quasars\footnote{A notable exception are systems
  with multiply lensed quasars.}).  We sort pairs of quasars into SDSS
plates. If a given quasar pair has two objects observed on two
different plates, the pair is uniquely associated with one of the two
plates. There are 817 plates in DR9 which parcels the problem of
computing the 3D correlation function into feasible computation
chunks, but also allows a suitable base for bootstrapping errors. In
particular, one could imagine a class of systematic errors that are
associated with plates and such errors would be ``discovered'' as
extra variance by the bootstrap algorithm.

However, even in the approximation where quasars are decoupled, the
computation of estimates is still too slow. Therefore, we reduce the
problem by optimally reducing the number of pixels by a factor of
three or four. This reduction is done \emph{after} multiplying the
data vector by the inverse covariance matrix as described in Appendix
\ref{sec:append-coarse-grain}.

Our model for the 3D correlation function is parametrized in terms of
completely observable coordinates that do not assume a fiducial
cosmology.  In particular, each bin of the measured correlation
function is characterized by the radial separation measured in $\Delta
\log \lambda$ (where $\lambda$ is the observed wavelength of a given
pixel), angular separation measured in radians and redshift (which is
just a proxy for the mean observed wavelength). In our parametrized
model for the correlation function that we measure, we interpolate
linearly in $\Delta \log \lambda$ and $\log (1+z)$. This is
particularly important in the redshift direction, where bins are quite
wide. In the angular direction, however, the bins are assumed to be
uncoupled (i.e.,top hat in shape).  The reason for this choice is that
it significantly simplifies the resulting Fisher matrix, which is zero
for elements corresponding to pairs of estimates at different
separations.  This point will be elaborated upon further in Section
\ref{sec:comp-bootstr-estim}.

The main advantage of this approach is that one is truly measuring
correlations without an assumption on cosmology. The main drawback is
the large number of correlation bins and an inevitable smoothing of
the measured correlation function, because the BAO peak position in
$(\delta \log \lambda,s)$ configuration space is a function of
redshift. We will show later that this is a negligible effect. We use
three redshift bins ($z=2.0$, $z=2.5$ and $z=3.0$), 18 separation bins
centered around 5,15,\ldots 175 arc minutes and 28 bins in $\Delta
\log \lambda$ spanning the relevant range with non-uniform bin spacing
(0, 0.001, 0.003, 0.005 \ldots 0.049, 0.059, 0.083); this procedure
results in 1512 measured bins.

The quadratic estimator does produce a Fisher matrix as an estimate of
error covariance of the measured data-points. In our case, this
assumes that the flux fluctuations correlation within individual
quasars dominate the covariance matrix. In practice we would like to
have a more robust measurement of the error-bars for our final
measurement; for this purpose we use various techniques that estimate
the covariances internally.

\subsection{Distortion of the three-dimensional correlation function}
\label{sec:dist-3d-corr}

It has been known since the 1970s \cite{1978SvA....22..523D} that the
na\"{i}ve estimates of 3D correlations from the \lyaf
result in distorted correlation function due to continuum fitting. 

The effect appears because continuum fitting inevitably entails
inferring information about continuum shape from the actual forest
data and the model will naturally try to minimize the variance of
residuals with respect to the predicted mean continuum. For example,
an unbiased estimates of $\delta_F$ will have a small but finite mean
due to presence of Fourier mode whose wavelength is much larger than
then the length of the forest. Continuum fitting cannot distinguish
between a true underlying offset and a change in quasar amplitude and
will typically set the amplitude of such mode to zero. The size of
this effect is surprisingly large and will result in a biased
correlation function measurement.

In this work, we would, in principle, not need to worry about
distortion effects, because we are trying to find an isolated bump in
the correlation function, while distortion is a smooth function that
could be absorbed with a sufficiently flexible
broadband. Nevertheless, prevention is always preferable to cure.

There are several ways to address this distortion. The simplest one,
used in \cite{2011JCAP...09..001S}, is to assume that the effect of
continuum fitting is to force a set of measured $\delta_F$ values to
have zero mean in each quasar line of sight, by subtracting its mean
$\delta_F \rightarrow \delta_F - \left<\delta_F\right>$. One can
propagate this assumption through a simple analytical model and obtain
a formula for the correction. Resulting expression is approximate, but
worked well for the signal-to-noise used in the above paper.  In this
work we can do better than that by removing the effect of the
continuum fitting by down-weighting linear combinations of datapoints
that we know are affected by the continuum fitting process, namely the
constant offset and slope in per-quasar data-vectors. This is done by
associating large variances with these particular modes as described
in Appendix \ref{sec:append-marg-out}. This technique is equivalent to
discarding a point by associating a very large variance with it, only
that we now associate a large variance with a particular linear
combination of points.  Of course, the contribution from a particular
linear combination can only be down-weighted if one uses matrix
weighting. The result is that, by construction, these marginalized
modes cannot have an effect on the final result.

In particular, we marginalize modes that appear as a constant offset
and those that appear as a linear function of $\log \lambda$ in the
$\delta_F$ field. We add these terms to the covariance matrix both for
the 1D P(k) measurements as well as the 3D
correlation function..

As a result of this marginalization, the estimator propagates the
uncertainty associated with the unknown modes into the
measurements of the 3D correlation function.  As an
unfortunate consequence of working in configuration space, this
approach results in large, but nearly completely correlated errors in
measurements. In Fourier space, this result corresponds to a few
poorly measured low-$k$ modes.

For plotting purposes, these modes can be self-consistently projected
out of the measured correlation function and theory predictions,
resulting in a distorted correlation function measurement whose
error-covariance is considerably more diagonal as we will do in the
Section \ref{sec:visualizing-peak}. However, the shape of the
distortion is now known from the eigen-vectors of the projected
modes. If one limits the analysis to inspection of $\chi^2$ values, this
is, of course, unnecessary.

\subsection{Determining the covariance matrix of the $\xi$ measurements}

There are two aspects of any analysis that can affect the results in
a detrimental way: the bias of the estimation procedure and the
uncertainty in the derivation of the covariance matrix of the errors.
The former can be most precisely determined by examining the results
on the synthetic data. As far as accuracy of the covariance matrix is
concerned, fifteen realizations is not enough if one wants to
consider events that are rarer than $1$ in $15$ (i.e., $>2\sigma$
events). Therefore one must rely on the internal checks of the data, and
we provide several such methods.

\subsubsection{Bootstrapped covariance matrix}
\label{sec:bootstr-covar-matr}

We start by creating a bootstrapped covariance matrix. This process
recreates bootstrap ``realizations'' of the data by randomly
selecting, with replacement, $N$ plates from the original set of
plates and combining them using their estimator provided weights. In
each realizations some plates appear more than once and sometimes
never.  The covariance matrix of these realizations can be assumed to
be a good approximation to the covariance of the estimation, if the
plates are sufficiently independent and if we have sufficient number
of them. Although the plates are not strictly independent (the pairs
of quasars spanning two plates are randomly assigned to one of the
plates), the approximation that these are independent should be a very
good one - as we will see later, not just plates, but even quasars can
be considered independent.  There is one subtlety, however. The
variance associated with the distortion of the correlation function
described in Section \ref{sec:dist-3d-corr} will not be accurately
represented in this bootstrapped matrix, because this variance will be
artificially low (by virtue of these modes being calibrated out) and
therefore the bootstrapped matrix cannot be used directly on the
measured correlation function.

The full bootstrapped covariance matrix from the dataset (or one
realization of the synthetic data) is non-positive definite, because
its size is $1512\times1512$ while the number of plates contributing
to the bootstrapped matrix is only 817. In other words, the 817
vectors do not span the full 1512 dimensional space and therefore
linear combinations of them cannot generate 1512 independent
eigen-vectors of the covariance matrix.  However, the bootstrapped
covariance matrix where elements corresponding to different angular
separations are assumed independent is positive-definite as we are now
determining eighteen $84\times84$ matrices from eighteen sets of 817
vectors of size $84$.  We show that this approach is a good
approximation in the next subsection.

The bootstrap matrices used here were created from 150,000 bootstrap
samples, which is well over what is required for convergence
(covariance matrix converges in the sense of  $\chi^2$ values changing
at less than unity at about 30,000 samples).

\subsubsection{Comparing bootstrapped and estimator covariance matrices.}
\label{sec:comp-bootstr-estim}

The covariance matrix derived from our estimator has, by construction,
structure that is diagonal in the angular separation.  This can be
understood as follows: our weighting matrix (discussed in Section
\ref{sec:estim-3d-corr}) approximates quasars as being
independent. This is the same as assuming that any given quasar pair
is completely independent from any other quasar pair. Since a quasar
pair contributes to the correlation function at only one angular
separation bin (i.e. the angular separation between quasar lines of
sight), it necessarily implies that different separation bins have
uncorrelated errors.  This assumption is not exactly true and we need
to investigate the accuracy of this approximation.

\begin{figure}[h!]
  \centering
  \includegraphics[width=1.0\linewidth]{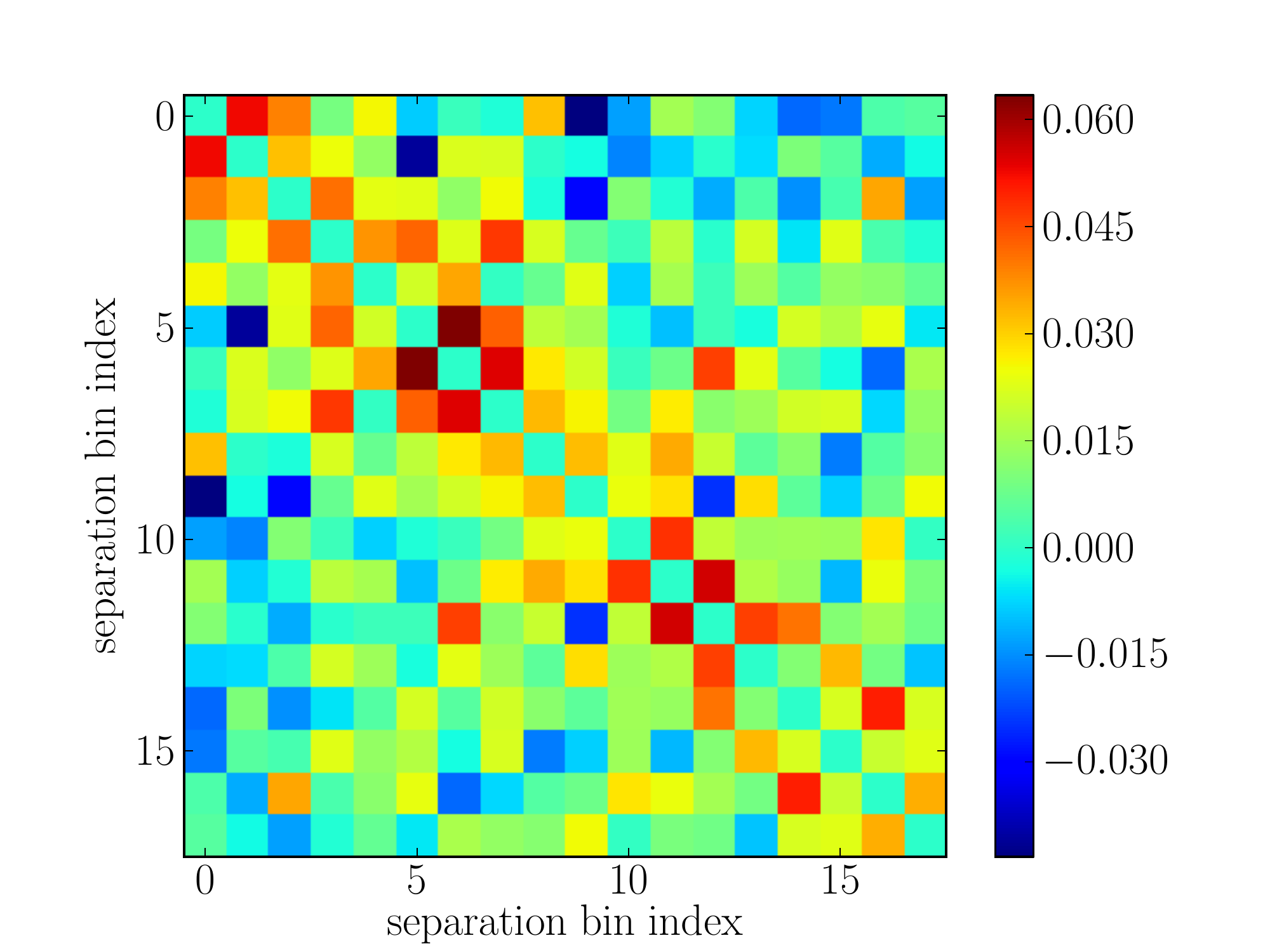}
  \caption{Cross-correlation coefficients for the bootstrapped matrix
    in the separation direction, where other two parameters are held
    at fixed position in the middle of our fitting range ($\dll =
    0.023$, $z=2.5$). The diagonal elements, which are unity by
    definition, have been set to zero to emphasize the off-diagonal
    structure.}
  \label{fig:shoja1}
\end{figure}

To this end we combined bootstrap matrices from 15 realizations of the
mock datasets, i.e.\ from the full $15\times817=12225$ plates. These
now hold enough information  to create a fully
positive-definite covariance matrix by bootstrap. We then inspect the
correlation elements in this matrix that span different
separations. These elements are small, typically less than $0.05$. We plot the
cross-correlation matrix in the angular separation direction for a
fixed choice of $\dll$ and separation bins in Figure
\ref{fig:shoja1}. While the eye can discern some coherent structures,
they do not dominate the matrix.

To make this statement more precise we turn to fitting the bias
parameters. We fit for parameters with their redshift evolution
with the full bootstrapped covariance matrix and the one in which we
have set the non-diagonal elements at different separations to
zero. The fitted parameters were unchanged, and the best fit $\chi^2$
changes by less than 10 units for combined 15 realizations,
implying a change of less than unity in $\chi^2$ per realization.  We
therefore conclude that for the present analysis, it is a good
approximation to assume correlation function measurement errors to be
uncorrelated across angular separations.

\begin{figure}[h!]
  \centering
  \includegraphics[width=0.49\linewidth]{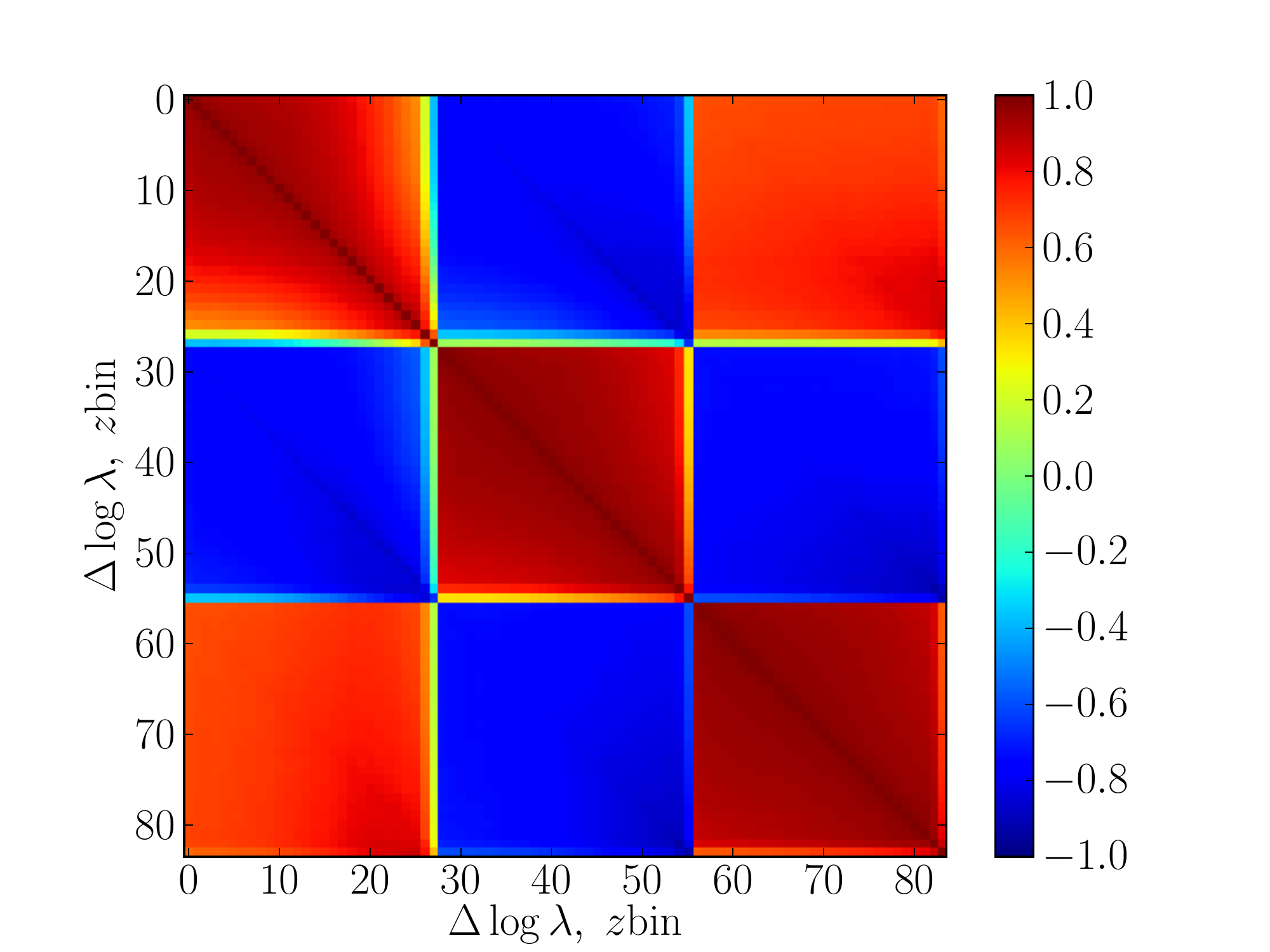}
  \includegraphics[width=0.49\linewidth]{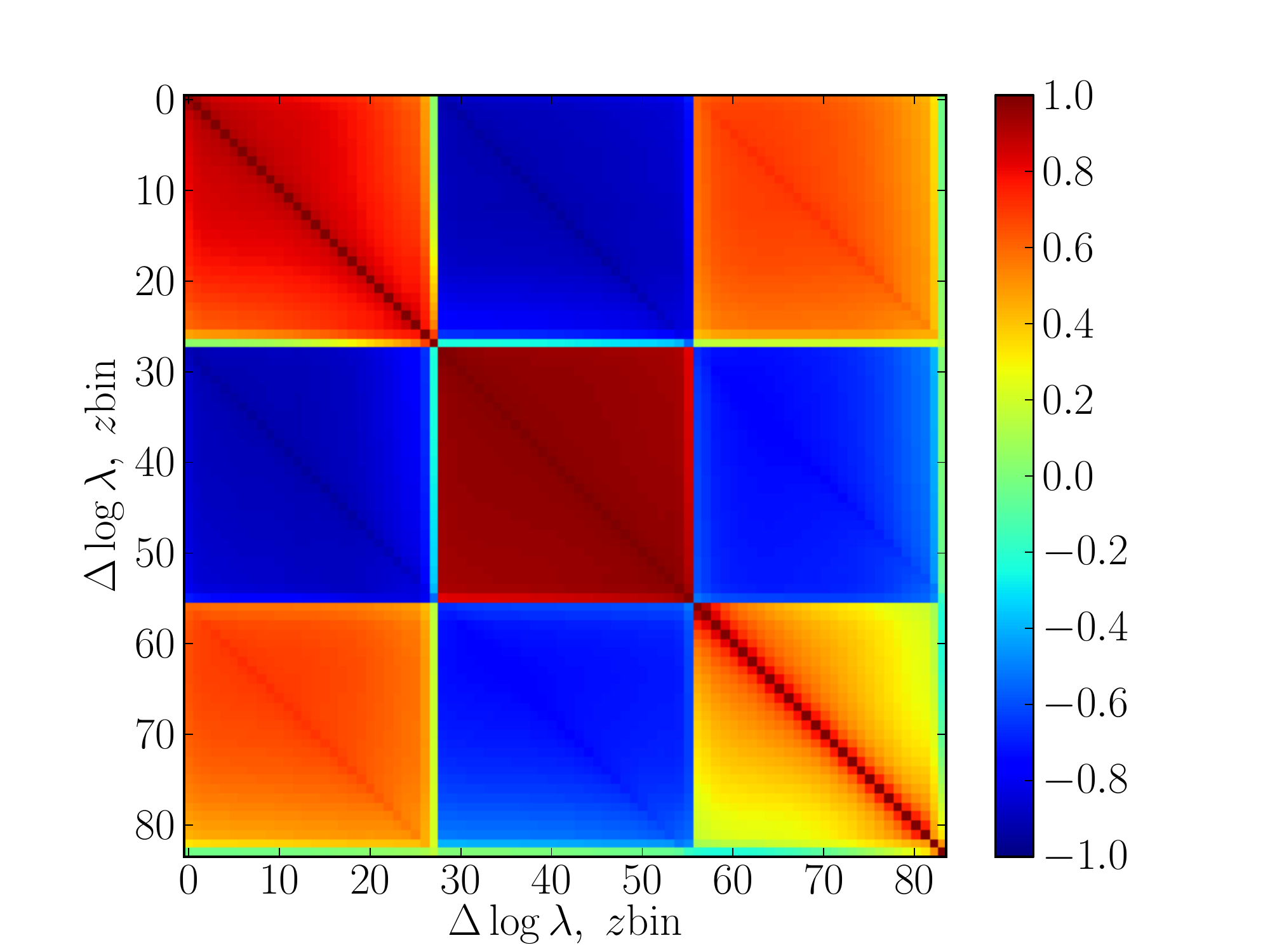}
  \caption{Bootstrapped correlation matrix at a fixed separation
    ($120'$, but plots for other separations are similar). The left
    panel shows the bootstrapped covariance matrix, while the right
    panel shows the quadratic estimator matrix. Three $3\times3$
    ``tiles'' in each plot correspond to the three redshift bins,
    while structure inside each ``tile'' correspond to the
    correlations between $\dll$ bins. See text for further discussion.
  }
  \label{fig:shoja2}
\end{figure}

Next we focus on the correlations inside a single angular separation
bin. The quadratic estimator covariance matrix has two large
eigenvalues corresponding to marginalization over amplitude and
spectral slope that were performed as described in Section
\ref{sec:estim-3d-corr} and Appendix \ref{sec:append-marg-out}. The
bootstrapped covariance matrix of course does not know about this
marginalisation, so we manually remove these from the quadratic
estimator covariance matrix. This is done by diagonalizing the matrix
and then reconstructing it with the two largest eigenvalues set to
zero. The resulting correlation matrices are displayed in Figure
\ref{fig:shoja2}. This result shows that the basic structure of the
correlation matrix is correct. However, it is difficult to judge these
matrices in detail from the plots and therefore we turn to another
test.

\subsubsection{The SK test}
\label{sec:sk-test}
We measure the correlation function in terms of plates. The quadratic
estimator provides a covariance matrix estimate for each individual
plate. These are then combined into the complete estimate using
\begin{eqnarray}
  C_{\rm tot}^{-1} \xi_{\rm tot} &=&  \sum_i C_i^{-1} \xi_i \\
  C_{\rm tot}^{-1} &=&  \sum_i C_i^{-1} 
\end{eqnarray}
where the subscript ``$\rm{tot}$'' denotes the total estimate and the
subscript $i$ the contribution arising from  plate $i$ (note that $i$
does \emph{not} denote the vector element). One could make an
approximation that within errors of a single plate, the survey mean is
a good approximation for the true mean.  However, the resulting
$\chi^2$ would be lower than expected, because any given plate has
contributed to the total mean. However, it can be shown that the
following pseudo $\chi^2$ \cite{peanutboy}
\begin{equation}
  \Delta \tilde{\chi}^2_i = \left( \xi_i - \xi_{\rm tot} \right)^T \left( C_i - C_{\rm tot}
  \right)^{-1} \left( \xi_i - \xi_{\rm tot} \right)
  \label{eq:4}
\end{equation}
has the expected mean for the given number of degrees of
freedom. Additionally, if the distribution of the $\xi$'s is Gaussian then these quantities
are also $\chi^2$ distributed.

\begin{figure}[h!]
  \centering
  \includegraphics[width=0.32\linewidth]{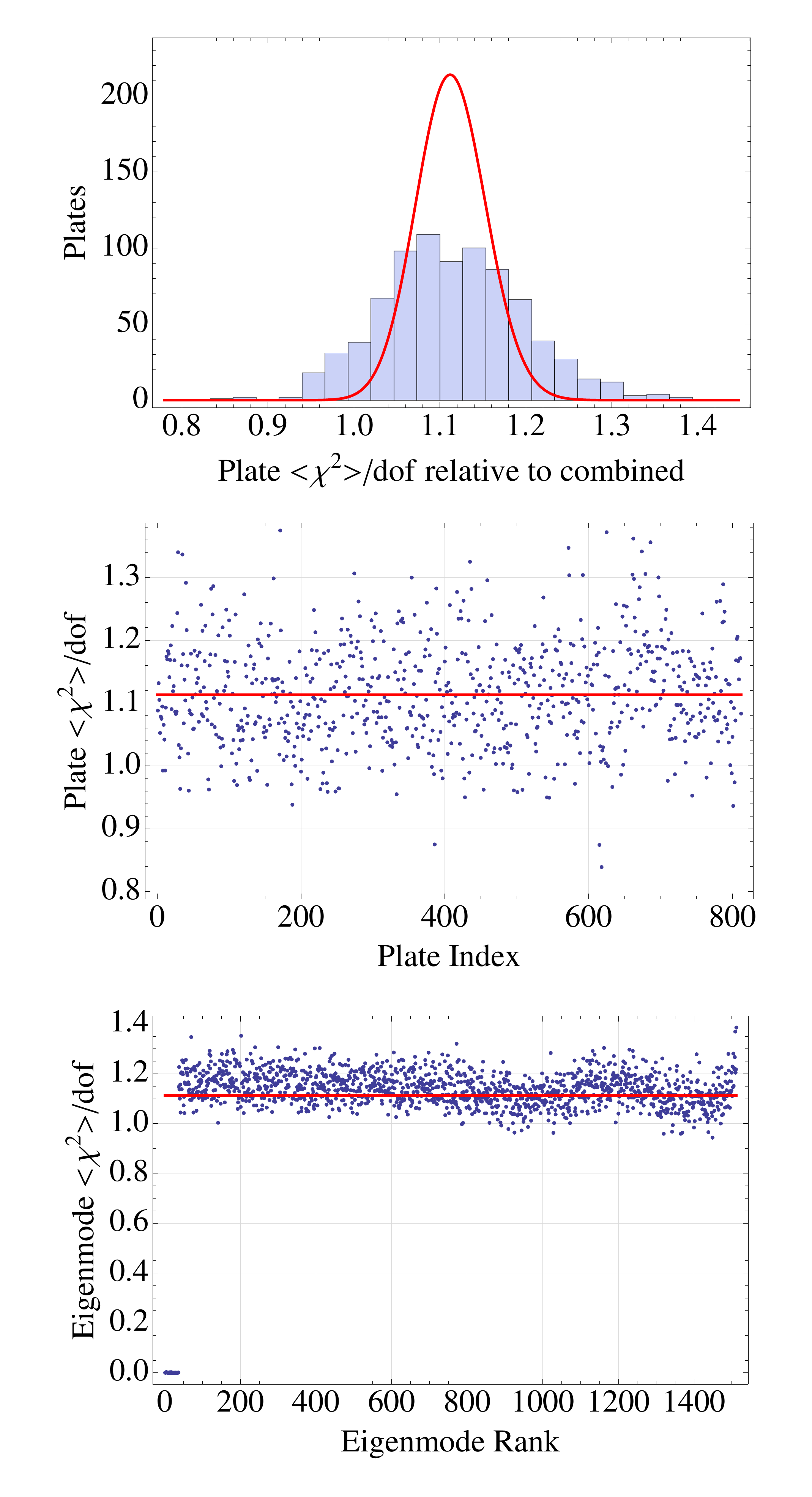}
  \includegraphics[width=0.32\linewidth]{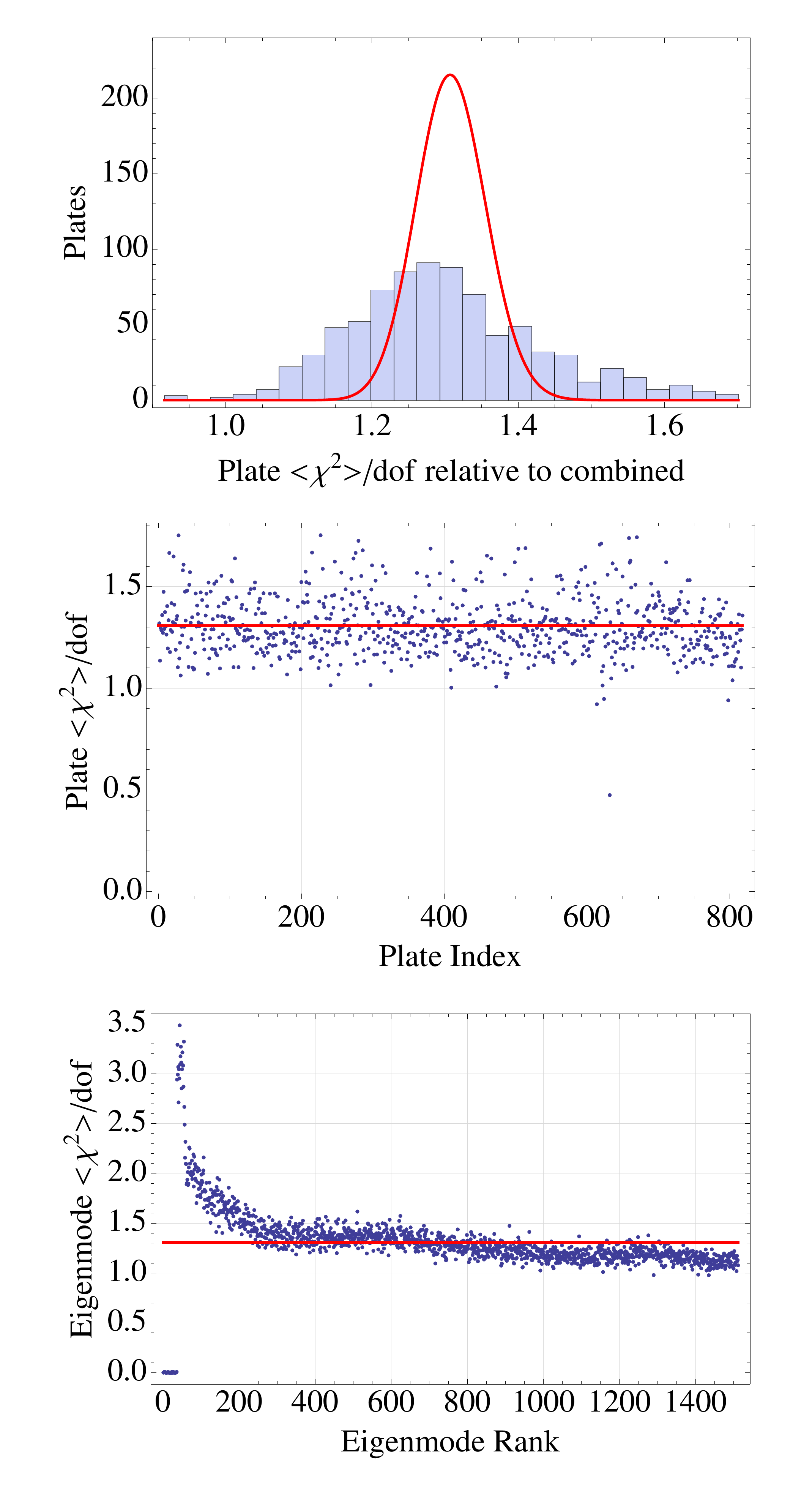}
  \includegraphics[width=0.32\linewidth]{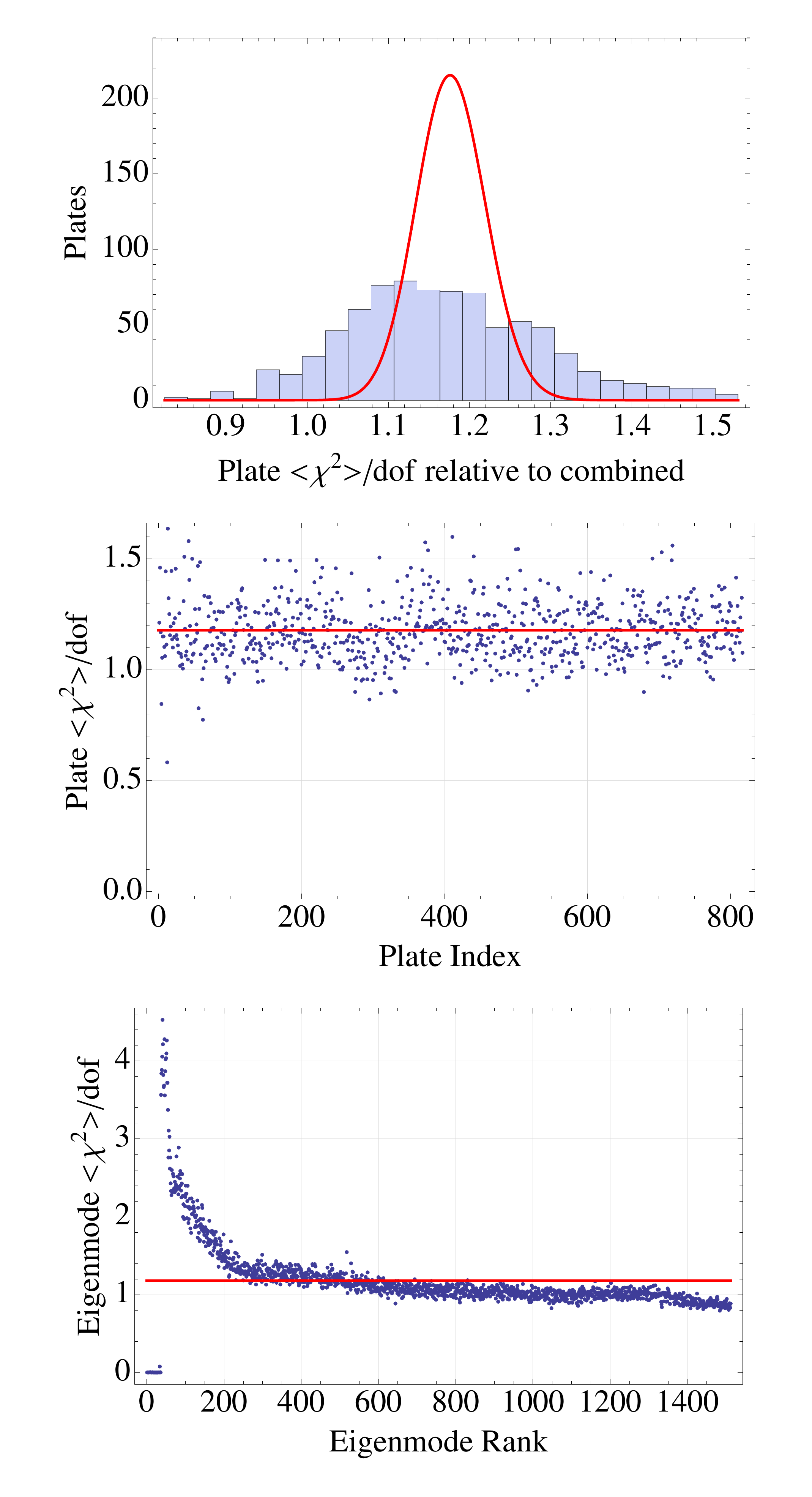}
  \caption{ Result of SK test done on the pure flux field in  one
    realization of synthetic data (left column), on realization of
    the fully reduced synthetic data (middle column) and the real
    data (right column). For each dataset we plot the distribution of
    the $\tilde{\chi}^2$ per d.o.f. values (top row), the $\chi^2$
    per unit d.o.f. as a function of plate number (middle row) and the relative
    contributions to this quantity coming from eigenvalues of
    $C_i-C_{\rm tot}$ sorted by eigenvalue rank and averaged over all
    plates (bottom row). See text for discussion.  }

  \label{fig:ks}
\end{figure}

We have performed this test on three sets of data: i)  on the
reduction where the continuum and other nuisance quantities were
assumed to be known perfectly, ii) on the fully analyzed synthetic
data and iii) on the data. Results of these tests are presented in 
Figure \ref{fig:ks}. 

From these plots we can make the following conclusions. First, even
when all nuisance quantities are known, the estimator covariance
matrix is not perfect as it underestimates the variance by about
10\%. This can be due to several reasons, such as the 1D power
spectrum not being well estimated, but most likely arises from the
fact that the field itself is non-Gaussian. However, the mis-estimate
does not seem to correlate with separation, redshift or the magnitude
of the eigenvalue contributing to it. When we move our attention to a
more realistic analysis in which we fit for the continuum, etc, we
find that the estimators make an overall underestimate of the variance
by about 35\%, but also that this underestimate is primarily produced
by the large eigenvalues of $(C_i-C_{\rm tot})$. In other words, the
noisier eigen-modes of the covariance matrix are more underestimated
than the less noisy ones. We find no structure if we slice the data in
other parameters (separation, redshift, etc.)

This finding suggests an alternative way of correcting the covariance
matrix. We fit the ratio of measured to expected scatter in Figure
\ref{fig:ks} as a function of eigenvalue rank with a smooth curve and
apply it to each covariance matrix separately. The so-constructed
matrix will have desired properties in terms of the SK test by
construction.

\subsubsection{Choice of covariance matrix}
\label{sec:which-covar-matr}

To recap this section, we have found that the estimator-provided
covariance matrix is a decent approximation, but not sufficiently
accurate to provide reliable error-bars. Hence we used three different
methods for estimating the errors on our BAO parameters:

\begin{itemize}
\item \emph{Method 1.} Perform the eigen-value decomposition of the
  estimator matrix into eigenvectors $\mathbf{v}_i$ and eigenvalues $\sigma_i^2$,
  and for each eigen-vector we determined the variance predicted by
  bootstrapped covariance matrix:
  \begin{equation}
    \sigma^2_{i,{\rm bs}} = \mathbf{v}_i^T C_{\rm bs} \mathbf{v}_i.
  \end{equation}
  We set the variance to the larger of the two: $\sigma_i^2
  \rightarrow {\rm max}(\sigma_i^2,\sigma_{i,{\rm bs}}^2)$ and
  reconstruct the covariance matrix.

\item  \emph{Method 2.} Use the eigenvalue-ranked fixed individual
  matrices as described in Section  \ref{sec:sk-test}

\item \emph{Method 3.} Use the bootstrap technique on the final BAO
  parameters. Generate boot-strap samples as described in Section
  \ref{sec:bootstr-covar-matr},  then proceed to fit the BAO
  parameters with it. Use the final distribution of these parameters
  as an estimate of the uncertainty in those parameters.

\end{itemize}

As we will see, these estimates provide compatible results in the
vicinity of high-likelihood regions. They do, however, differ in the
tails of distributions; we will return to this point later.

It might be argued that one should take the version of Method 1 in
which the roles of bootstrap and estimator covariance matrices are
reverse, i.e.  bootstrap matrix is expanded eigenvalue-by-eigenvalue
and small eigenvalues corrected by their variances in the estimator
matrix.  The problem with this approach is that in general
eigen-vectors of the bootstrapped matrix are not orthogonal to the 36
large-eigenvalue eigen-vectors in the estimator matrix. We attempted
various ways to address this issue, but none produced a satisfactory
matrix (i.e.\ one that would pass tests described in Section
\ref{sec:fitt-bary-acoust}). We are therefore omitting this approach
in the paper.

\section{Fitting Baryonic Acoustic Oscillations parameters}
\label{sec:fitt-bary-acoust}

The process of fitting the BAO oscillation parameters is described in
 detail in the companion paper \cite{peanutboy}. In short, we
model the data as 
\begin{equation}
  \xi_{\rm observed} (\Delta \log \lambda, s, z)  = \xi_{\rm cosmo}
  (r_\parallel, r_\perp,  \apar,  \aperp) (1+B_m(r,\mu,z)) + B_a(r,\mu,z).
\end{equation}
That is, using the standard cosmological model, we convert the
observed coordinates ($\Delta \log \lambda$, $s$, $z$; difference in
logarithm of observed wavelength, separation and redshift) into
cosmological coordinates ($r,\mu,z$) (radial distance in correlation
function, angle with respect to the radial direction and
redshift). Functions $B_m$ and $B_a$ are multiplicative and additive
broadband model discussed below. Our model for $\xi_{\rm cosmo}$ is
given by
\begin{equation}
  \xi_{\rm cosmo} = \xi_{\rm no\ peak}  (r_\parallel, r_\perp ) +
  a_{\rm peak} \cdot \xi_{\rm peak} (r_\parallel \apar, r_\perp \aperp),
\end{equation}
where we have decomposed the linear theory correlation function into a
smooth component and the peak, but as opposed to most other BAO work, \emph{we
  only dilate the peak part of the correlation function.} As discussed
in \cite{peanutboy}, this is the most conservative way of ensuring no
information is received from the broadband shape of the correlation
function.  The cosmological model has an adjustable ``peak amplitude''
parameter $a_{\rm peak}$. This amplitude is usually set to unity
(recovering the full model), unless we explicitly state that we are
fitting for it.

The two isotropic dilation parameters measure the ratio of the BAO
standard ruler size, namely sound speed $r_s$, to the appropriate
distance scale
\begin{eqnarray}
  \apar &=& \frac { \left[ (r_s H(z))^{-1} \right] }{\left[ (r_s H(z))^{-1}\right]_{\rm
    fid}} \\
  \aperp &=& \frac { \left[D_a/r_s(z)\right] }{\left[D_a/r_s\right]_{\rm
    fid}} 
\end{eqnarray}
with respect to our fiducial model, which is flat $\Lambda$CDM with
$\Omega_b h^2=0.0227$, $\Omega_m=0.27$ and $h=0.7$. In the above
equations $H(z)$ is the Hubble parameter and $D_a(z)$ is the comoving
angular diameter distance.

For most of the test, we perform fits with the isotropic dilation
factor $\aiso$ which corresponds to the fits where we set
$\apar=\aperp$ ($=\aiso$). These isotropic fits fit a single parameter
and hence have higher signal-to-noise and are similar to fits to the
monopole of the correlation function often performed in the case of
galaxy BAO. The reason why our measurement cannot be usefully
compressed to $D_v=[((1+z) D_A)^2 (cz/H(z))]^{1/3}$ quantity lies in
the fact that the redshift-space distortion parameter $\beta$ is
considerably larger in the case of \lya forest, considerably
increasing the signal-to-noise ratio in the radial direction. One
cannot therefore assume that the relative amount of information coming
from $D_a(z)$ and $H^{-1}(z)$ will simply follow the geometrical factors of
$2/3$ and $1/3$ respectively.

When fitting the synthetic data we use a linear model, because this is
what the synthetic data assumed. However, when fitting the real data,
we use the model that has been smoothed to model the weakly non-linear
dark matter power spectrum at the redshift of interest. Strictly
speaking, this approach is not necessary as it negligibly affects our
$\chi^2$ and only marginally increases our error-bars, as we discuss
further in section \ref{sec:non-linear-effects}.  We assume that the
interpolation in the estimate of the correlation function takes care
of the finite bin-size in the correlation function measurement.  The
linear cosmological model at a given redshift has two parameters: the
bias $b$ and the redshift-space distortion parameter
$\beta$. Following \cite{2011JCAP...09..001S}, we use the independent
parameters $\beta$ and $(1+\beta)b$.

We also use a parameter that governs the redshift evolution of
the clustering amplitude, $\gamma_b = d \log ( b^2g^2) / d
\log(1+z)$, where $g(z)$ is the growth factor. This means that we
never explicitly use the growth-factor: we calculate the linear
prediction at the reference redshift and then scale the linear power
spectrum (together with non-linear corrections) using $\gamma_b$.
Because the inclusion of broadband parameters significantly degrades
our ability to measure these parameters we institute a broad Gaussian
prior on $\beta=1.4\pm 0.4$ and $\gamma_{b}=3.8 \pm 1$.

We additionally fit for a multiplicative broadband ($B_m$) and an
additive one ($B_a$). These two broadband functions are smooth and are
designed to remove any non-peak BAO information as described
\cite{peanutboy}. They are defined to be of the form
\begin{equation}
  B_{a/m}= \sum A_{i,\ell,k} r^i L_\ell(\mu) \left(\frac{1+z}{1+z_{\rm ref}}\right)^k,
\label{eq:bbd}
\end{equation}
where $L_\ell$ is the Legendre polynomial of the $\ell$-th order and
where we fit around 10 $A_{i,\ell,k}$ parameters simultaneously. In
Table \ref{tab:bb} we tabulate which combinations of $i,\ell,k$ terms
we float in the 6 broadband models considered in this paper. We have
also investigated less physically motivated broadband models such as
those with odd powers of $\mu$ and stronger angular dependencies. All
models gave consistent results. For more information see the companion
paper \cite{peanutboy}.

\begin{table}
  \centering
  \begin{tabular}{c|ccccc}
    Name & Type & $i$ & $\ell$ & $k$ &  \# free parameters \\
    \hline
    BB1 & Additive & 0\ldots 2 & 0,2,4 & 0 & 9 \\
    BB2 & Additive & -2\ldots 0 & 0,2,4 & 0 & 9 \\
    BB3 & Additive & 0\ldots 2 & 0,2 & 0,1 & 12 \\
    BB4 & Multiplicative & 0\ldots 2 & 0,2,4 & 0 & 8 \\
    BB5 & Multiplicative & 0\ldots 2 & 0,2 & 0,1 & 10 \\
    BB6 & Additive + Multiplicative & 0,1 & 0,2,4 & 0 & 11 \\
  \end{tabular}
  \caption{Broadband models used in this work. This table shows span
    of power indices $i$,$\ell$ and $k$ used for different broadband
    model defined in Equation \ref{eq:bbd}. When counting free parameters, one must take
    into account that some are perfectly degenerate with standard bias
    parameters (i.e.\ monopole in multiplicative broadband is
    equivalent to bias change).  See text for discussion.}
  \label{tab:bb}
\end{table}

Our fit uses data in the range $50\mpch-190\mpch$. This choice
contains sufficiently large buffers at both ends of the fitting range
to be able to determine the broadband parameters while preventing the
sharp upturn in the correlation function at smaller separations from
affecting our fit. We also always cut the data at $\Delta \log \lambda
< 0.003$ as these data are the most affected by the sky subtraction
residuals \cite{2011JCAP...09..001S}.

For each model, we minimize $\chi^2$, varying all parameters except
the parameter of interest, e.g., $\aiso$ or $\apar$/$\aperp$, and then
gridding the minimum $\chi^2$ over the parameters of interest. This
procedure is equivalent to marginalizing over all the other parameters
in the limit of those marginal likelihoods being Gaussian. This
approach is a good approximation when the broadband parameters are not
completely degenerate. We have found that combining all broadband
models into  general broadband models (with $\sim 100$ parameters)
results in the constraints that are very degenerate (in the sense that
the data cannot distinguish between terms in the broadband model). In
such case, our explicit Monte Carlo tests have shown that the prior
volume completely dominates the posterior, or equivalently, the
$\Delta \chi^2$ distribution in minimization becomes very non $\chi^2$
distributed (in the sense that incorrect solutions can be found that are
up to $\Delta \chi^2>25$ away from the true solution).

We calculate uncertainty on the parameters using a Bayesian approach:
we convert $\Delta \chi^2$ values into probabilities and then measure
the median and relevant points in the cumulative probability
distribution. This is an important point as the distribution is
non-Gaussian with the confidence limits being asymmetric at a 10\%
level for our data.

\subsection{Tests on synthetic data.}

We start by discussing the goodness of fit for our basic fit.  These
results are shown in Table \ref{tab:chi2mocks}. This table shows that
the 15 synthetic mocks on average produce a good $\chi^2$ that is
independent of the broadband model used. We show results for both
Method 1 and Method 2 covariance matrices. Table
\ref{tab:chi2mocks} also presents the results of combining fits to the
isotropic BAO from all mock realizations;  results are consistent with
no bias in the method and are consistent between the two methods for
estimating the covariance matrix. There is weak evidence that Method 2
might be performing slightly better than Method 1.

\begin{table}
  \centering
  \tiny
  \begin{tabular}{c|cccccc}
    Broadband & BB1 & BB2 & BB3 & BB4 & BB5 & BB6  \\
    \hline
    % ./printtable.py 2
    Best fit $\chi^2$ Me1  & $904 \pm  42$ & $904 \pm 42.3$ & $898 \pm 39.5$ & $904 \pm 42.4$ & $901 \pm 41.3$ & $899 \pm 42.5$  \\
    Best fit $\chi^2$ Me2  & $908 \pm 40.6$ & $908 \pm 41.2$ & $899 \pm 36.9$ & $908 \pm 40.8$ & $903 \pm 38.3$ & $902 \pm 41.2$  \\
    \hline
    $100\times(\aiso-1)$\\
    % ./printtable.py 4
    Comb. r. Me1 & $-0.795 \pm 0.749$ & $-0.544 \pm 0.761$ & $-0.641 \pm 0.748$ & $-0.818 \pm 0.705$ & $-0.862 \pm 0.675$ & $-1.06 \pm 0.728$ \\
    Comb. r. Me2 & $-0.677 \pm 0.727$ & $-0.303 \pm 0.735$ & $-0.514 \pm 0.726$ & $-0.551 \pm 0.675$ & $-0.518 \pm 0.648$ & $-0.905 \pm 0.684$ \\
  \end{tabular}

  \caption{Mean and variance (printed as error-bar) 
    of best fit $\chi^2$ for different
    choices of covariance matrix (labels Me1 and Me2 refer to Method 1
    and Method 2 covariance matrix) and broadband model (BB1\ldots BB6) for 
    our 15  synthetic datasets. The effective
    number of degrees of freedom is 894 for BB1. We also show the result
    of fitting for the BAO position when all 15 $\Delta \chi^2$
    contributions are summed. 
  }
  \label{tab:chi2mocks}
\end{table}

\begin{figure}[h!]
  \centering
  \includegraphics[width=\linewidth]{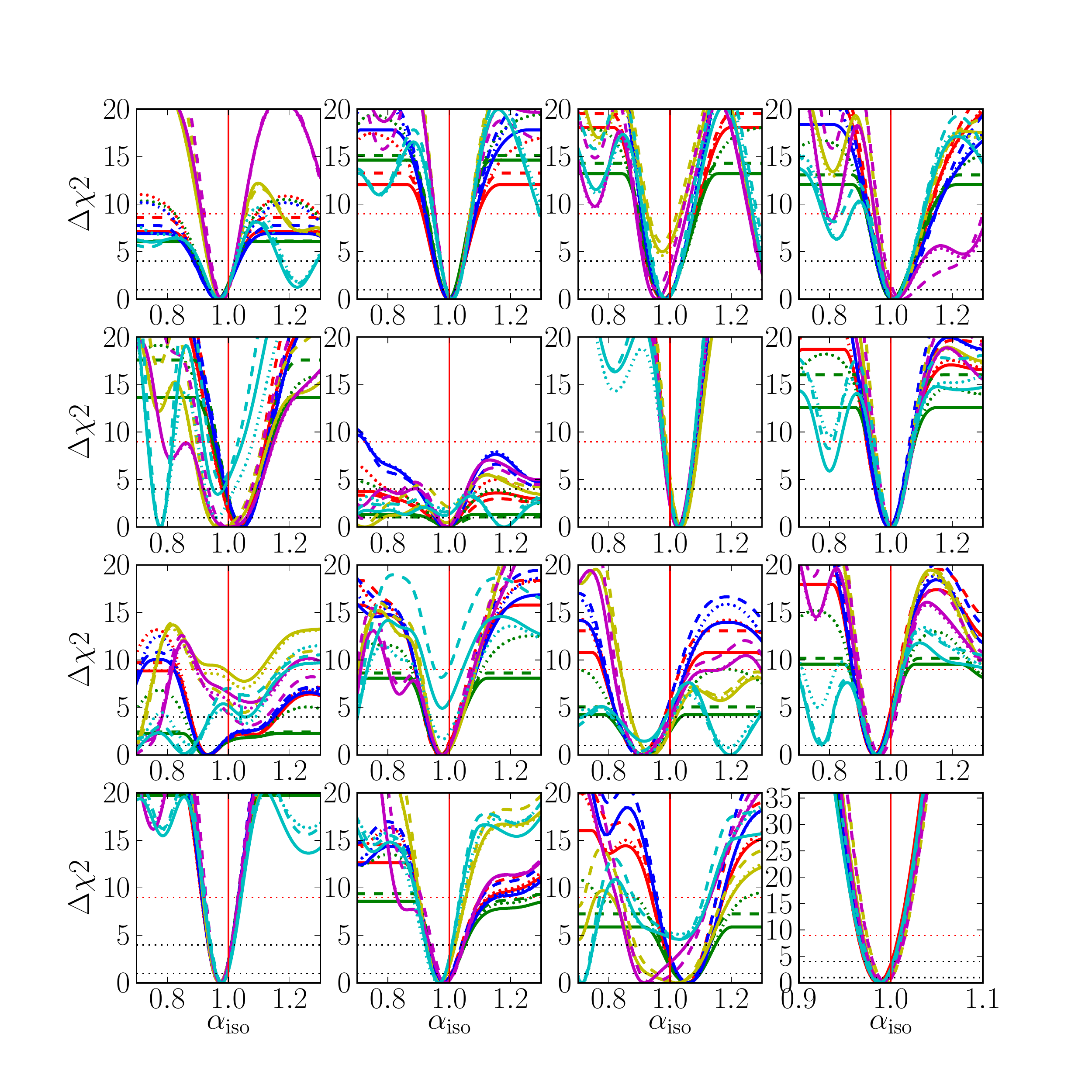}
  % ./plotbaopos.py 1
  \caption{$\Delta \chi^2$ curves for 15 synthetic realizations of the
    full dataset and the results of combining those curves (bottom
    right panel). Different colors corresponds to different broadbands:
    red (BB1), green (BB2), blue (BB3), cyan (BB4), magenta (BB5),
    yellow (BB6). Dashed lines are the same for the Method 2
    covariance matrix. Dotted lines are the same upon applying the bias
    prior. Note the different scale on the axes for the case of
    combined realizations.   }
  \label{fig:fluxpos}
\end{figure}

Figure \ref{fig:fluxpos} displays the $\chi^2$ contours for different
broadbands for our synthetic data for the isotropic fit for the Method
1 covariance matrix. We also show the results of the combined fit,
which was calculated by summing $\Delta \chi^2$ contributions from
each individual measurement. The curves are parabolic in the vicinity
of the best fit position, but far from this point, they yield a
constant for the additive broadband.  In those cases, the inferred
bias is always close to zero: the model decides that the penalty for
having a peak at the wrong position is too large and instead replaces
most of the correlations with the broadband model which then makes the
fit independent of $\aiso$. This is not the case for broadbands with a
multiplicative component, and in this case we indeed see structure
outside the most likely region. We also find that for some ``lucky''
realizations, all broadbands agree on the position of the peak and have
the same error-bars, while for some the differences can be
staggering. In particular, multiplicative broadbands often latch onto
noise features.

The ability of the broadband model to completely replace the true
cosmological model when fitting the peak at an undesirable position is
unphysical. In other words, one should be able to use the information
that there is no evidence for a peak at certain dilations, not just
the evidence that there is a peak at others. We can cure this problem
by instituting a weak prior on $(1+\beta)b = 0.336\pm 0.12$. This
prior essentially states that at two sigma, at least one third of the
total signal must be cosmological in origin. Note that while the
central value of this prior is the same as the value measured in
\cite{2011JCAP...09..001S}, the error is \emph{ten} times larger,
i.e. this is a really weak prior compared to what measurements say
about this quantity. $\chi^2$ curves for those fits are plotted as
dotted lines in the Figure \ref{fig:fluxpos}.  The bottom line is that
outside the favored region, the $\Delta \chi^2$ difference increases
further, but in the region of high-likelihood it is not affected, so
this prior will not be used for the final results.

Results for Method 2 are quite similar to those of Method 1. In Figure
\ref{fig:fluxpos} we also plot comparisons between methods for two
ways of estimating covariance method (dashed vs solid lines). We see
that Method 2 in general gives higher significance, but not uniformly
so.

% \begin{figure}[h!]
%   \centering
%   % ./plotbaopos.py 2
%   \includegraphics[width=\linewidth]{fluxposK}
%   \caption{Same as Figure \ref{fig:fluxpos}, but showing the difference
%     between the covariance matrices from Method 1 and Method 2 (dotted).
%   }
%   \label{fig:fluxposK}
% \end{figure}

\begin{figure}[h!]
  \centering
  % ./cumhistos.py 1/2

  \includegraphics[width=0.49\linewidth]{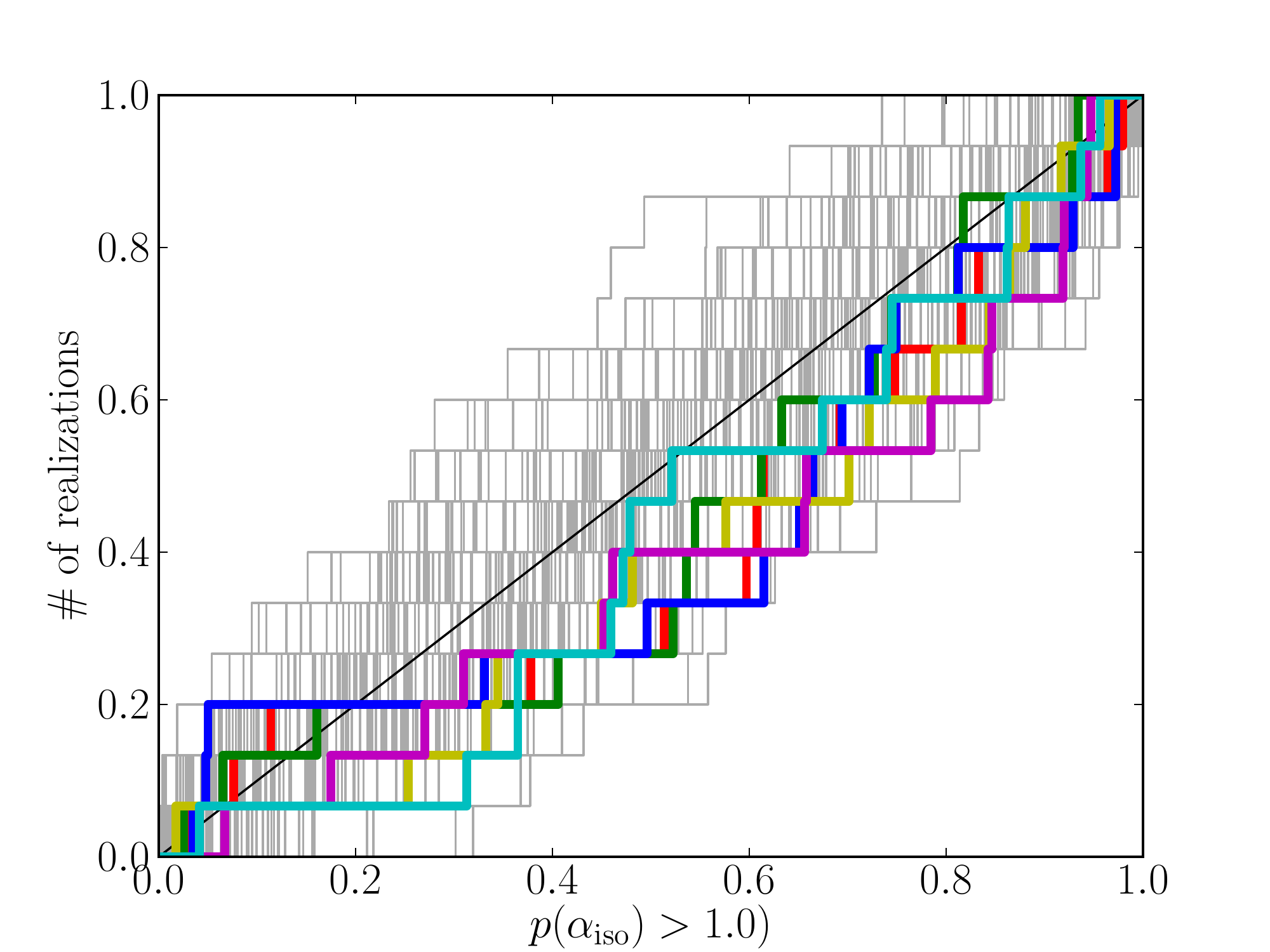}
  \includegraphics[width=0.49\linewidth]{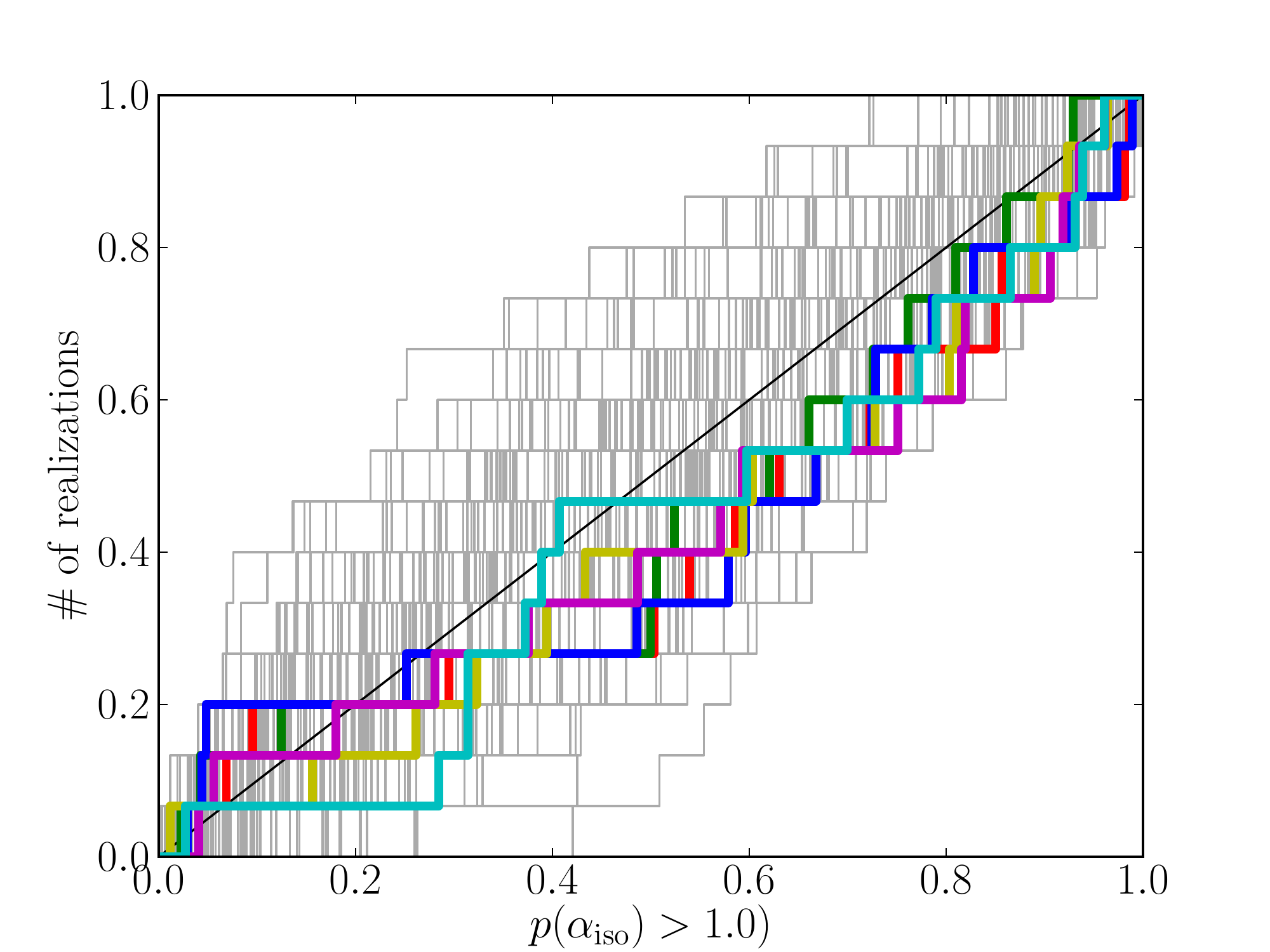}
  \caption{The cumulative distribution of the fraction of realizations
    with value of $\alpha_{\rm iso}$ above the fiducial value when
    fitting synthetic data. For correctly estimated errors, this
    distribution should be flat between zero and unity, producing a
    straight line for cumulative distribution, plotted as the thick
    black line. Colors correspond to different broadband models and are the
    same as in Figure \ref{fig:fluxpos}. The left plot is for Method 1
    covariance matrix while the right plot is for Method 2 covariance
    matrix. Faint gray lines are 100 realizations of 15 points drawn
    from uniform distribution.}

  \label{fig:grey}
\end{figure}

Next we investigate the probability that $\aiso>1.0$ on each individual
mock. If our estimates are unbiased and errors correctly estimated,
this probability should be uniformly distributed between $0$ and $1$.
We plot the cumulative distribution of this quantity in  Figure
\ref{fig:grey}. 

This figure deserves some attention. From the figure it appears that
there might be a mild tension with the expected cumulative
distribution.  The minimum and maximum values of $p(\aiso>1)$ are
around 2-3\% and 93-99\%, which are consistent with no bias. Plotting
the distribution with  $p(\aiso>0.995)$ or $p(\aiso>1.005)$
produces outliers in the 1 percentile range.  Our synthetic datasets
share the same continuum realizations and are thus not completely
independent, with continuum errors not canceling out when one sums
realizations.  This results in any one realization being consistent
with the correct solution (i.e., no outliers), but they are somewhat
correlated leading to skewed plots in Figure \ref{fig:grey}.

In the absence of a larger number of and more independent synthetic
datasets, we associate a $0.5$\% systematic uncertainty with our BAO
peak position fitting.

Next we move to the anisotropic fit and relax the assumption of
$\aperp=\apar$.  This is not a purely statistical exercise as the two
quantities probe different underlying physical parameters: $\aperp$ is
a measure of the comoving angular diameter distance to the effective
redshift, while $\apar$ is the measure of the local expansion rate at
the redshift of interest. If one wants to use our results to estimate
cosmological parameters, it is important to distinguish between the
two parameters.

\begin{figure}[h!]
  \centering
  % ./plotbaoani.py 11
  \includegraphics[width=\linewidth]{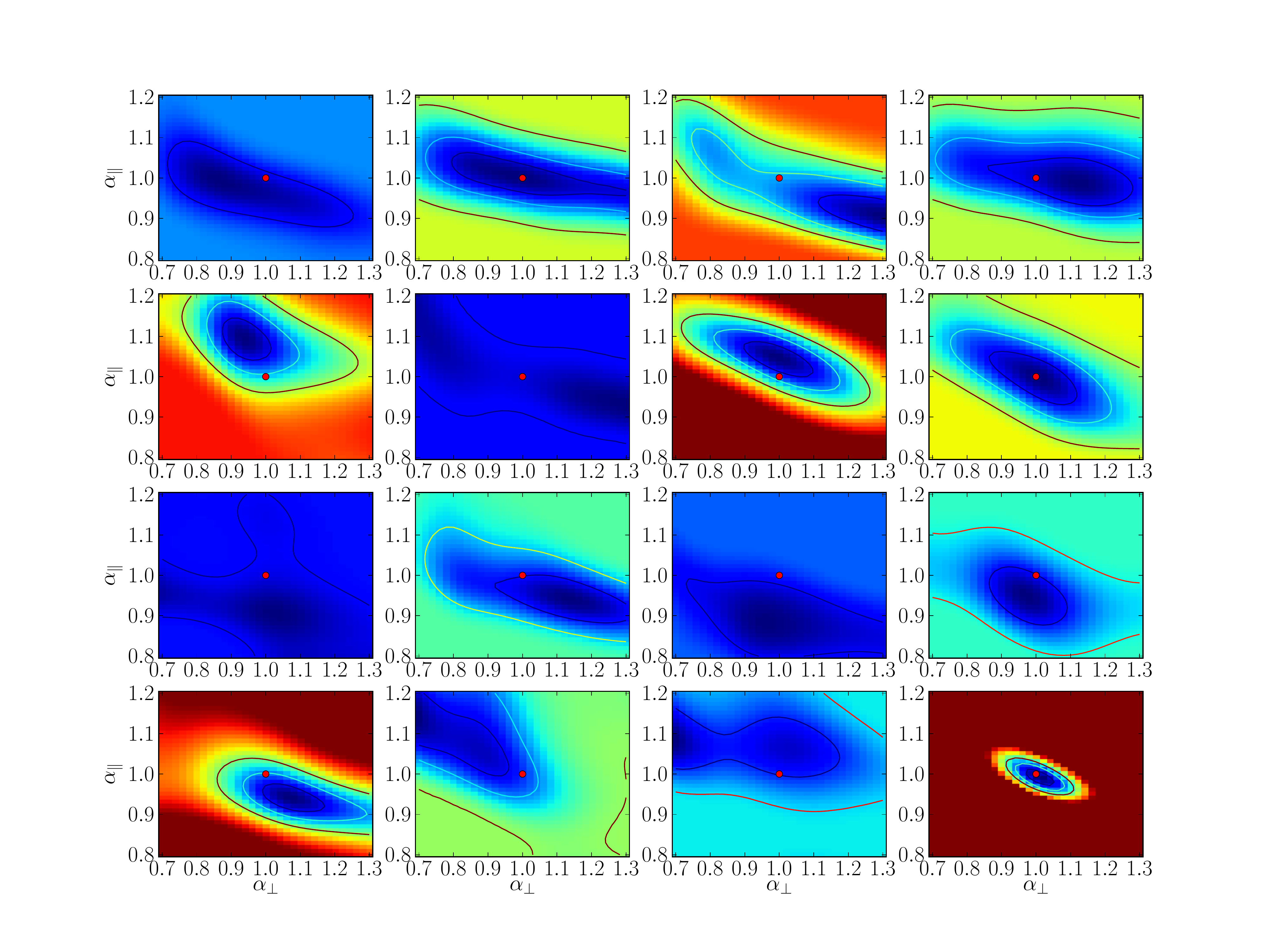}
  \caption{$\Delta \chi^2$ contours for 15 synthetic realizations of
    the full dataset and the results of combining those curves (bottom
    right plot). Contours are plotted at probabilities that enclose
    68\%, 95\% and 99.7\% probability. This plot is for broadband
    model 3 and the Method 2 covariance matrix. Color scale saturates
    at $\Delta \chi^2=25$. }
  \label{fig:fluxani}
\end{figure}

We begin by plotting the equivalent of Figure \ref{fig:fluxpos} for
the anisotropic fit.  This is presented in Figure \ref{fig:fluxani}.
We show plots for one broadband (BB3) and the method 2 covariance
matrix. Plots for the method 1 covariance matrix look very similar with
larger error-bars. Plots for other broadband models again are similar,
although the contours  do move for realizations where BAO is only weakly
detected. For the strong detections and for the combined fit, the
contours are essentially the same.

\subsection{Determination of uncertainties}
\label{sec:determ-uncert}

It is clear that for realizations for which the detection of BAO is
strong, the likelihood is Gaussian and the error-bars symmetric and
Gaussian. In this work we calculate error-bars by applying a uniform
prior of $\aiso$ between 0.8 and 1.2 and calculating the confidence
limits from the appropriate points in the cumulative distribution
function. We compare these results with errors derived from a Gaussian
expansion of the likelihood around the maximum-likelihood point in 
Table \ref{tab:NG}. The $1$-sigma error-bars are always
adequately determined by the Gaussian approximation (although also
systematically underestimated at a percent level). However, for
$3$-sigma error-bars, the Gaussian approximation can fail
catastrophically (i.e.\ the 3-sigma error-bars are limited by the
prior). This result is to be expected; when the significance of the BAO detection
is low, the information becomes prior dominated at higher levels of confidence \cite{2010arXiv1005.1664B}.

\begin{table}
  \centering
  \begin{tabular}{c|ccc|c}
    Set & +$N\sigma$ & $-N\sigma$ & $N\sigma$ symmetrized & $N\sigma$ Gaussian approx \\
    \hline
    % ./esterr.py bao_pos_allz_bb3/Flux_aln.7.scan
    Re 7, N=1 & 1.59 & 1.58 & 1.58 & 1.53 \\
    Re 7, N=2 & 3.28 & 3.22 & 3.25 & 3.06 \\
    Re 7, N=3 & 5.22 & 5.07 & 5.15 & 4.59 \\
    \hline
    Re 8, N=1 & 2.31 & 2.27 & 2.29 & 2.19 \\
    Re 8, N=2 & 5.24 & 4.94 & 5.09 & 4.39 \\
    Re 8, N=3 & 16 & 15.7 & 15.9 & 6.58 \\
    \hline
    Re 12, N=1 &2.79 & 2.59 & 2.69 & 2.45 \\
    Re 12, N=2 &8.57 & 5.88 & 7.22 &  4.9 \\
    Re 12, N=3 &23.7 & 14.2 & 18.9 & 7.35 \\
    \hline
    Re 13, N=1 &1.81 & 1.76 & 1.78 & 1.74 \\
    Re 13, N=2 &3.83 & 3.59 & 3.71 & 3.48 \\
    Re 13, N=3 &6.43 & 5.79 & 6.11 & 5.22 \\
  \end{tabular}
  \caption{Gaussianity of errors for select realizations for broadband
    model 2. Middle section correspond to Bayesian errors determined
    from cumulative PSF while the right column is the Gaussian
    expansion around the most likely point.}
  \label{tab:NG}
\end{table}

\section{Results on Baryonic Acoustic Oscillation parameters}
\label{sec:results-bary-acoust}

We start with the goodness of fit for our data.  Fixing BAO parameters
to the fiducial cosmological model and using no broadband, we can
attempt to fit for the basic bias parameters. With the method 1
covariance matrix we find $\beta=1.7\pm0.9$,
$b(1+\beta)=-0.336\pm0.019$ and $\gamma_b=2.7\pm1.4$ and $\chi^2=896$
with 941-3 degrees of freedom. With the method 2 covariance matrix we
find $\beta=1.3\pm0.5$, $b(1+\beta)=0.350\pm0.016$ and
$\gamma_b=2.3\pm1.3$ and $\chi^2=958$. The simplest model of \lyaf
flux fluctuations, namely that we are measuring a linearly biased
tracer of the underlying dark-matter field with Kaiser-like
redshift-space distortions, is thus a satisfactory fit with just three
floating parameters and is consistent with \cite{2011JCAP...09..001S}
even when a completely different radial scale range is employed
($10\mpch<r<90\mpch$ in \cite{2011JCAP...09..001S} and
$50\mpch<r<190\mpch$ in this work; lack of small scales is also the
reason for increased error-bars in this work despite quadrupling of
the number of quasars). Since we have not tested these measurements on
synthetic data, we mention them here just to demonstrate that the
values are not wildly incompatible with previous measurements.

Table \ref{tab:chi2data} is the equivalent of Table
\ref{tab:chi2mocks} for the data.  All broadband models lead to an
improvement in $\chi^2$ compared to the pure cosmology fit, beyond
what is expected from extra degrees of freedom. This was not the case
in the synthetic data.  The $\chi^2$ values are low but not
pathologically so compared to expectation.

\begin{table}
  \centering
  \begin{tabular}{c|cccccc}
    Broadband & BB1 & BB2 & BB3 & BB4 & BB5 & BB6 \\ 
    \hline
    % ./printtable.py 1
    Me1 & 855.3 & 857.3 & 830.7 & 864.2 & 849.6 & 856.9  \\
    Me2 & 897.5 & 900.2 & 863.7 & 908.4 & 890.5 & 900.1  \\
  \end{tabular}

  \caption{Best fit values of  $\chi^2$ for different
    choices of covariance matrix (labels Me1 and Me2 refer to Method 1
    and Method 2 covariance matrix) and broadband model (BB1\ldots BB6) for 
    data. Table \ref{tab:chi2mocks} but for data. See text
    for discussion.}
  \label{tab:chi2data}
\end{table}

\begin{figure}[h!]
  \centering
  % ./plotbaoposx.py 1
  \includegraphics[width=\linewidth]{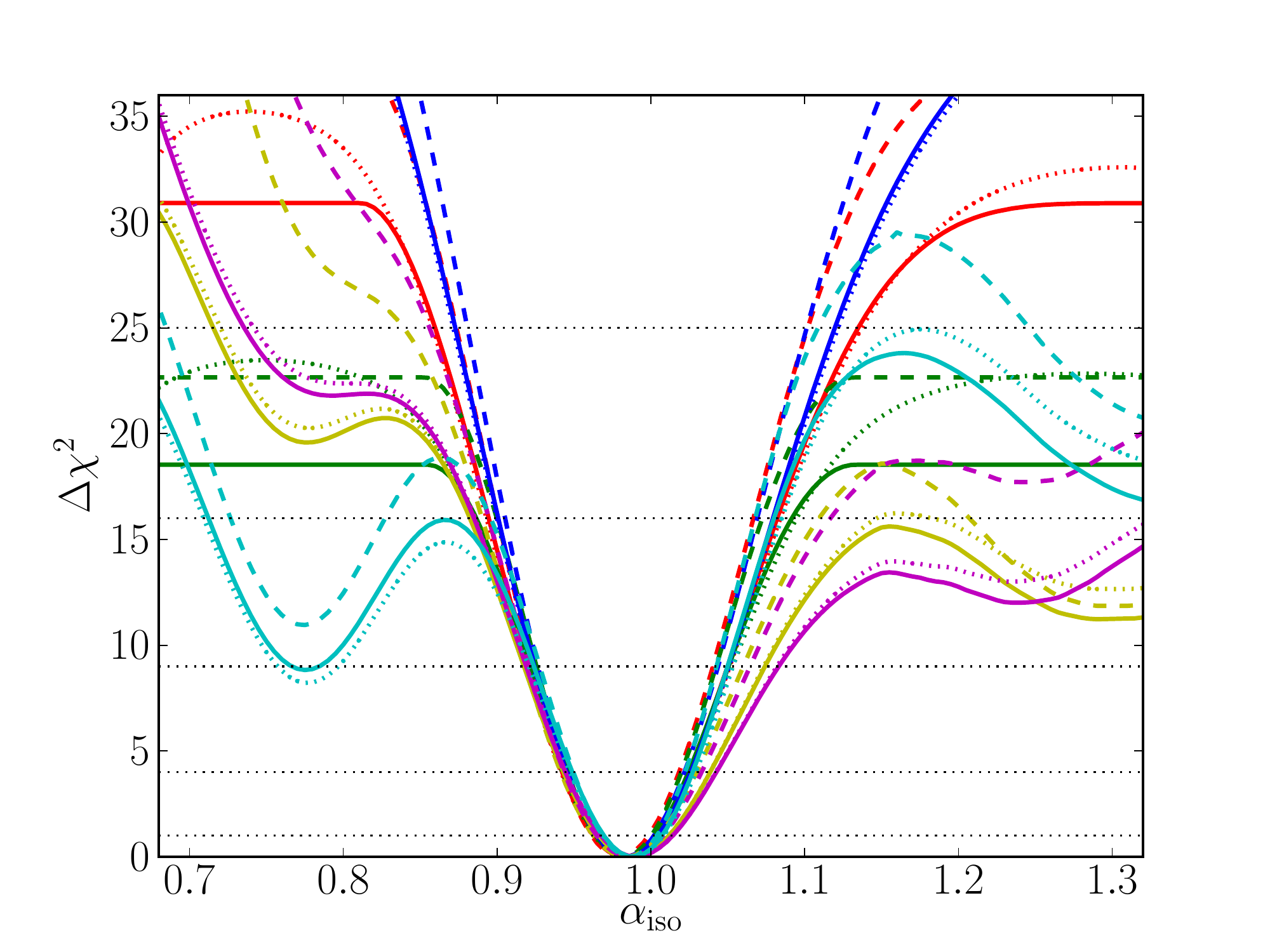}
  \caption{Variation of $\chi^2$ as a function of $\aiso$ for the
    data. Lines of different colors are for different broadband
    models, following the same color scheme as in Figure
    \ref{fig:fluxpos}. Solid lines are for Method 1 covariance matrix
    without applied prior. Dotted lines are upon applying the bias prior
    and dashed lines are for the Method 2 covariance matrix. }
  \label{fig:prioreff}
\end{figure}

\begin{figure}[h!]
  \centering
  % ./plotbaoposx.py 4
  \includegraphics[width=\linewidth]{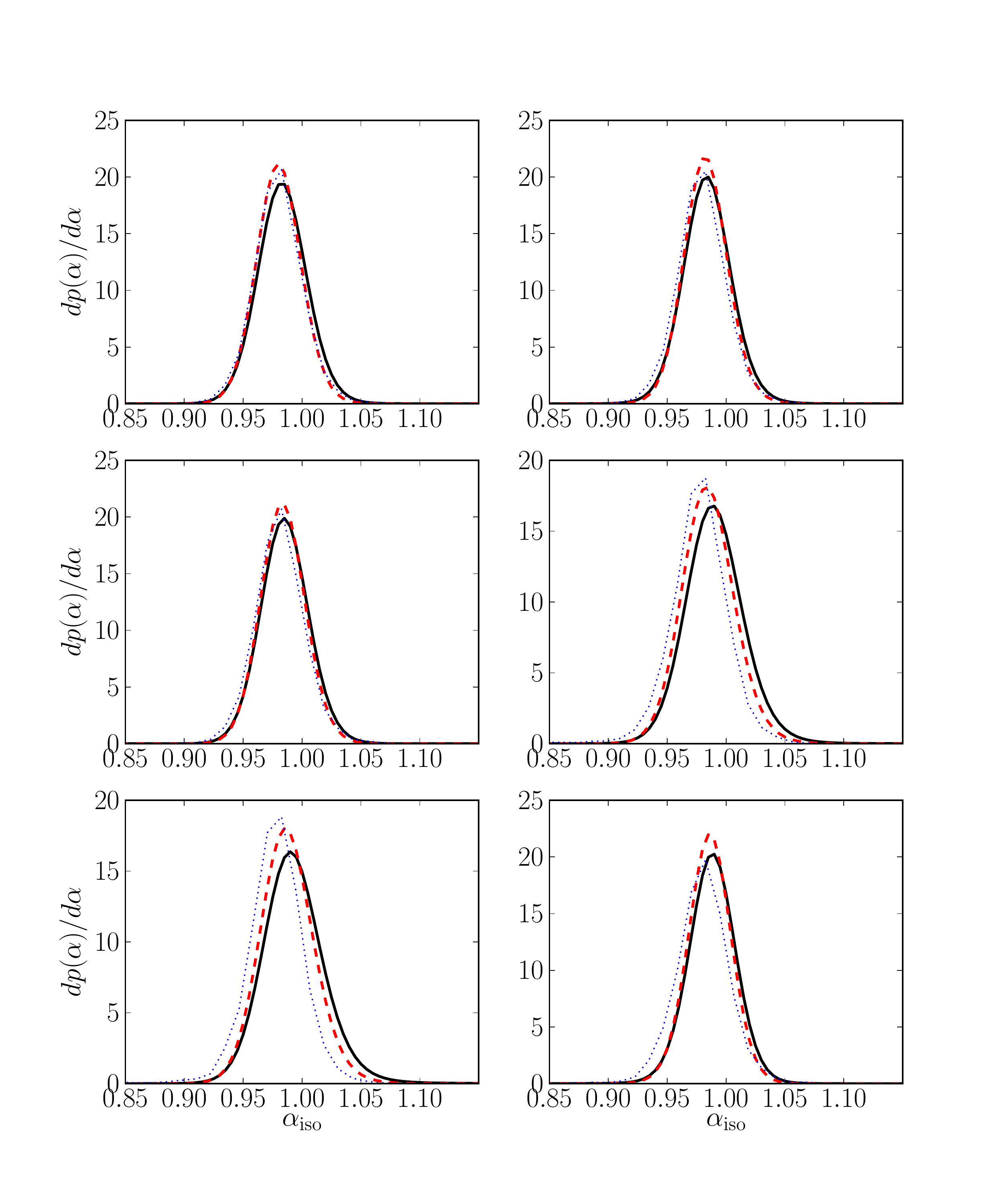}
  \caption{Probabilities for isotropic parameter $\aiso$.  The panels
    correspond to six broadband models. For each model we plot the
    Method 1 inferred probability with the solid black line, Method 2
    with the dashed red line and the bootstrap number density with
    the dotted blue line. }
  \label{fig:m3}
\end{figure}

\begin{figure}[h!]
  \centering
  % ./plotbaoposx.py 41
  \includegraphics[width=\linewidth]{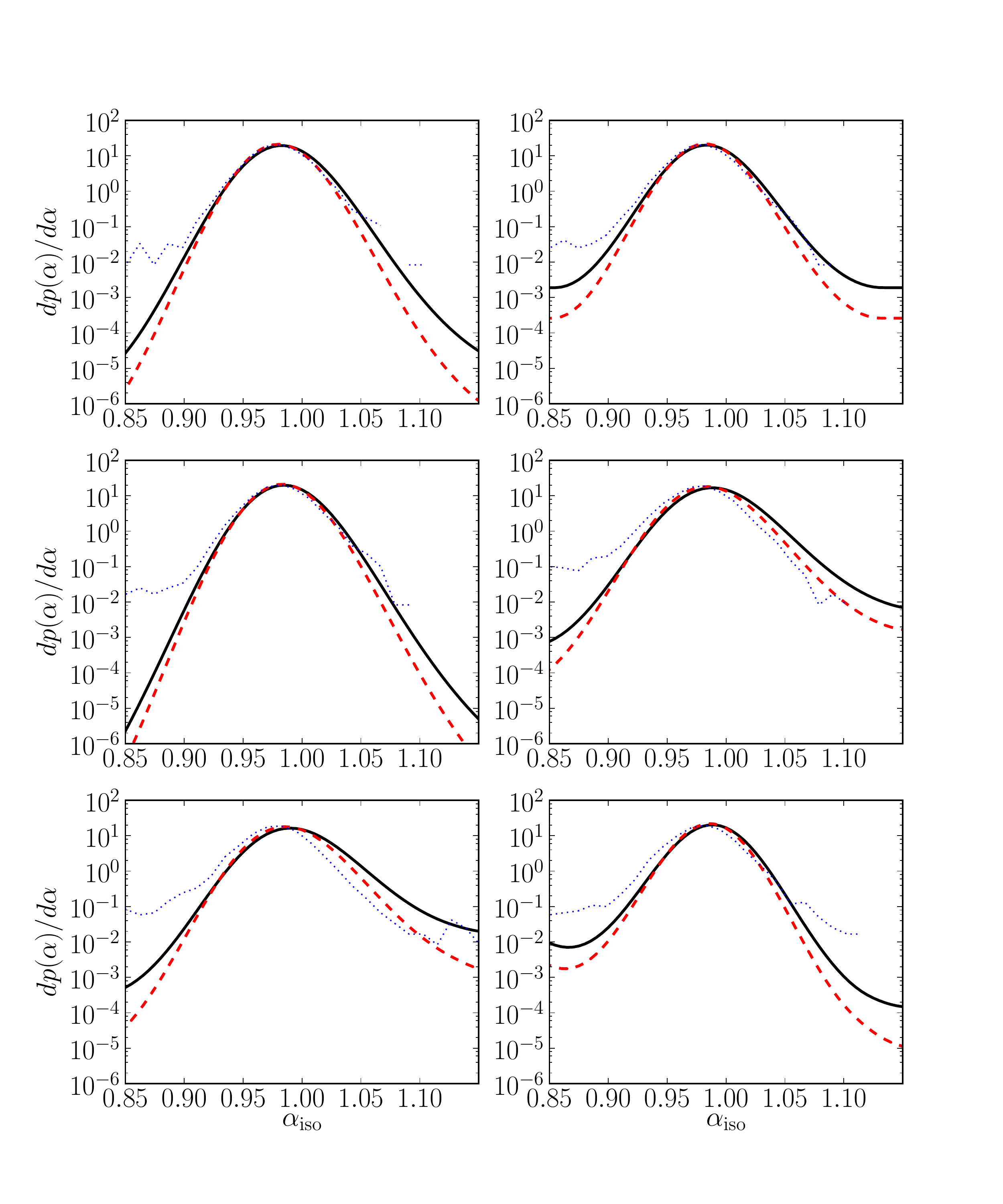}
  \caption{Same as Figure \ref{fig:m3}, but with a logarithmic
    vertical axis.}
  \label{fig:m3l}
\end{figure}

In Figure \ref{fig:prioreff} we show the $\chi^2$ curves for the
data. Even without the prior on bias (dotted lines), we have a solid
4-sigma detection even in the case of the most conservative BB3 prior.
The same figure also displays how these constraints depend on the
choice of covariance matrix (solid vs dashed lines). Perhaps more
relevant are the probability plots, which are shown in Figures
\ref{fig:m3} and \ref{fig:m3l}. Here we have converted the same
$\chi^2$ values shown before into probabilities and have also added
the third method described in Section \ref{sec:which-covar-matr}:
bootstrapping on the final parameters. This test treats the bootstrap
samples of the BAO parameters as if they were the MCMC samples, i.e.,
we inferred the probability by treating the number density of the
samples as relative probability. Whether this is a statistically sound
procedure is not clear as it mixes frequentist and Bayesian
statistical approaches.  Figure \ref{fig:m3} demonstrates excellent
agreement between the three methods for most broadband models, except
for broadband model 5. The agreement likely stems from the fact that all
statistical methods agree in the Gaussian limit. However, when we make
the vertical axis logarithmic as we do in Figure \ref{fig:m3l}, we see
that bootstrap samples have considerably more outliers than predicted
by the Bayesian method.  Perhaps the correct conclusion is that our
error-bar estimates are sound only to some 3-$\sigma$ distance from
the most likely point.

% \begin{figure}[h!]
%   \centering
%   % ./plotbaoposx.py 2
%   \includegraphics[width=\linewidth]{covariances}
%   \caption{Same as figure \ref{fig:prioreff}, except that the
%     dashed line show the Method 2 covariance matrix.}
%   \label{fig:datacovs}
% \end{figure}

\begin{table}
  \centering
  {\tiny

    % % ./printtable.py 3
    \begin{tabular}{c|ccccccc}
      Broadband & BB1 & BB2 & BB3 & BB4 & BB5 & BB6 \\
      \hline
      Me1  & $-1.7 \pm 2.06$ & $-1.62 \pm 2.01$ & $-1.49 \pm 2.02$ & $-1.04 \pm 2.4$ & $-0.813 \pm 2.49$ & $-1.19 \pm 1.97$ \\
      Me1 w $\gamma_\beta$ & $-1.72 \pm 2.04$ & $-1.62 \pm   2$ & $-1.57 \pm 2.01$ & $-1.1 \pm 2.41$ & $-0.885 \pm 2.47$ & $-1.24 \pm 1.94$ \\
      Me1 w prior  & $-1.69 \pm 2.1$ & $-1.63 \pm 2.09$ & $-1.49 \pm 2.04$ & $-1.04 \pm 2.39$ & $-0.821 \pm 2.47$ & $-1.17 \pm 2.06$ \\
      Me1 packed obs. & X  & X  & X  & X  & X  & X  \\
      Me1 w Lin. th.  & $-1.87 \pm 1.86$ & $-1.88 \pm 1.91$ & $-1.73 \pm 1.78$ & $-1.53 \pm 2.17$ & $-1.57 \pm 2.07$ & $-1.78 \pm 1.85$ \\
      Me1 w alt. NL  & $-1.6 \pm 2.06$ & $-1.57 \pm   2$ & $-1.45 \pm 2.01$ & $-0.984 \pm 2.38$ & $-0.716 \pm 2.46$ & $-1.18 \pm 1.95$ \\
      Me2  & $-1.99 \pm 1.9$ & $-1.75 \pm 1.85$ & $-1.64 \pm 1.89$ & $-1.55 \pm 2.22$ & $-1.3 \pm 2.25$ & $-1.4 \pm 1.82$ \\
      Me2 w prior  & $-1.98 \pm 1.92$ & $-1.77 \pm 1.91$ & $-1.64 \pm 1.91$ & $-1.55 \pm 2.21$ & $-1.3 \pm 2.24$ & $-1.28 \pm 1.9$ \\
      Bootstrap  & X  & X  & X  & X  & X  & X  \\
      Bootstrap w prior & X  & X  & X  & X  & X  & X  \\
    \end{tabular}
  }
  \caption{The best fit $100\times(\aiso-1)$ (that is
    percentage deviation from fiducial model) for various choices of
    broadbands and covariance matrices.}
  \label{tab:yog}
\end{table}

These results are numerically condensed in Table \ref{tab:yog}. The
bottom line is that the fit is quite stable with respect to the
choice of covariance matrix and the broadband model. Results vary
mostly by around $\sim$ \nicefrac{1}{4} and in a few cases by $\sim$
\nicefrac{1}{2} sigma.  This result implies that choice of broadband
model and covariance matrix are not crucial and that our dominant
uncertainty is statistical rather than systematic.

\begin{figure}[h!]
  \centering
  % ./plotbaoani.py 2
  \includegraphics[width=\linewidth]{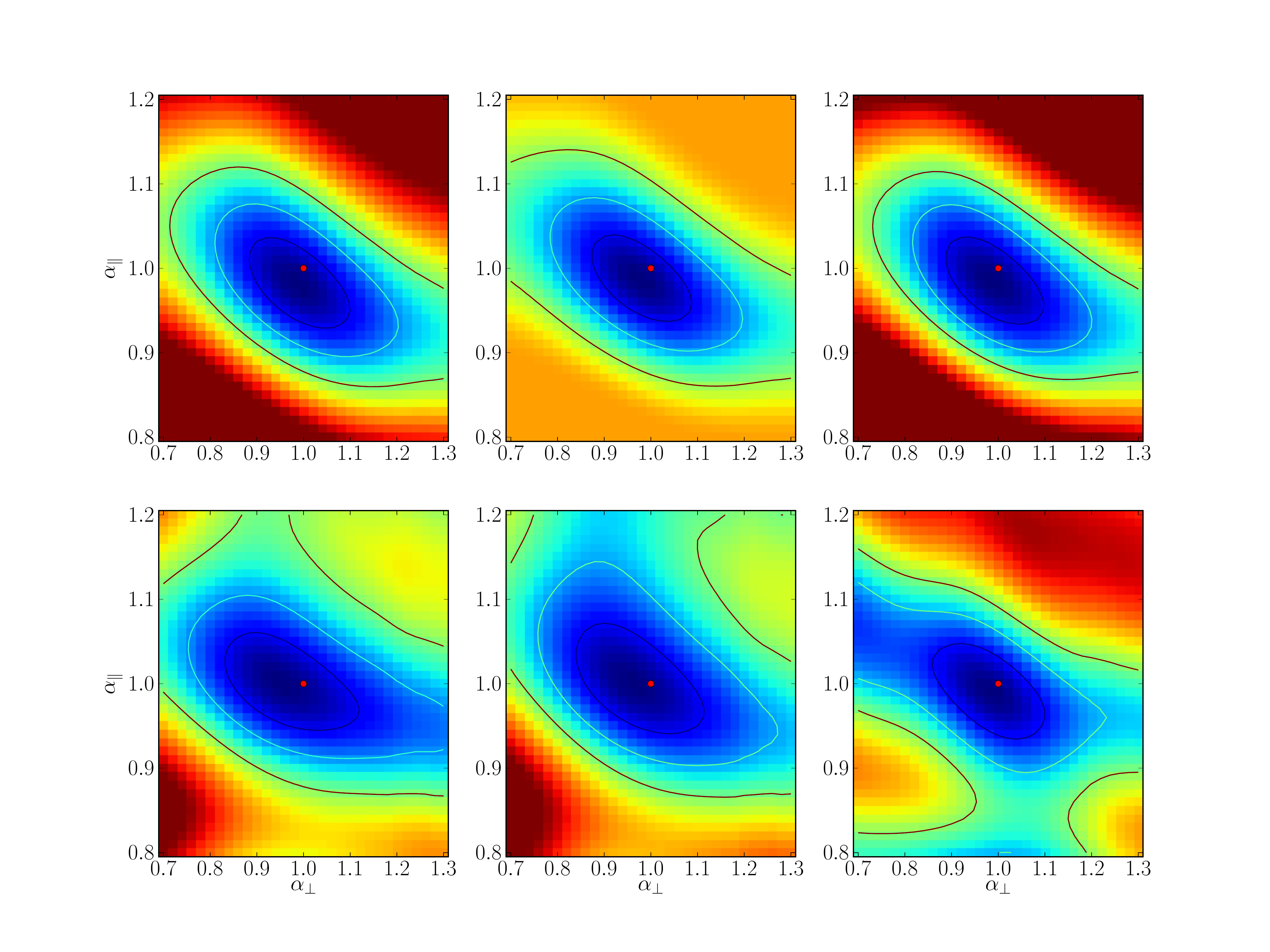}
  \caption{Contours for the anisotropic fit on the data for the Method
    1 covariance matrix. Plots show different broadband models,
    starting with BB1 (top left) to BB6 bottom right. Contours enclose
    68\%, 95\% and 99.7\% probability. The color scale
    saturates at $\Delta \chi^2=25$. }
  \label{fig:dataani}
\end{figure}

\begin{figure}[h!]
  \centering
  % ./plotbaoani.py 21
  \includegraphics[width=\linewidth]{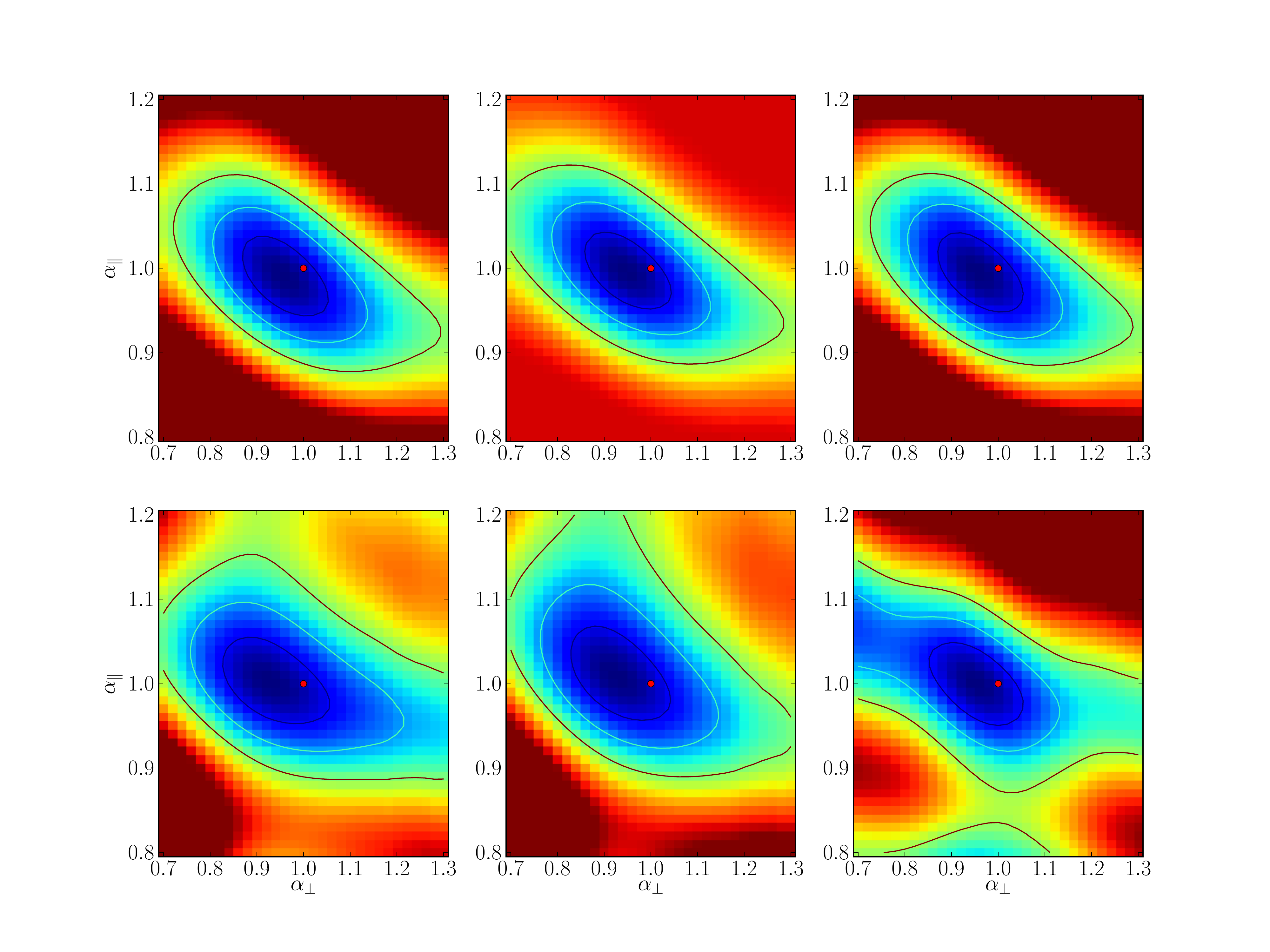}
  \caption{Same as Figure \ref{fig:dataani} but for Method 2
    covariance matrix. The contours at the same confidence level are tighter.}
  \label{fig:dataaniK}
\end{figure}

Next we turn to the anisotropic fit. The resulting contours for
different broadband models are plotted in Figures \ref{fig:dataani}
and \ref{fig:dataaniK}. The data are consistent with our fiducial
model and stable with respect to the broadband model used for up to
two-sigma confidence limits. Beyond that limit, the choice of
broadband significantly affects where the contours close. Therefore,
given the current signal-to-noise ratio, it is impossible to give a
broadband independent statement for allowed range for confidence
limits above 95\%.  We also see that constraints stemming from the
method 2 covariance matrix are essentially the same in the shape, but
tighter.

\subsection{Redshift dependence of BAO position and determination of $z_{\rm eff}$}
We measure the correlation function in three redshift bins at $z=2.0$,
$2.5$ and $3.0$. The estimator interpolates between these redshift
bins and hence the redshift evolution information is not lost.  In
order to make cosmological inferences from our dataset, it is
important to determine what is the effective redshift at which we
measure the position of the BAO.

\begin{figure}[h!]
  \centering
  % ./plotbaoaniz.py
  \includegraphics[width=\linewidth]{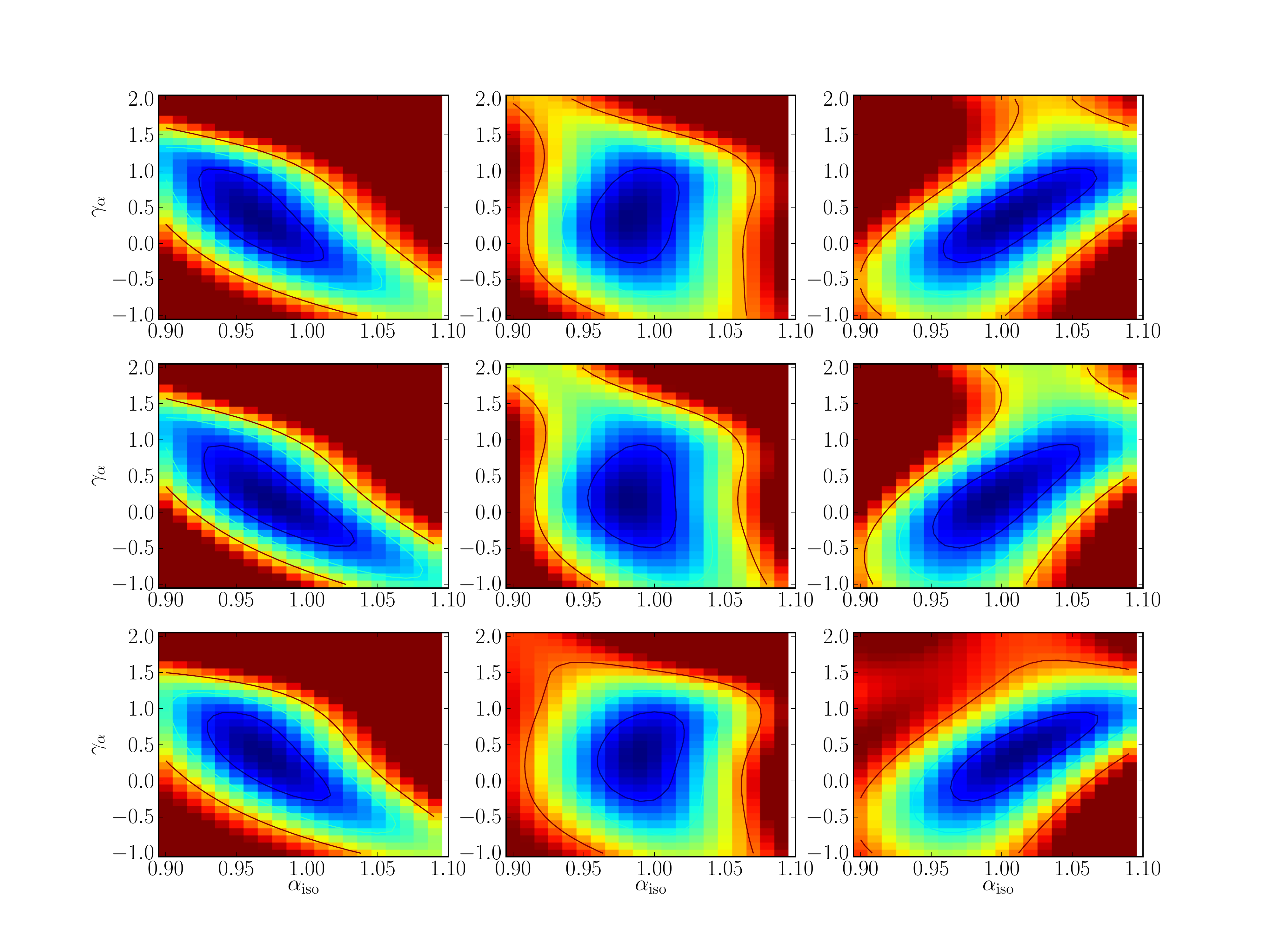}
  \caption{Two-dimensional constraints on the
    $\aiso$-$\gamma_{\alpha}$ plane. We plot this only for broadbands
    BB2 (top tow), BB3 (middle row) and BB6 (bottom row) for reference redshifts
    $z_{\rm ref}=2.2$ (left column), $z_{\rm ref}=2.4$ (middle column), $z_{\rm
      ref}=2.6$ (right column). See text for further discussion.}
  \label{fig:zeff}
\end{figure}

To find the effect redshift, we add a parameter to our fit that
governs the evolution of the BAO scale with redshift: $\gamma_{\alpha}
= d \log \aiso/d \log (1+z)$. In other words, we assume that the
isotropic scale factor varies as
\begin{equation}
  \aiso(z) = \left(\frac{1+z}{1+z_{\rm ref}}\right)^{\gamma_\alpha}.
\end{equation}
One can expect that measurements of $\aiso$ and $\gamma_\alpha$ are
correlated so that $\aiso$ is best constrained at the position where
data are the most constraining.  We illustrate this property in Figure
\ref{fig:zeff}, where we present the two-dimensional constraints on
the $\aiso$-$\gamma_\alpha$ plane for three different choices of
$z_{\rm ref}$. We see the expected behavior as the contours turn their
direction as we move from a low reference redshift to a high reference
redshift. However, contours are not completely Gaussian; this is
expected, when the detection in any one redshift is not very
strong. This feature also means that the precise effective redshift is
a poorly defined quantity. Different broadbands produce tightest
constraints on $\aiso$ between $z\sim2.3$ and $z\sim2.5$. Therefore we
set our $z_{\rm eff} \sim 2.4$. This should not matter for any
realistic model that we might want to test using our data. If a
cosmological model has a widely varying $\alpha$ across small redshift
range, then our approach of wide redshift bins fails. In such cases
one would expect the peak to be smeared, but, as discussed in the
following section, we do not detect any evidence for such effect.

Finally, this figure also demonstrates that the parameter governing the
redshift evolution of the BAO peak is consistent with zero, which is
by itself a useful systematics check.

\subsection{Discussion of systematic effects}
\label{sec:disc-syst-effects}
The \lyaf\ correlation is a difficult field: systematic effects that
affect the measurements of two-point functions are many and closely
spaced. However, the BAO measurement is particularly robust. Once the
existence of the peak is established in the data, there are few
possible systematic effects that can affect its position. Perhaps a
suitable analogy is measuring the redshift of a distant object: it is
very difficult to spectro-photometrically calibrate spectra even for a
well designed modern telescope, but if line features are available in
the target spectra, one can nevertheless measure the object's redshift
with considerable precision.

\subsubsection{Non-linear effects on the correlation function.}
\label{sec:non-linear-effects}
The dark matter field at the redshift relevant for our analysis
($z\sim2.4$) is already weakly non-linear, although the effect is
considerably smaller than that due to non-linearities (and scale
dependent biasing)  for galaxy surveys at considerably smaller
redshifts.  In particular, one expects $\sim 6\mpch$ smoothing of the
correlation function in the radial direction and $\sim 3 \mpch$ smoothing
in the transverse direction \cite{peanutboy}.

We implemented this smoothing in an approximate manner as described in
\cite{peanutboy}. In general, the anisotropic smoothing generates
multipoles beyond hexadecapole, that depends on both the smoothing
parameters and the redshift-space distortion parameters.  However,
when the smoothing can be treated perturbatively, these multipoles can
be ignored and the analysis restricted to $\ell=0,2,4$ multipoles.

Following \cite{2012arXiv1203.6594A}, we only smooth over the peak
part of the correlation function, but we have shown explicitly that
smoothing the entire linear correlation function affects $\chi^2$
negligibly. This is expected since we are using only distances
$r>50\mpch$.

To further explore this effect, we have also tried a simple isotropic
smoothing with kernel size of $3$, $6$ and $12$ $\mpch$. The
error-bars and best-fit $\chi^2$ in Table \ref{tab:NL}.  Based on
$\chi^2$, we cannot distinguish the models. We also see that using a
smoothed model increases the position error-bars at $\sim$10\% level
and changes the best-fit peak position by a small fraction of standard
deviation. Both effects are negligible, but we use the non-linear
model for consistency.

Finally, we note that using a model which is unphysically smooth (the
$12\mpch$ isotropic smoothing) decreases the goodness of fit by $\sim
7$ units. Systematics in the data and processing would typically
smooth the peak rather than sharpen it, suggesting that our
interpolation scheme is sufficient and that the measured peak has the
expected width. The sharpness of the measured peak also precludes fast
variations of the peak position with redshift, which would result in
further smoothing of the BAO peak.

\begin{table}[h!]
  \centering
  % printtableNL.py
  \begin{tabular}{c|cc}
    Model  & Best Fit $\Delta \chi^2$ & $\aiso$ \\
    \hline
    linear  & 831 & $-1.73^{+1.81\ +3.78\ +6.04}_{-1.75\ -3.52\ -5.39}$\\ 
    non-linear  & 830.698 & $-1.49^{+2.05\ +4.25\ +6.77}_{-  2\ -4.02\ -6.18}$\\ 
    Alt. non-linear & 830.73 & $-1.45^{+2.04\ +4.21\ +6.68}_{-1.99\ -4.01\ -6.17}$\\ 
    3 $\mpch$ iso. smooth & 830.775 & $-1.73^{+1.89\ +3.94\ +6.27}_{-1.82\ -3.66\ -5.59}$\\ 
    6 $\mpch$ iso. smooth & 831.29 & $-1.7^{+2.08\ +4.3\ +6.86}_{-2.03\ -4.07\ -6.24}$\\ 
    12 $\mpch$, iso. smooth & 838.446 & $-1.09^{+2.95\ +6.2\ +9.99}_{-2.88\ -5.78\ -8.86}$\\ 
  \end{tabular}
  \caption{Results of fitting smoothed and unsmoothed model for the
    Method 1 with broadband model 3. See text for discussion.}
  \label{tab:NL}
\end{table}

\subsubsection{Data cuts}

We explicitly investigated a few possible systematic effects, by
restricting the analysis to a subset of data and rerunning our
correlation function analysis to see if the peak position is affected
more than one would anticipate given the amount of the data
removed. In particular, we have performed the following tests:

\begin{itemize}
\item \emph{Removing the data at the ultraviolet end of the
    spectrograph.} We normally use the data redward of $3600$\AA.  The
  spectra at the ultraviolet end of the wavelength range are the least
  understood part of the data. They are difficult to calibrate
  spectro-photometrically due to nature of the calibration stars.
  This is exacerbated by the fact that quasar fibers were drilled in a
  plate position which, given atmospheric dispersion, maximizes the
  blue throughput relevant for the \lya forest rather than the red
  throughput used for other objects and calibration stars. Hence for
  this test we remove the data between $3600$\AA and $3700$\AA.

\item \emph{Spectro-photometric variability}. This cut is another way
  of examining the effect of poor spectro-photometry at the blue-end
  of the spectrograph.  For every quasar, we calculate two variability
  scores (``slope'' and ``normalization'') as follows. For each
  individual exposure, the mean inverse variance weighted flux is
  calculated in two wavelength bands (4000-4500 \AA and 5500-6000
  \AA). The ``slope'' (``normalization'') score is calculated using a
  weighted average of the deviations of the differences (sums) of the
  mean flux in the two bands for an individual exposure to the mean
  from the differences (sums) of the mean flux in the two bands for
  all the exposures corresponding to a single observation. We then
  remove 10\% of the quasars that show the most exposure-to-exposure
  variability in the sum of ``slope'' and ``normalization''
  variability scores.

\item \emph{A more conservative upper rest-frame cut}. In our standard
  analysis, we use an aggressive upper rest-frame cut of
  $1210$ \AA, going nearly up to the actual \lya emission line.
  We check the results by restricting to a much more conservative $1185$\AA

\item  \emph{DLA cut.} Normally, we use quasars that have an
  identified DLA system, but we exclude the data from 1.5
  equivalent widths around the DLA.  In this test we have 
  removed all quasars that contained one or more DLA systems.

  \begin{figure}[h!]
    \centering
    % ./plotbaoposx.py 3
    \includegraphics[width=\linewidth]{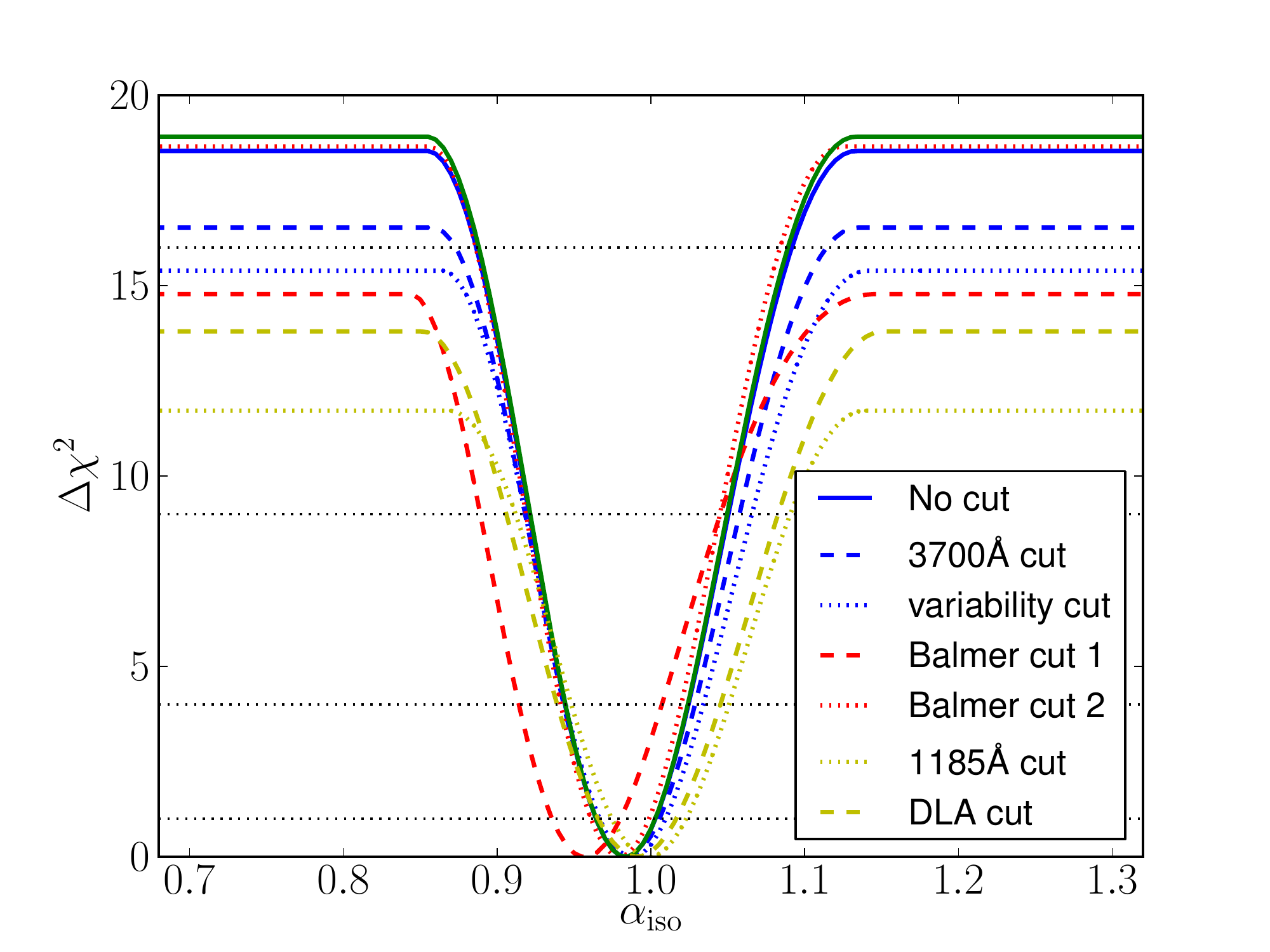}
    \caption{Results of the systematics test. The blue solid line is
      our default data reduction. We additionally plot results from
      the $3700$\AA cut (blue dotted), variability cut (blue dashed),
      DLA cut (yellow dashed), $\lr=1185\AA$ cut (yellow dotted) and
      Balmer line cuts (red dotted and dashed). Solid green line shows
      result upon addition of $\gamma_\beta$ parameter. 
      % Dashed green
      % line corresponds to completely decoupling the BAO peak. 
      All tests are
      done for the entire data for broadband model BB2 with method 1
      covariance matrix. Other fits shows similar change. See text for
      discussion.}
    \label{fig:system}
  \end{figure}

\item \emph{Balmer line cuts.}  We investigated the effect of cutting
  around the position of the observed frame Balmer lines. As discussed
  in the Section \ref{sec:continuum-fitting}, we see features due to
  mis-calibration.  If these mis-calibration features were correlated,
  these could produce correlated systematic signal at fixed radial
  distances. In order to check for the effect, we have reanalyzed our
  data with two separate cuts eliminating data in  $4102\pm 15$\AA and $4341\pm 15$\AA.

\end{itemize}

The results of these tests are shown in Figure \ref{fig:system}. None
of the cuts affect the data significantly except potentially the
$4102$\AA Balmer line cut. This effect has been analyzed in 
\cite{FPG} and we show explicitly that data are statistically
consistent with the full sample in Appendix \ref{sec:balmy}.

The same figure also shows the effect of adding a parameter describing
the redshift evolution of redshift-space parameter $\beta$, namely
$\gamma_\beta=d \ln \beta / d\ln (1+z)$; this result also appears in
the Table \ref{tab:yog}. This parameter has negligible effect,
regardless of the broadband being employed.

\subsubsection{Other systematics}

There are additional potential sources of systematics which we argue
are negligible in this work:

\begin{itemize}

\item \emph{Metal contamination.} Another important systematic effect
  might be metal contamination \cite{2010ApJ...724L..69P}. The
  strongest is \sit, which absorbs at $1206.5$\AA, producing a
  correlated absorption separated by 2271 km/s from the \lya
  absorption by the same gas, corresponding to around $20\mpch$ of
  radial separation. The largest contribution is the \lya-\sit\
  contamination, which is suppressed by a factor of $\sim 20$ with
  respect to the \lya -- \lya correlation. In other words, the
  measured correlation function in the fiducial cosmology is the true
  cosmology plus a suppressed shadow with radial distances shifted by
  $\sim 20\mpch$. We thus expect the contribution from \sit\ to be
  negligible unless our amplitude accuracy begins to approach $O(\sim
  1/20)$ of the total peak height. This warrants further investigation
  along with the impact of weaker metal lines and we will do so in a
  future publication.

\item \emph{Binning artifacts.} In principle, when one uses flat bins
  to estimate the correlation function, it is possible that the actual
  bin centers do not correspond to the true bin centers, thus skewing
  the measurement of the BAO position. For example, if the number of
  pairs is a steeply falling/increasing function of separation, then
  the actual mean separation of pairs contributing to a given bin of
  correlation function will be less/more than the nominal bin
  center. This is alleviated by linearly interpolating between any two
  bins in radial and redshift directions (instead of using top hat
  bins). We have also tested our procedures on synthetic datasets
  with exactly the same geometry as the real dataset.

\item \emph{Wavelength and astrometric calibration.} Radially, the BAO
  distance corresponds to a separation in wavelength. The relative
  wavelength calibration is better than $10^{-4}$
  \cite{2012arXiv1208.2233S} and therefore well under the required
  precision. Transverse, the BAO distance corresponds to an angular
  separation of the order $1\deg$. The angular separation between any
  two quasars is known to arc-second precision and in addition these
  errors are not coherent. This is another completely negligible
  effect.

\end{itemize}

\subsection{Consensus result}

We have presented a number of results which are reasonably consistent,
but vary in precise values, because of different assumed covariance
matrix and broadband model. However, it is useful to reduce this
variety to a single consensus value.

We show the marginalized confidence limits for all broadband models
and both covariance matrices in Table \ref{tab:fin}.  These limits
were derived from the cumulative distribution functions of
marginalized likelihoods as described in the Section
\ref{sec:determ-uncert}. The marginalized error-bars on the $\aperp$
and $\apar$ are significantly larger than the errors on the isotropic
fit. This degeneracy is due to large cross-correlation coefficient
between the two parameters, which is estimated to be $\sim -0.55$ by
fitting a Gaussian to the best-fit model.

\begin{table}
  \centering
  \begin{tabular}{c|ccc}
    % ./printtablefinal.py
    Model & $100\times(\aiso-1)$ & $100\times(\apar-1)$ & $100\times(\aperp-1)$\\
    \hline
    BB1 Me1  &  $-1.7^{+2.09\ +4.36\ +7.03}_{-2.04\ -4.13\ -6.42}$ &  $-1.9^{+3.58\ +7.51\ +11.9}_{-3.43\ -6.88\ -10.1}$&$-1.07^{+7.53\ +17.4\ +28.9}_{-6.88\ -14.2\ -23.3}$ \\

    BB1 Me2  &  $-1.99^{+1.91\ +3.95\ +6.23}_{-1.88\ -3.79\ -5.83}$ &  $-1.08^{+3.18\ +6.56\ +10.3}_{-3.11\ -6.3\ -9.63}$&$-4.05^{+6.47\ +14.6\ +27.1}_{-5.95\ -12.3\ -19.9}$ \\

    BB2 Me1  &  $-1.62^{+2.04\ +4.32\ +7.42}_{-1.99\ -4.11\ -6.77}$ &  $-1.31^{+3.51\ +7.58\ +12.3}_{-3.3\ -6.69\ -10.2}$&$-2.24^{+7.44\ +17.3\ +29.7}_{-7.06\ -15.3\ -25.2}$ \\

    BB2 Me2  &  $-1.75^{+1.87\ +3.9\ +6.34}_{-1.83\ -3.73\ -5.9}$ &  $-0.389^{+3.1\ +6.56\ +10.6}_{-2.97\ -6.03\ -9.45}$&$-5.1^{+6.4\ +14.4\ +27.2}_{-6.08\ -12.9\ -21.3}$ \\

    BB3 Me1  &  $-1.49^{+2.05\ +4.25\ +6.77}_{-  2\ -4.02\ -6.18}$ &  $-1.65^{+3.46\ +7.21\ +11.4}_{-3.33\ -6.67\ -9.91}$&$-0.95^{+7.39\ +17.2\ +28.7}_{-6.71\ -13.8\ -22.4}$ \\

    BB3 Me2  &  $-1.64^{+1.91\ +3.94\ +6.2}_{-1.87\ -3.77\ -5.77}$ &  $-0.628^{+3.14\ +6.47\ +10.1}_{-3.07\ -6.2\ -9.49}$&$-3.97^{+6.4\ +14.5\ +26.9}_{-5.92\ -12.2\ -19.8}$ \\

    BB4 Me1  &  $-1.04^{+2.48\ +5.46\ +10.7}_{-2.33\ -4.72\ -7.53}$ &  $-0.214^{+3.86\ +8.29\ +12.4}_{-3.61\ -7.23\ -10.9}$&$-3.23^{+10.8\ +25.6\ +32.7}_{-8.4\ -16.6\ -24.6}$ \\

    BB4 Me2  &  $-1.55^{+2.28\ +4.89\ +8.55}_{-2.16\ -4.35\ -6.77}$ &  $0.0392^{+3.4\ +7.25\ +11.5}_{-3.25\ -6.59\ -10.2}$&$-6.15^{+8.58\ +22.4\ +34.7}_{-6.84\ -13.5\ -20.5}$ \\

    BB5 Me1  &  $-0.813^{+2.59\ +5.93\ +14.8}_{-2.38\ -4.79\ -7.55}$ &  $0.206^{+4.38\ +9.51\ +12.5}_{-3.9\ -7.73\ -11.5}$&$-3.1^{+9.4\ +21.7\ +31.9}_{-8.07\ - 16\ -24.2}$ \\

    BB5 Me2  &  $-1.3^{+2.32\ +5.01\ +8.78}_{-2.17\ -4.35\ -6.7}$ &  $0.995^{+3.77\ +8.11\ +11.5}_{-3.46\ -6.89\ -10.5}$&$-6.86^{+7.35\ +16.9\ +30.5}_{-6.49\ - 13\ -19.8}$ \\

    BB6 Me1  &  $-1.19^{+1.95\ +4.03\ +6.48}_{-1.98\ -4.27\ -17.4}$ &  $-0.664^{+3.73\ +8.15\ +12.2}_{-3.42\ -7.09\ -11.2}$&$-2.26^{+7.35\ +17.7\ +30.9}_{-8.76\ -24.5\ -27.6}$ \\

    BB6 Me2  &  $-1.4^{+1.81\ +3.72\ +5.86}_{-1.83\ -3.83\ -6.88}$ &  $0.159^{+3.23\ +7.01\ +10.8}_{-3.02\ -6.21\ -10.2}$&$-4.94^{+6.24\ +14.6\ +31.9}_{-7.09\ -21.3\ -24.9}$ \\

  \end{tabular}

  \caption{Marginalized 1,2,3-sigma confidence limits for different
    broadband models and covariance matrices.}
  \label{tab:fin}
\end{table}

For the final results, we would like to quote a consensus result that
captures the statistical and systematic uncertainty. To this end we
select the broadband 2 model with the less constraining (method 1)
covariance matrix. Inspecting the table \ref{tab:fin} we see that it
has values and errors that are not at the edges of distributions.
Figure \ref{fig:prioreff} shows that it is the most conservative of
the additive broadbands.  We could use the most relaxing broadband 6,
however, this model latches onto a noise feature at $\aiso\sim 0.8$,
which breaks its 3-sigma errors assuming the noise feature is indeed
not real (no other broadband can fit it and it becomes weaker with the
more constraining method 2 covariance matrix). Therefore we absorb
these uncertainties by specifying systematic errors that should
contain them. Hence we quote our fit

\begin{equation}
  \text{\caiso}
\end{equation}
for the isotropic fit and
\begin{eqnarray}
  \text{\capar}\\
  \text{\caper}
\end{eqnarray}
for the anisotropic fit.  We are not quoting the marginalized 3-sigma
error in the $\aperp$, as it is dominated by the top-hat prior on the
$\apar$-$\aperp$ plane. Note, however, that this is not true for the
un-marginalized likelihood, as large parts of the $\apar$-$\aperp$ plane
are excluded by our measurement, even for the less constraining
models.

The systematic error should be understood as having a top-hat
probability distribution function, and thus while significantly
affecting the 1-sigma confidence limits, its effect on the 3-sigma
confidence limits are considerably smaller.

The fact that our quoted systematic errors are of the same order as
one-sigma statistical errors is not coincidental. As the
signal-to-noise ratio will improve with additional data, so will our
ability to constrain the broadband model, and the differences between
best-fit values for different broadband models will naturally
decrease.

\subsection{Comparison to Busca et al.}
\label{sec:comparison-busca-et}

In \cite{FPG} we presented an analysis of the BAO in the \lyaf from a
very similar dataset. The two analyses are complementary and the
present analysis builds upon the previous work. The main
methodological differences between the two can be summarized as
follows:

\begin{itemize}
\item The analysis in \cite{FPG} uses more conservative cuts in their
  choice of data. They exclude all quasars in which one or more DLAs
  are present as indicated by the visual inspection, while we remove
  only 1.5 equivalent widths around the position of the DLA. We
  also use a slightly different DLA catalog from \cite{billinprep}.
  Moreover, we aggressively use as much of the forest as available
  ($1036$\AA-$1210$\AA\ rest-frame for non-BAL quasars), while
  \cite{FPG} restrict the analysis to $1054$\AA-$1184$\AA\ rest-frame.
  These different restrictions are responsible for most of the
  increase in the signal-to-noise in the present paper.

\item The continuum fitting procedure is different in the two
  analyses.

\item We use considerably finer binning in our correlation
  function. In particular we measure the correlation function in three
  redshift bins and for each redshift bin we measure it in 18
  transverse and 28 radial bins. The analysis in \cite{FPG} uses 2D
  measurement as an intermediate measurement, but they eventually
  compress all data into 90 measurements of the monopole and
  quadrupole of the correlation function for a given fiducial
  cosmology. The finer binning has an advantage that it is in
  principle easier to catch systematic errors that are localized in
  particular bins of the measured correlation function; however, the
  larger number of bins makes measuring and understanding of the
  covariance matrix considerably more difficult.

\item The analysis in \cite{FPG} uses diagonal weighting with a fixed
  assumption about the amplitude evolution of the forest correlations
  with redshift, but allow for a completely unconstrained $\beta$
  parameter (we institute a weak prior on $\beta$ to regularize fits)

\item The analysis in \cite{FPG} makes stronger assumptions about the
  broadband and in particular do not consider a multiplicative broadband.

\item The analysis in \cite{FPG} uses a wider fitting range $20-200 \mpch$
  vs $50-190 \mpch$ employed in this paper.

\item The analysis in \cite{FPG} does not take into account the
  broadening of the peak by non-linear evolution. We show that this
  has a negligible effect on the position and only small effect on the
  errors.

\end{itemize}

In \cite{FPG} the isotropic dilation factors has been determined to be
$100\times(\aiso-1)=1\pm3$, to be compared with our result
\caiso. This is a one sigma difference, from what might appear to be
essentially the same dataset. We use looser cuts on data, but our
detection significance has improved more than one would expect based
on the increase in the number of data-points alone. The same has been
observed, when more inclusive cuts have been applied to the \cite{FPG}
analysis (N. Busca, private communication), so this result seems to be
a feature of the sample. Given the increase in the signal-to-noise
ratio, the difference is not statistically significant.

\subsection{Consistency with SPT and BOSS galaxy BAO results}

Recently, the measurements of the acoustic oscillations in the
Cosmic Microwave Background by the SPT experiment \cite{2012arXiv1210.7231S},
combined with WMAP7 results \cite{2011ApJS..192...18K} predicted a
position for the BAO peak at $z=0.57$ in the simplest flat $\Lambda$CDM
model, which is in tension with the measurements of the BAO in the
CMASS galaxies from the BOSS experiment \cite{2012arXiv1203.6594A}. We
briefly discuss this tension in the context of our measurement.

We show results for the quantities of interest for several standard
models in Table \ref{tab:sptcmass}.  In the cosmology of the CMASS
fiducial model, the SPT result can be cast as a measurement of
$\alpha=0.971\pm0.020$, while the ``consensus value'' (after
reconstruction) for the CMASS measurement is
$\alpha=1.033\pm0.017$. The difference $\delta \alpha=0.062$ is
discrepant at 2.3-sigma assuming the errors can be added in
quadrature.  We will not discuss the details of this discrepancy (see
e.g.\ \cite{2012arXiv1212.6267H}), but instead ask if \lyaf BAO can
add anything to this discussion. To this end we added two modification
of the CMASS fiducial model to Table \ref{tab:sptcmass} that have the
same value of $r_s/D_v$ as the CMASS best fit, either by changing the
value of $\theta$ or $\omega_{\rm dm}$.

Inspecting  Table \ref{tab:sptcmass} show that the SPT best and
CMASS best models predict values of $(\apar,\aperp)$ that differ by
less than our uncertainties. Differences in $\apar$ are of the
order of 0.5\%, while changes to $\aperp$ are of the order of 2\%, both
small compared to our error-bars. When correctly taking into account
the covariance between $\apar$ and $\aperp$, the size of the effect is
larger, but BOSS \lyaf BAO is currently unable to distinguish between
the models in a statistically significant manner.

Therefore, \lyaf BAO will not be able to distinguish between these
models in the $\Lambda$CDM scenario. This is because the Kaiser
redshift-space distortion parameter $\beta$ is large for the \lya
forest, pushing the statistical weight of the BAO signal towards the
measurements of Hubble parameter (where the effect is small) rather
than angular diameter distance (where the effect is large). This is an
expected result\cite{2012arXiv1201.2434W}. \lyaf BAO measurements do
not improve the constraints to the minimal flat $\Lambda$CDM model. It
does, however, help constrain models with non-zero curvature or dark
energy that is dynamically significant at high redshift.

\begin{table}[h!]

  \begin{tabular}{c|ccccc}
    Model & $\omega_b$ & $\omega_{dm}$ & $h$ & $\theta$ & $r_s$ \\% ($z_\star$) & $r_s$ ($z_{\rm drag}$) \\
    
\hline
    \lya fid & 0.0227 & 0.1096 & 0.700 & 1.0347 & 146.90 \\%& 149.70 \\
    SPT best & 0.0229 & 0.1093 & 0.731  & 1.0396 & 146.83 \\%& 149.56 \\
    CMASS fid & 0.0224 & 0.1119 & 0.7 & 1.0382 &  146.51 \\%&  149.40 \\
    CMASS best 1 & 0.0224 & 0.1119 & 0.668 & 1.0287 &   146.51 \\%& 149.40\\  
    CMASS best 2 & 0.0224 & 0.1192 & 0.673 & 1.0382 &  144.57 \\%& 147.42 \\
\hline
  \end{tabular}

\begin{tabular}{c| cccc|cc}
 & \multicolumn{4}{c}{$z=2.4$ Lyman-$\alpha$} &
 \multicolumn{2}{|c}{$z=0.57$ BOSS galaxies} \\
 &  $Hr_s$ & $100\times r_s/D$ & $\apar$ & $\aperp$ & $100\times r_s/D_v $ & $\alpha$\\
\hline
\lya fid & 35290 & 2.5484 & 1 &        1          & 7.5576 & 1.001\\
SPT best & 35384 & 2.5975 & 0.997 & 0.981  & 7.7869 & 0.971 \\ 
CMASS fid & 35464 & 2.5544 & 0.995 & 0.997  & 7.5625 & 1.000 \\ 
CMASS best 1 & 35326 & 2.5001 & 1.00 & 1.022 & 7.3201 & 1.033 \\  
CMASS best 2 & 35738 & 2.5122 & 0.992 & 1.017 & 7.3198 & 1.033 \\ 
\end{tabular}
\caption{Cosmological models relevant for this work and the
  CMASS/SPT tension. Columns are predictions of various quantities
  for the selected models. After specifying $\omega_b$ and
  $\omega_{\rm dm}$ one needs to specify either the value of
  the Hubble parameter $h$ or equivalently $\theta=r_s/D_{\rm LSS}$,
  the ratio of the sound horizon. to the distance of the last
  scattering.     The \lya fid refers to the fiducial cosmology
  used in this work. SPT best is the best fit model coming from
  SPT+WMAP7. The CMASS fid is the fiducial model used in
  \cite{2012arXiv1203.6594A}. 
  The $\alpha$ values in the $z=2.4$ case are calculated
  with respect to the \lya fid model (and can be compared to $\alpha$
  constraints in this paper), while those for the $z=0.57$ model were
  calculated in the CMASS fid model (and can be compared to $\alpha$
  constraints in the \cite{2012arXiv1203.6594A}).
  CMASS best 1 and 2 are changes to the CMASS fiducial model that
  reproduce the best-fit position of the measured BAO position by
  either changing $\theta$ or $\omega_{\rm dm}$ }
  \label{tab:sptcmass}
\end{table}

\section{Visualizing the peak}
\label{sec:visualizing-peak}

Errors in the correlation function measurements are always significantly
more correlated than the errors in a more natural space for two-point
function such as Fourier space measurement. In our case, the
marginalization over the unwanted modes makes this issue particularly acute
- one obtains unbiased estimates at the price of large, but nearly
perfectly correlated uncertainties. We also measure the  correlation
function in many points, each of which is rather noisy. Moreover, some
of the BAO signal is in the monopole while some is in the quadrupole.
Nevertheless, one would still wish to simply ``see'' the BAO peak and
produce the peak plots in an objective manner.

To this end we compress the measured correlation function as
follows. We fit for a cosmological model with broadband. We then fix
the parameters of this model, set the BAO amplitude to zero and fit
for the binned deviations of the real space correlation function. The
real space correlation function is piece-wise linearly interpolated
between bins in the $\xi(r) r^2$. Given the best-fit $\beta$ value we
propagate this fit to the monopole to higher multipoles. The result of
this exercise is an optimally estimated real-space correlation
function that has Kaiser-like redshift-space distortions. Since we are
fitting around the model without the BAO peak, this fit has no way of
knowing about the existence or the position of the BAO peak.

However, the resulting correlation function will still have large
correlated errors. We can remove these, by diagonalizing the
covariance matrix of our fitted binned correlation function and
setting the amplitude of the largest eigen-mode to zero. At the same
time, we must also project out the same eigen-vector from the
theory. We have thus made a full circle; we end up with a distorted
correlation function and a theory that accounts for this distortion.

\begin{figure}[h!]
  \centering
  \begin{tabular}{cc}
    %plotpeak.py 1
    %plotpeak.py 2 in py/peak
    \includegraphics[width=0.45\linewidth]{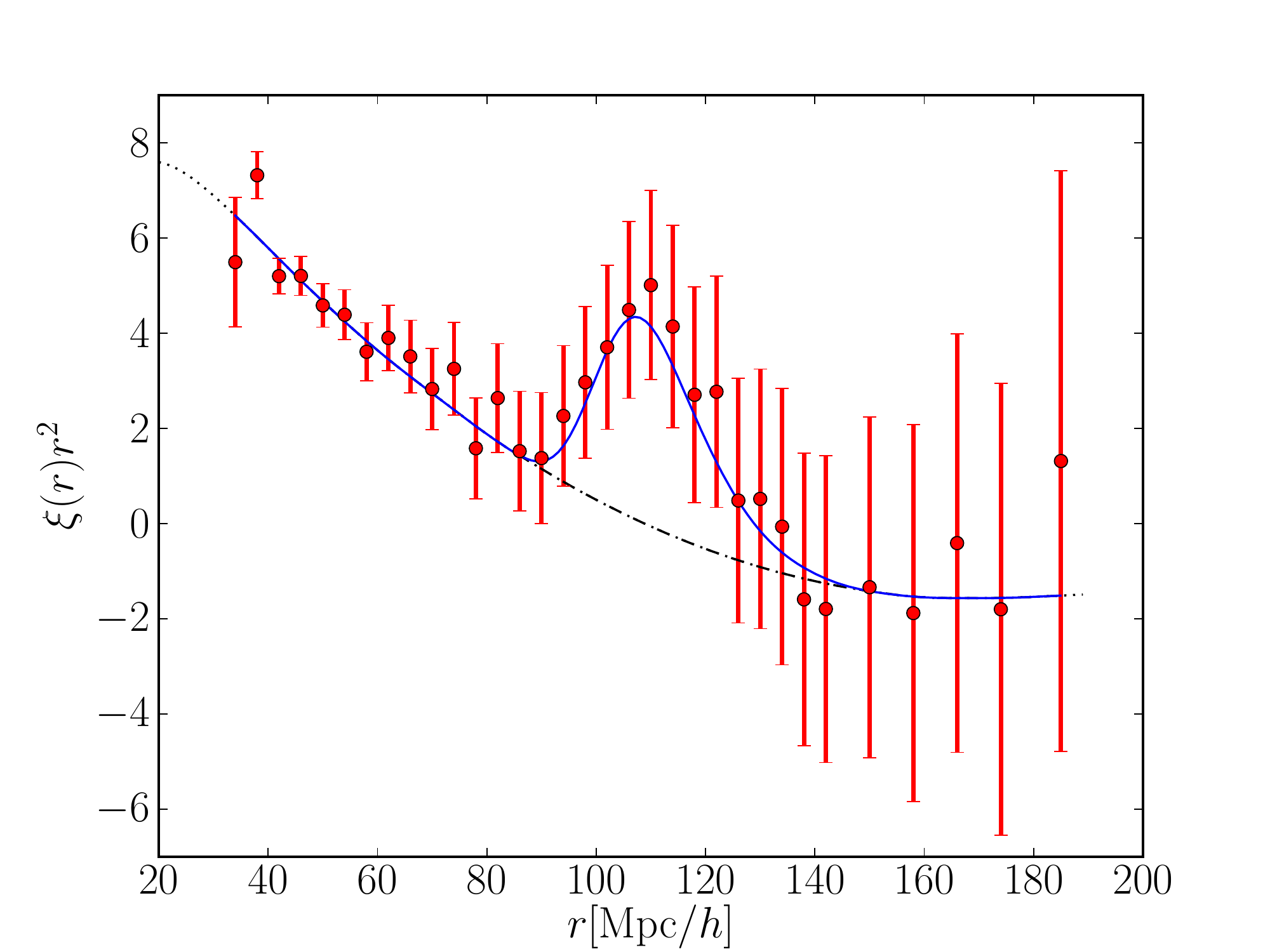} &
    \includegraphics[width=0.45\linewidth]{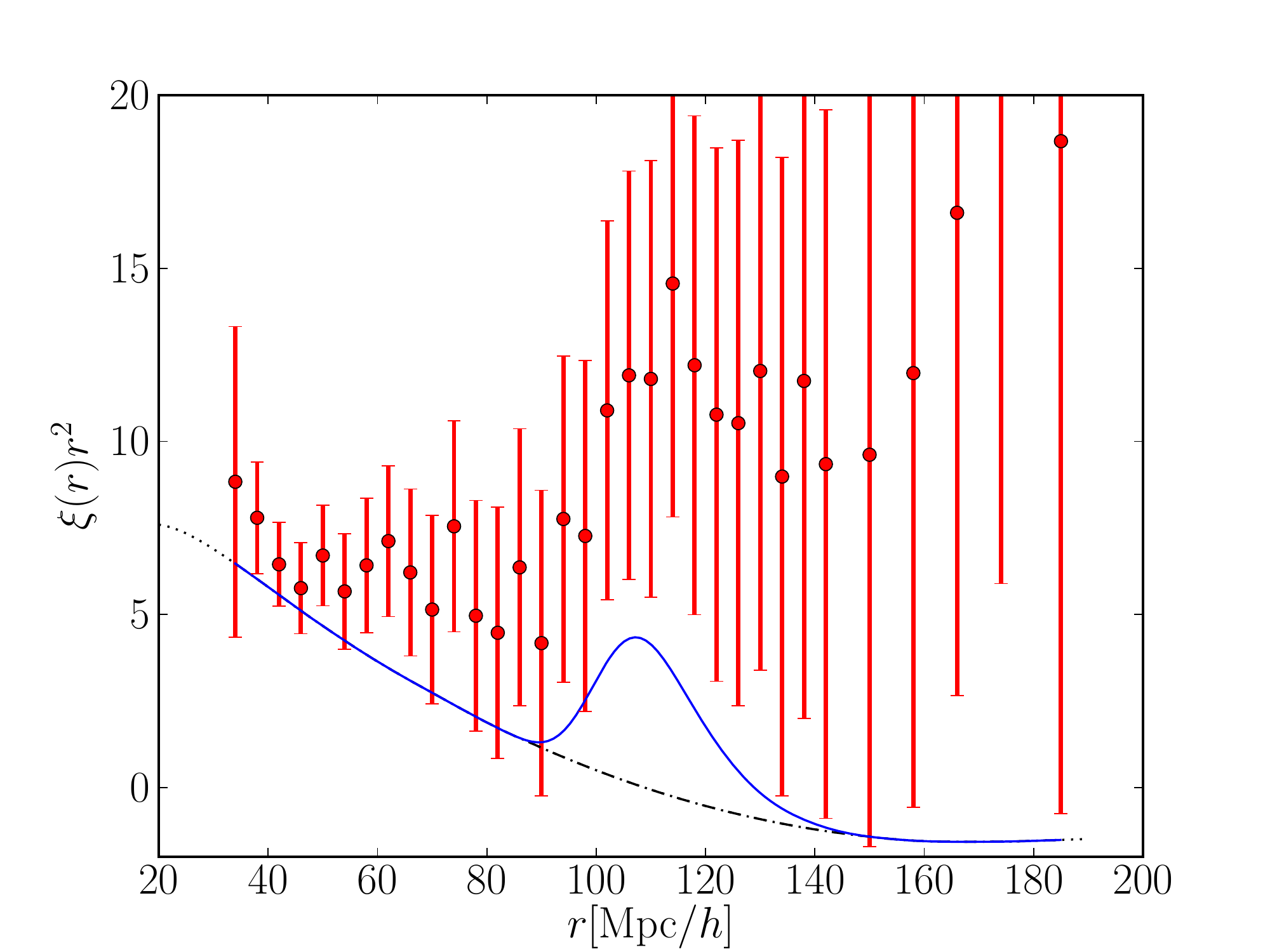} \\
    \includegraphics[width=0.45\linewidth]{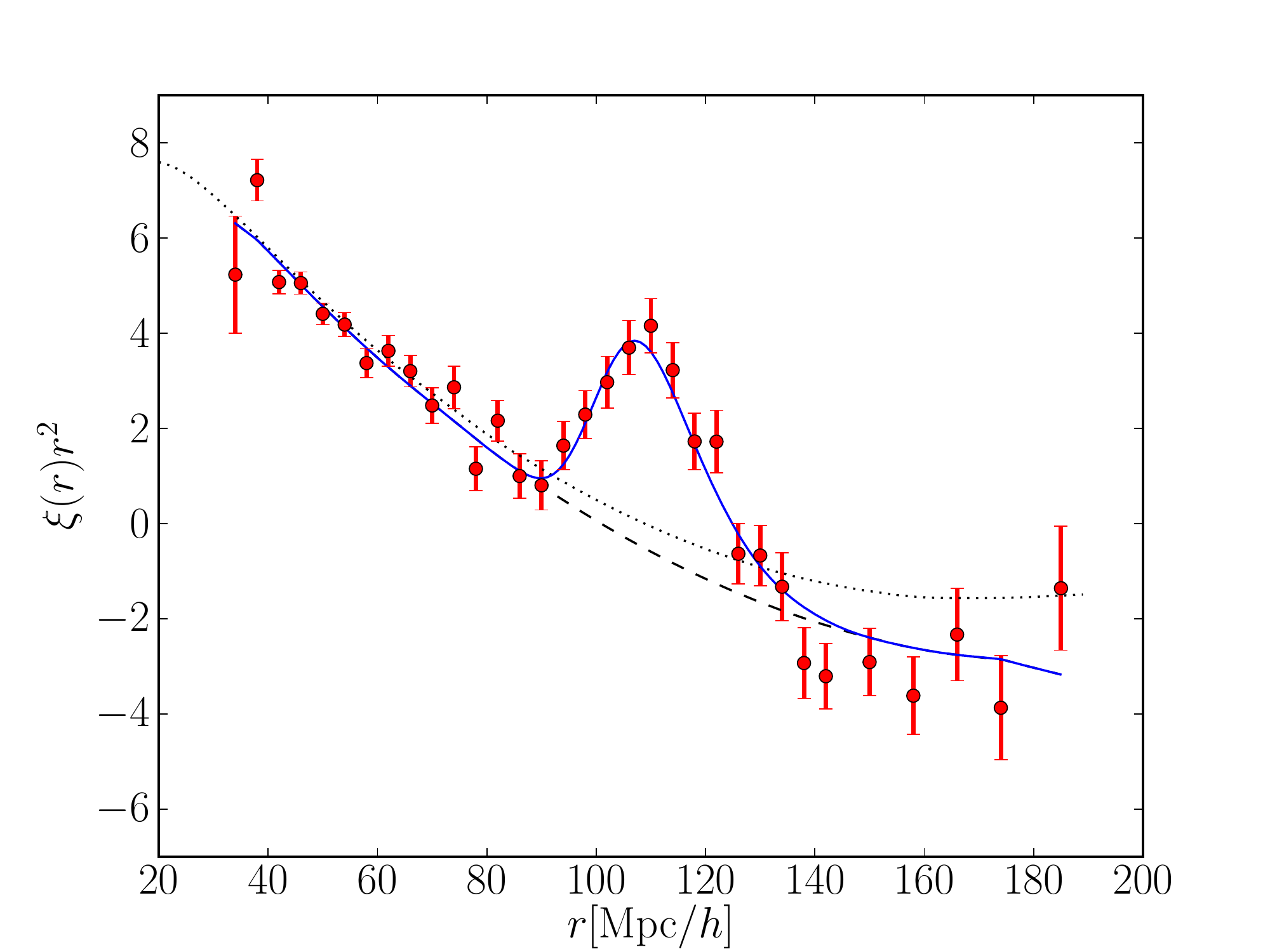} &
    \includegraphics[width=0.45\linewidth]{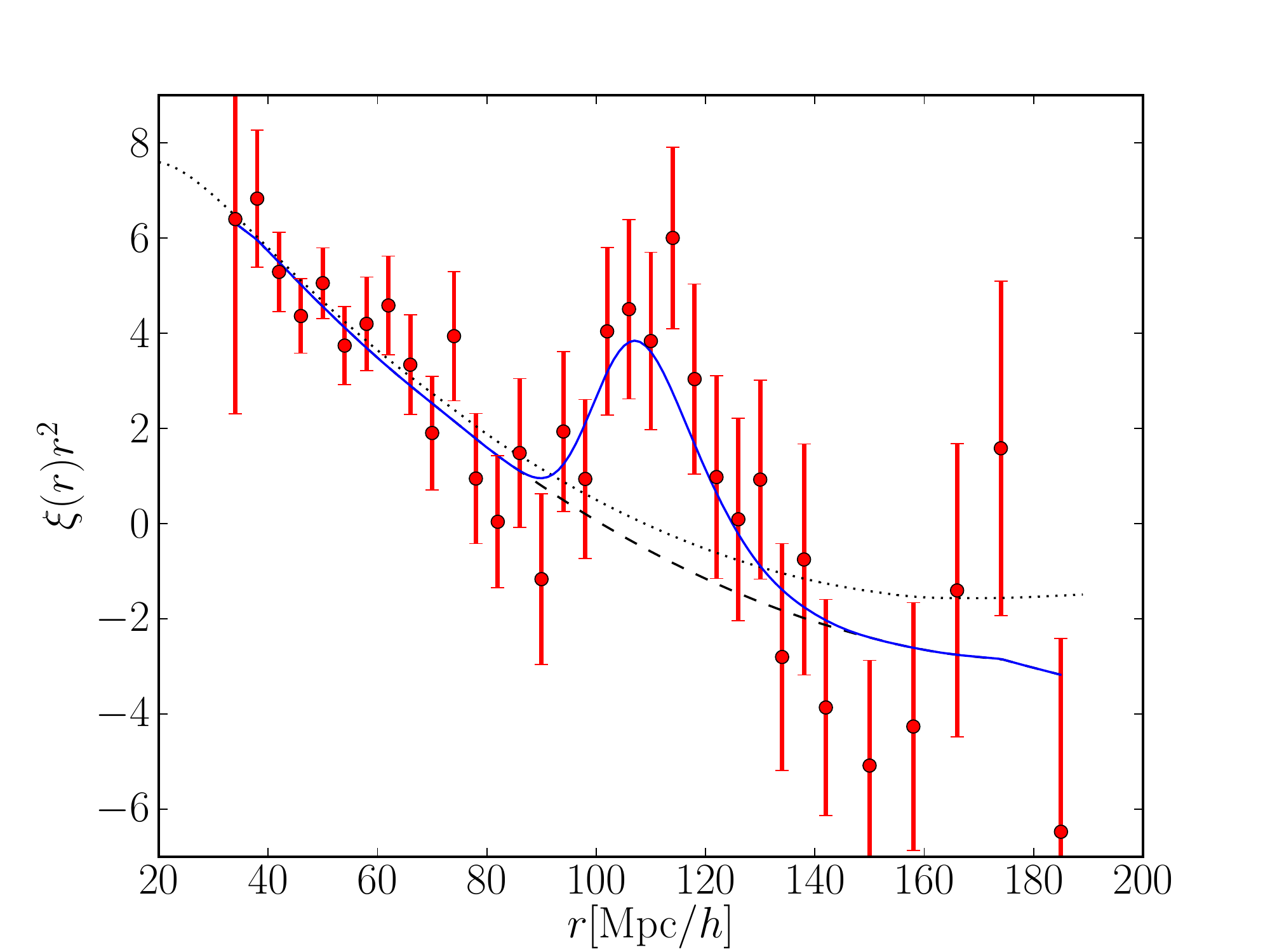} \\
  \end{tabular}

  \caption{Correlation functions under the assumption of Kaiser-like
    redshift-space distortions for all fifteen realizations of
    synthetic data combined (left column) and the real data (right
    column). The top row are bins as inferred, the bottom row is with
    the largest eigenvalue removed.  The dotted line is the model
    around which we measure the binned correlation function, the
    dashed-line is after taking into account the eigen-vectors
    projected out. The blue line is the theoretical peak added back to
    the dashed-line model to show agreement. The correlation functions
    are normalized to the underlying dark-matter correlation function
    with the bias factor inferred for our fiducial cosmological model
    divided out.}
  \label{fig:peak}
\end{figure}

Results of this exercises are plotted in Figure \ref{fig:peak} for
data and for the mean of all realizations of synthetic data. Removing
more than one eigenvalue produces essentially the same plots.

\section{Discussion \& Conclusions}
\label{sec:conclusions}

In this paper we have revisited the measurement of the \lyaf BAO. We
have detected the baryonic acoustic oscillations in the flux
fluctuations in the \lyaf of distant quasars which allows a
measurement of the expansion rate at a redshift of $z\sim2.4$. We have
confirmed an earlier result with a significantly different method and
data cuts.

Our results are consistent with but more stringent than \cite{FPG}.
By far the most important reason for this improvement is that we use
more data by imposing looser cuts (which we think is valid when one
tries to find an isolated feature such as BAO peak). Second, we use an
optimal estimator, which should increase the available signal-to-noise
ratio by 10-20\% as indicated by tests on fields with known continuum.
Finally, we use a much finer binning, resulting in less information
loss when simultaneously fitting for the BAO and broadband
parameters.

Our results are more precise than the theoretical expectations for
what the BOSS experiment with the present dataset is expected to
achieve with the \lyaf technique \cite{2007PhRvD..76f3009M}. The
covariance of $\aperp$ and $\apar$ measurements have the same
correlation coefficient as predicted by the Fisher matrix forecast
($\sim -0.55$), but our signal-to-noise is about 40\% better. The
Fisher matrix calculation assumed bias parameters from
\cite{2003ApJ...585...34M} and the forecasted expectations for the
BOSS spectrograph signal-to-noise.  In fact we have now achieved the
statistical position accuracy that is consistent with what would be
expected at the end of the survey. We saw with the synthetic data that
realization-to-realization scatter is large so it is possible that we
might just have been fortunate. On the other hand, it is also possible
that the forecasts were pessimistic and that the signal is larger than
expected. To distinguish between these possibilities, one needs to
perform the analysis on a larger dataset and/or carefully compare the
achieved spectrograph signal-to-noise and measured bias parameters
with values that were used in the Fisher matrix analysis.

We fitted different broadband models and found that these produce
shifts in the best-fit values of BAO position of the order of the
1-sigma statistical errors. This effect is similar to the difference
between the correlation function analysis and the power spectrum
analysis in \cite{2012arXiv1203.6594A} -- there is just one two-point
function and hence different results reflect how a broadband model
parametrized in Fourier or configuration space is simply a different
model leading to different results.  As the signal-to-noise increases,
the likelihoods will become more Gaussian, which will make it easier
to make sense of differing results from different broadband models.

We attempted to be conservative when quoting our results: we used the
least constraining broadband model and the less constraining of our
two covariance matrices. We also assigned differences between best-fit
models from different broadbands to systematic error.  Using these
criteria, we measure \caiso\ with respect to the fiducial model (flat
$\Lambda$CDM with $\Omega_m=0.27$, $h=0.7$) for the isotropic dilation
factor and we find marginalized constraints \capar\ and \caper\ for
the anisotropic fit.

We do not attempt to present a precise number for the statistical
significance of the peak detection. Assuming our most optimistic
broadband model, our significance is well over 5-sigma. Believing that
the main solution is the correct one, but choosing our most
conservative broadband models reduce it to $\sim 4$-sigma.  Taking
seriously the secondary solution at $\aiso\sim0.8$ found by the
broadband model 6, the significance drops to $\sim 3$-sigma.

To a large extent, the question of the BAO peak significance is a moot
point. One is measuring the BAO position in order to learn about
possible and impossible cosmological models. Plausible cosmological
models will not deviate from concordance by more than a few percent
and hence stay in the region where all methods give consistent results
on the peak position.

An important lesson learned in this work is that it is essential to
have a method that is computationally feasible enough to analyze not
just the real dataset, but many (hundreds) of realizations of the mock
data to gain full understanding of the estimator and possible
systematics effects arising from the analyses. What we might have
gained in statistical errors was lost in our ability to fully
characterize the estimator using a large number of realizations.
However, when going beyond measuring BAO and attempting to use the
broadband power to constrain cosmology, an approach similar to the one
employed in this work will be necessary to deal with the
continuum-fitting induced distortion. In addition, a better analysis
requires a more sophisticated continuum fitting and instrument model
and will require a red-side systematics subtraction.  In short, it is
clear that many improvements can be made in the \lyaf data reduction
methodology.

We defer deriving cosmological constraints from the position of the
BAO peak to a future publication. The current measurement of the BAO
peak position is a striking confirmation of our basic understanding of
the universe, as pointed out in \cite{FPG}; and we expect it to be very
competitive for models with early dark energy and curvature.

\section*{Acknowledgments}

Funding for SDSS-III has been provided by the Alfred P. Sloan
Foundation, the Participating Institutions, the National Science
Foundation, and the U.S. Department of Energy Office of Science. The
SDSS-III web site is \texttt{http://www.sdss3.org/}.

SDSS-III is managed by the Astrophysical Research Consortium for the
Participating Institutions of the SDSS-III Collaboration including the
University of Arizona, the Brazilian Participation Group, Brookhaven
National Laboratory, University of Cambridge, Carnegie Mellon
University, University of Florida, the French Participation Group, the
German Participation Group, Harvard University, the Instituto de
Astrofisica de Canarias, the Michigan State/Notre Dame/JINA
Participation Group, Johns Hopkins University, Lawrence Berkeley
National Laboratory, Max Planck Institute for Astrophysics, Max Planck
Institute for Extraterrestrial Physics, New Mexico State University,
New York University, Ohio State University, Pennsylvania State
University, University of Portsmouth, Princeton University, the
Spanish Participation Group, University of Tokyo, University of Utah,
Vanderbilt University, University of Virginia, University of
Washington, and Yale University.

%\clearpage

\bibliographystyle{JHEP}
\bibliography{cosmo,cosmo_preprints,local}

\appendix
\section{Appendix: Optimal Quadratic Estimators}
\label{sec:append-optim-quadr}

Optimal quadratic estimators are a mature technology used extensively
to estimate the two-point function of correlated data
\cite{1998PhRvD..57.2117B,2000ApJ...533...19B,1998ApJ...503..492S}.
We will provide a general derivation in the next subsection. This
formulation is what we use to measure the 1D power spectrum of
quasars. However, for measuring the 3D power spectra, we use a
slightly modified procedure which halves the sizes of matrix
operations in our case: an optimal cross-correlation estimator -- this
is discussed in subsection \ref{sec:appr-covar-matr}.

\newcommand{\vd}{\mathbf{d}}
\newcommand{\vm}{\mathbf{\mu}}
\newcommand{\vD}{\mathbf{D}}
\newcommand{\mC}{{\rm C}}
\newcommand{\mW}{{\rm W}}
\newcommand{\mG}{{\rm G}}
\newcommand{\mS}{{\rm S}}
\newcommand{\mN}{{\rm N}}

\subsection{Optimal auto-correlation estimator}
\label{sec:optim-auto-corr}

Optimal quadratic estimators can be derived in several ways. The most
common method is to write the Gaussian likelihood function, take its
derivatives and then cast it as a Newton-Raphson approach to the
minimum. There are subtleties to this derivation in that it produces
the optimal unbiased estimator only if the second derivative is
approximated as the Fisher matrix and not if one uses the exact second
derivative in the Newton Raphson step.  An alternative approach, which we
follow in this paper, is to write a plausible ansatz for the estimator
and then show in what limit it is optimal.

\renewcommand{\mN}{\mathcal{N}}

Let $\vd$ be a data vector of size $N$, whose correlation properties we
wish to estimate. The expectation value of pairs of pixels is
\begin{equation}
  \left< \vd \vd^T \right> = \mC = \mC_{,i} \theta_i + \mN,
\end{equation}
where $\mC$ is the $N \times N$ covariance matrix, which we assume can
be written as a sum over some component matrices $\mC_{,i}=d \mC / d
\theta_i$ with coefficients $\theta_i$. We implicitly sum over
repeated indices. If the covariance matrix is written in this manner,
we can identify $\theta_i$ as the estimate of the two-point function
in bin $i$. In general, $\theta_i$ can be values describing either
the measurements of the correlation function or of the power-spectrum
(depending in $\mC_{,i}$).  The $\mN$ matrix is the noise matrix which
we take to be provided and has no free parameters.

We start by considering an estimator of the form

\begin{equation}
  E_i = (\mW\vd)^TC_{,i}(\mW\vd),
\end{equation}
where $\mW$ is a weighting matrix which is at the moment a completely
general symmetric matrix.

The expectation value of $E_i$ is given by
\begin{equation}
  \left< E_i \right> = \Tr \left( \mC \mW \mC_{,i}\mW \right) = G_{ij} \theta_j + b_i,
\end{equation}
where 
\begin{eqnarray}
\label{eq:6}
  b_i= \Tr \left(\mN \mW \mC_{,i} \mW\right)\\
  G_{ij} = \Tr \left(\mC_{,i} \mW \mC_{,j} \mW\right)
\end{eqnarray}

One can therefore construct an unbiased estimator
\begin{equation}
  \tilde{\theta}_i = G^{-1}_{ij} (E_j-b_j), \label{eq:est}
\end{equation}
where, by construction $\left<\tilde{\theta}_i\right> = \theta_i$.

The (co)variance of this estimator is 

\begin{multline}
K_{ij}= \left<\tilde{\theta}_i\tilde{\theta}_j\right> -
\theta_i \theta_j = 
G^{-1}_{ik} G^{-1}_{jl} \bigg[
\left<  \left(\vd^T \mW \mC_{,k} \mW \vd - b_k  \right)\left(\vd^T
    \mW \mC_{,l} \mW \vd - b_l  \right) \right>\\
-
\left<  \left(\vd^T \mW \mC_{,k} \mW \vd - b_k  \right)\right> \left<\left(\vd^T
    \mW \mC_{,l} \mW \vd - b_l  \right) \right>
\bigg],
\end{multline}

Applying Wick's theorem to the average in the first term yields three
terms. One of these terms decouples the first two brackets and thus cancels
the second term and the other two are equivalent. One thus obtains

\begin{equation}
K_{ij}= 
2 G^{-1}_{ik} G^{-1}_{jl} \Tr \left( \mC \mW \mC_{,k} \mW \mC \mW \mC_{,l} \mW\right)
\label{eq:err}
\end{equation}

Although these equations are complicated, they in fact reduce to the
well-known cases for suitable choices of $\mW$, which we will discuss
next. 

For simplicity, we will consider a toy model, where we try to measure
the correlation function for a set of pixels. In this case $C_{,i}$ is
unity for pairs of pixels corresponding to bin $i$ and zero otherwise
(in other words, we approximate correlation function as a binned
function whose value is constant across each bin). Consider the case
of a trivial estimator first. In that case $\mW$ is an identity matrix
and $E_i$ becomes the sum over pixels pairs corresponding to
correlation function in bin $i$. $G_{ij}$ becomes a diagonal matrix,
whose elements are simply the number of pixel pairs corresponding to
correlation function in bin $i$. Equation \ref{eq:est} thus reduces to
\begin{equation}
  \tilde{\theta}_i = \frac{\sum_{\rm pairs} d_a d_b}{N_{{\rm pairs},i}}.
\end{equation}

In other words, we calculate the average of product of data pairs
corresponding to the relevant bin - exactly as one would naively
expect. What is the error of this estimator? Equation \ref{eq:err} can
be rewritten as
\begin{equation}
  K_{ij} = \frac{2}{N_{{\rm pairs},i} N_{{\rm pairs},j}} \sum_{pairs (a,b)\in i,(c,d)\in j} \mC_{ac} \mC_{bd}.
\end{equation}
where sum is over all pairs (where $a,b$ pair is counted as distinct
from $b,a$ pair) of data-points with indices $a$ and $b$ that
contribute to bin $i$ and pairs of data-points with indices $c$ and $d$ that
belong to bin $j$.  The error estimation using this path is thus an
$N^4$ operation, although the memory requirements are trivial.

Optimal weighting can be obtained by setting $\mW=\mC^{-1}$.  In this
case, the equation simplifies considerably. Both $G_{ij}$ and $K_{ij}$ can be
expressed in term of Fisher matrix:
\begin{equation}
 F_{ij} = \frac{1}{2}\Tr \left( \mC_{,i} \mC^{-1} \mC_{,j} \mC^{-1}\right) =
 \frac{1}{2}G_{ij} = K_{ij}^{-1}.
\end{equation}
Since the covariance matrix of estimator is the inverse of the Fisher
matrix, the Crammer-Rao inequality stipulates that this must indeed be
the optimal weighting. 

The estimator itself simplifies to 
\begin{equation}
  \tilde{\theta}_i = \frac{1}{2} F^{-1}_{ij} (E_j-b_j). \label{eq:est2}
\end{equation}
Often this inversion is unstable, especially if one is estimating bins
that are highly-correlated. In that case, one can stabilize this
inversion by instead citing the so-called filtered estimates \cite{1998ApJ...503..492S}:
\begin{equation}
  \tilde{\theta}_i = \frac{1}{2} \frac{E_i-b_i}{\sum_j F_{ij}}.
\label{eq:filt}
\end{equation}
These estimates are now weighted with respect to the underlying valyes
through the window function given by $(\sum_j F_{ij})^{-1}
F_{jk}$. For power spectrum measurement, these window functions are
bell shaped around the central value, but since the power spectrum
varies smoothly with $k$ and $z$, the resulting power spectrum
estimate is essentially unbiased.

Intuitively, the closer the weighting matrix is to the inverse
covariance, the closer to the minimum variance our estimator
becomes. The inverse variance weighting is just an approximation that
the off-diagonal elements of the covariance matrix are negligible.

In this work we use the same formalism for both 1D and 3D two-point
function measurement. For the 1D power spectrum measurement, the
$C_{,i}$ matrices contain the response of the 1D correlation function
to a change in a single power spectrum band-power bin. We use the
filtered estimate of Equation \ref{eq:filt} and iterate until the
measurement converges: at each iteration our weighting matrix uses the
power spectrum estimated at the previous iteration. For the
three-dimensional correlation function measurement, the $C_{,i}$
matrices contain the response of the covariance matrix to a change in
one correlation function bin (linearly interpolated in redshift) and
we do not need to iterate: the weighting comes purely from the 1D
power spectrum measurement.

\subsection{Approximating covariance matrix as block-diagonal}
\label{sec:appr-covar-matr}

The fully optimal quadratic estimator described above requires matrix
operations on matrices of size $N \times N$, which is impractical
for the full \lyaf dataset, where $N = N_{\rm quasars} N_{\rm pixels}
\sim 10^7$. However, while pixels from a single quasar are very
strongly correlated, they are only weakly correlated between
quasars. Therefore it is sensible to approximate the covariance matrix
as block diagonal, where each block corresponds to one quasar. In
other words, we are assuming that quasars are independent as far as
weighting is concerned.

To make progress, let us collate the data vectors from the quasars
into a single vector of size $N_t=N_1+N_2$ (where $N_1$ and $N_2$ are
the sizes of the datavectors from two quasars under
consideration). The covariance matrix for this system can be written

\begin{equation}
  \mC_{\rm tot} = \begin{bmatrix}
    \mC_1 & \sum_i \mC_{,i} \theta_i  \\
    \sum_i \mC_{,i}^T \theta_i  & \mC_2  \\
  \end{bmatrix}
\label{eq:5}
\end{equation}
where we assume that the covariance matrix between pixels from a
single quasar is known and we are only measuring the
cross-correlations. $C_{,i}$ matrices are of size $N_1\times N_2$ and
so neither symmetric nor square.  The derivatives with respect to the
covariance matrix are then given by

\begin{equation}
  \mC_{\rm tot,i} = \begin{bmatrix}
    0 & \mC_{,i}  \\
    \mC_{,i}^T  & 0  \\
  \end{bmatrix}.
\end{equation}

If the two lines of sight are so distant that the correlation between
them can be neglected, then the inverse of the correlation matrix can
be approximated as

\begin{equation}
  \mC_{\rm tot}^{-1} \approx \begin{bmatrix}
    \mC_1^{-1} & 0  \\
    0  & \mC_2^{-1}  \\
  \end{bmatrix}
\end{equation}

Plugging this result into optimal quadratic estimator, results in 
\begin{equation}
  \tilde{\theta}_i = \mG^{-1}_{ij} E_j.
  \label{eq:est2}
\end{equation}
where now we have
\begin{equation}
  E_i =  2\vd_1^T \mC_1^{-1} \mC_{,i} \mC_2^{-1} \vd_2,
 \label{eq:est21}
\end{equation}
with Fisher matrix
\begin{equation}
  F_{ij} = \frac{1}{2} \mG_{ij}= \Tr \left(\mC_1^{-1} \mC_{,i} \mC_2^{-1} \mC_{,j}^T\right).
 \label{eq:est22}
\end{equation}

There is no noise contribution, because we assume noise has no
contribution that spans quasars (i.e.\ in the matrix of Eq. \ref{eq:5},
the noise resides in the $\mC_1$ and $\mC_2$ matrices and is assumed
to be diagonal in this work).

These expressions are what one would naively expect for
generalization of optimal quadratic auto-correlation equations.

\section{Appendix: Marginalizing out unwanted modes}
\label{sec:append-marg-out}
To marginalize out templates, we use a standard technique from the
Cosmic Microwave Background studies. Consider modifying the inverse of
weighting matrix $W$ by
\begin{equation}
  \mW'^{-1} = \mW^{-1} + A \vt \vt^T,
  \label{eq:marg}
\end{equation}
where $A$ is a large number and $\vt$ is the template we wish to
marginalize over. It is clear that $\vt$ is an eigen-vector of the
matrix $A \vt \vt^T$ with an eigenvalue of $A |\vt|^2$. As
$A\rightarrow \infty$ it is true that $\vt$ is also an
eigen-vector of $\mW'^{-1}$ and hence also an eigen-vector of $\mW'$ with
an eigenvalue of $(A |\vt|^2)^{-1}$. 

Therefore, by rotating $\vd$ into the eigen-space of the weighting matrix
$\mW$, it is clear that the product $\mW' (\vd + \alpha \vt) =\mW' \vd
+ \alpha (A |\vt|^2)^{-1} \sim \mW \vd$ if $A$ is sufficiently large. 

In other words, when dealing with a set of data-points, we can
effectively discard a certain point by making its errors very
large. Here we effectively do the same, but instead of discarding a
single data-point, we discard a given linear mode by making its
variance very large. This action is equivalent to a rotation into a
space in which the offending mode would be a basis function, setting a
large error on this one data-point that we want to discard and then
rotating back.

By performing the transformation in Equation (\ref{eq:marg}),  our
estimator becomes insensitive to changes in the template, effectively
marginalizing over it. The price paid, however, is the loss of
signal-to-noise ratio in that linear mode.

We wish to marginalize over the leading contributions from continuum
fitting - the continuum amplitude and slope, namely $\vt = {\rm
  const.}$ and $\vt = \log(\lambda/\lambda_0)$.

\section{Appendix: Coarse-graining}
\label{sec:append-coarse-grain}

When calculating the correlation function at large distances, it might
prove useful to coarse grain neighboring pixels into larger
pixels. To do this correctly, let us write the optimal linear
estimator for the coarse-grained pixels. We will, for the
moment, stop treating the field as random and write the
likelihood for the coarse grained pixels as

\begin{equation}
  L = \log {\mathcal L} = - \frac{N}{2}\log (2\pi) - \log |C| -
  \frac{1}{2}\left[ (\vd - \mS \vD)^T \mC^{-1} (\vd - \mS \vD)\right],
  \label{eq:cg}
\end{equation}
where $\vd$ are now the fine pixels in one given line of sight,
$\vD_i$ are the values of coarse pixels and $\mS$ is the
coarse-graining matrix which has dimension $N\times N_c$, where $N$ is
the number of fine pixels and $N_c$ is the (lower) number of coarser
final pixels. For example, when coarse-graining 6 pixels into 2, the
matrix $\mS$ would be of the form

\begin{equation}
  \mS =   \begin{bmatrix}
    1 & 0 \\
    1 & 0 \\
    1 & 0 \\
    0 & 1 \\
    0 & 1 \\
    0 & 1 \\
  \end{bmatrix}
\end{equation}

Taking the first and second derivatives of $L$ with respect of $\vD_j$, we obtain

\begin{eqnarray}
  \frac{\partial L}{\partial \vD_i} =
  -\left[S^T\mC^{-1}(\vd-\mS\vD)\right]_i \\
  \frac{\partial^2 L}{\partial \vD_i \partial \vD_j} = \left[ \mS^T
    \mC^{-1} \mS\right]_{ij}
\end{eqnarray}
The second derivative reveals that the covariance matrix of coarse
pixels is given by the coarse-graining of the inverse of the
fine-pixel matrix:

\begin{equation}
  \mC_{\rm coarse}^{-1} =  \left< \vD \vD^T\right> = S^T  \mC^{-1} \mS
\end{equation}

Solving the first derivative for $\vD$, we find that
\begin{equation}
  \mC_{\rm coarse}^{-1} \vD = \mS^T \mC^{-1} \vd 
\end{equation}

In other words, the optimal coarse-graining is done by:
\begin{itemize}
\item Calculating $\mC^{-1}$ and $\mC^{-1} \vd$
\item Coarse-graining both quantities to get $\mC_{\rm coarse}^{-1}$
  and $\mC_{\rm coarse}^{-1} \vD$
\item Inverting the coarse covariance matrix again and multiplying with
  the vector to get values of $\vD$.
\end{itemize}

A trivial example is when assuming $C=\delta_{ij} \sigma^2_i$. In that
case, coarse graining reduces to the usual inverse-variance weighted
mean.

\section{Appendix: Statistical significance of the Balmer cut shift}
\label{sec:balmy}

Figure \ref{fig:system} shows the effect of cutting the data in the
vicinity of the Balmer $4102$\AA\ line on the position of the BAO
peak. How significant is this shift?

In order to answer this question given our significantly non-Gaussian
$\chi^2$ , we subtract the cut-data $\Delta \chi^2$ from the full
$\Delta \chi^2$. This difference is essentially the ``pull'' of the
data that we cut to the total signal. The question is whether these
are compatible.

\begin{figure}
  \centering
 \includegraphics[width=\linewidth]{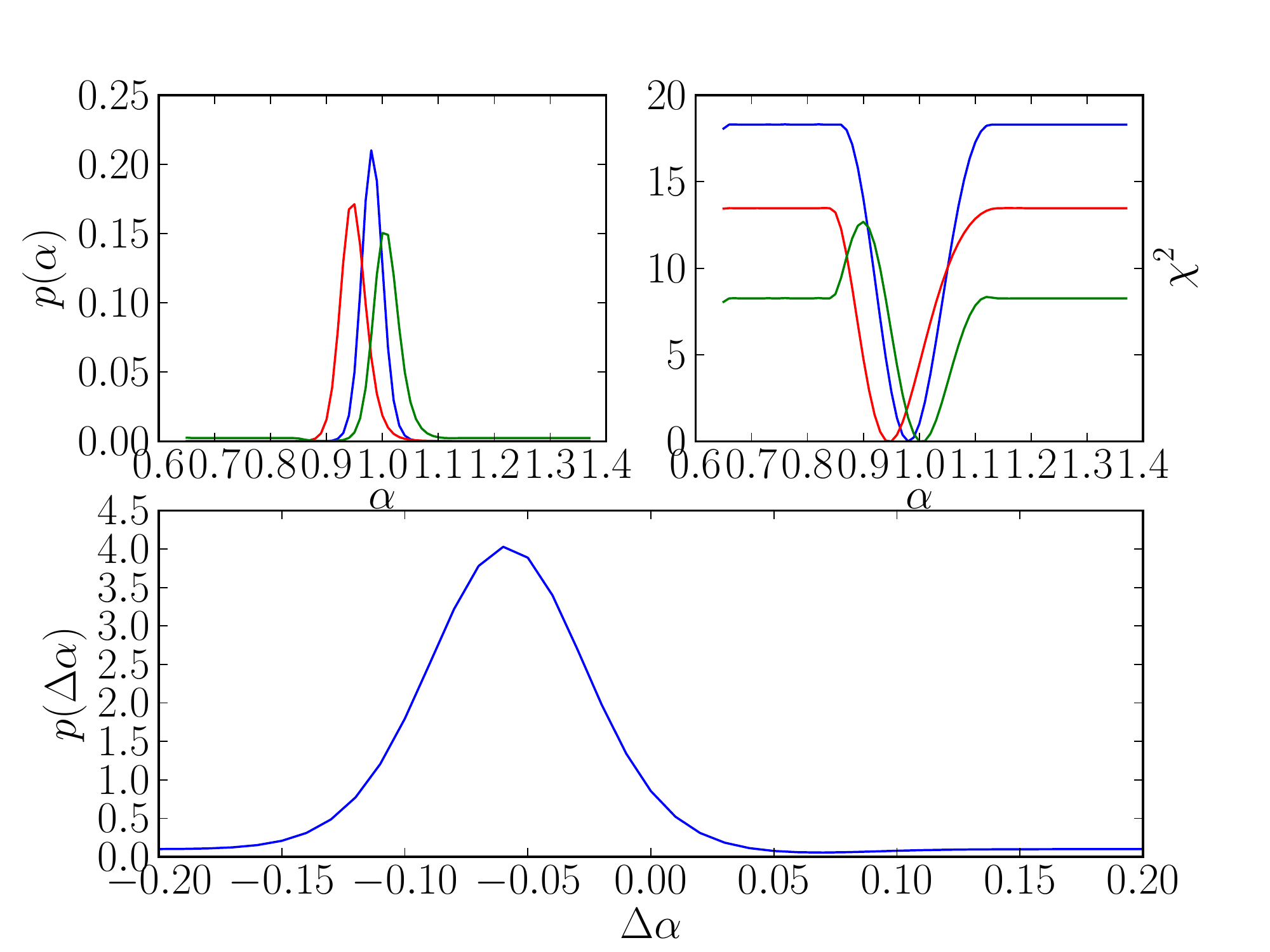} 
 \caption{Values of $\Delta \chi^2$ for the cut data (red), complete
   data (blue) and difference (green) are plotted in the upper right
   corner. The implied probability distributions are plotted in the
   upper left corner. The inferred probability for $\Delta \aiso$ is
   plotted in the lower panel. See text for discussion.}
  \label{fig:diffplot}
\end{figure}

Assuming these two contributions to be independent, we can infer the
probability that $\aiso$ of the two distributions are different.  In
order to do this, we infer the marginalized probability distribution
for the difference between two $\aiso$. We plot results in Figure
\ref{fig:diffplot}. The probability distribution for $\Delta \alpha$
has been calculated under the assumption that $0.8<\alpha<1.2$ and does
not go to zero away from the regions on high probability, because the
cut data are consistent with any $\aiso$ at a fixed $\Delta \chi2 \sim
9$. This plot shows that data are consistent with $\Delta \aiso=0$ at
$\sim 10\%$ level.

\end{document}